\documentclass[12pt]{article}

\usepackage{amssymb}
\usepackage{amsmath}
\usepackage{amscd}
\usepackage{latexsym}

\usepackage{graphicx}

\newcommand{\be}{\begin{equation}}
\newcommand{\ee}{\end{equation}}
\newcommand{\Dlt}{\Delta}
\newcommand{\dlt}{\delta}
\newcommand{\prt}{\partial}
\newcommand{\br}{{\bf r}}
\newcommand{\bk}{{\bf k}}
\newcommand{\bfe}{{\bf e}}
\newcommand{\bq}{{\bf q}}
\newcommand{\bj}{{\bf j}}
\newcommand{\bp}{{\bf p}}
\newcommand{\bP}{{\bf P}}
\newcommand{\bv}{{\bf v}}
\newcommand{\bS}{{\bf S}}
\newcommand{\bB}{{\bf B}}
\newcommand{\bH}{{\bf H}}
\newcommand{\bt}{\beta}
\newcommand{\vp}{\varphi}
\newcommand{\ep}{\varepsilon}
\newcommand{\al}{\alpha}
\newcommand{\ra}{\rightarrow}
\newcommand{\sgm}{\sigma}

\newcommand{\gm}{\gamma}
\newcommand{\om}{\omega}
\newcommand{\Om}{\Omega}
\newcommand{\Gm}{\Gamma}
\newcommand{\dgr}{\dagger}
\newcommand{\lbd}{\lambda}
\newcommand{\Lbd}{\Lambda}

\newcommand{\cH}{{\cal H}}

\newcommand{\cM}{{\cal M}}
\newcommand{\rgl}{\rangle}
\newcommand{\lgl}{\langle}

\begin{document}

\begin{center}
{\Large{\bf Basics of Bose-Einstein Condensation} \\ [5mm]

                    V.I. Yukalov} \\ [3mm]

{\it Bogolubov Laboratory of Theoretical Physics, \\
   Joint Institute for Nuclear Research, Dubna, Russia \\
                        and \\
      National Institute of Optics and Photonics, \\
University of S\~ao Paulo, S\~ao Carlos, Brazil}
\end{center}

\vskip 2cm

\begin{abstract}

The review is devoted to the elucidation of the basic problems 
arising in the theoretical investigation of systems with 
Bose-Einstein condensate. Understanding these challenging problems 
is necessary for the correct description of Bose-condensed
systems. The principal problems considered in the review are as 
follows: (i) What is the relation between Bose-Einstein condensation 
and global gauge symmetry breaking? (ii) How to resolve the 
Hohenberg-Martin dilemma of conserving versus gapless theories?
(iii) How to describe Bose-condensed systems in strong spatially 
random potentials? (iv) Whether thermodynamically anomalous 
fluctuations in Bose systems are admissible? (v) How to create 
nonground-state condensates? Detailed answers to these questions
are given in the review. As examples of nonequilibrium condensates, 
three cases are described: coherent modes, turbulent superfluids,
and heterophase fluids.
\end{abstract} 

\vskip 3cm


\newpage

\begin{center}
{\Large{\bf Contents}}
\end{center} 

\vskip 3mm
{\bf 1. Principal Theoretical Problems}  

\vskip 3mm
{\bf 2. Criteria of Bose-Einstein Condensation}

   2.1  Einstein Criterion

   2.2  Yang Criterion

   2.3  Penrose-Onsager Criterion

   2.4  Order Indices

   2.5  Condensate Existence

\vskip 3mm
{\bf 3. Gauge Symmetry Breaking}

\vskip 3mm

   3.1  Gauge Symmetry

   3.2  Symmetry Breaking

   3.3  Ginibre Theorem

   3.4  Bogolubov Theorem

   3.5  Roepstorff Theorem

\vskip 3mm
{\bf 4. General Self-Consistent Approach}

\vskip 3mm

   4.1  Representative Ensembles

   4.2  Bogolubov Shift

   4.3  Grand Hamiltonian

   4.4  Variational Principle

   4.5  Equations of Motion

\vskip 3mm
{\bf 5. Superfluidity in Quantum Systems}

\vskip 3mm

   5.1  Superfluid Fraction

   5.2  Moment of Inertia

   5.3  Equivalence of Definitions

   5.4  Local Superfluidity

   5.5  Superfluidity and Condensation

\vskip 3mm
{\bf 6. Equilibrium Uniform Matter}

\vskip 3mm

   6.1  Information Functional

   6.2  Momentum Representation

   6.3  Condensate Fraction

   6.4  Green Functions

   6.5  Hugenholtz-Pines Relation

\vskip 3mm
{\bf 7. Hartree-Fock-Bogolubov Approximation}

\vskip 3mm

   7.1  Nonuniform Matter

   7.2  Bogolubov Transformations

   7.3  Uniform Matter

   7.4  Local-Density Approximation

   7.5  Particle Densities

\vskip 3mm
{\bf 8. Local Interaction Potential}

\vskip 3mm

   8.1  Grand Hamiltonian

   8.2  Evolution Equations

   8.3  Equilibrium Systems

   8.4  Uniform Systems

   8.5  Atomic Fractions

\vskip 3mm
{\bf 9. Disordered Bose Systems}

\vskip 3mm

   9.1  Random Potentials

   9.2  Stochastic Decoupling

   9.3  Perturbation-Theory Failure 

   9.4  Local Correlations

   9.5  Bose Glass

\vskip 3mm
{\bf 10. Particle Fluctuations and Stability}

\vskip 3mm

   10.1  Stability Conditions

   10.2  Fluctuation Theorem

   10.3  Ideal-Gas Instability

   10.4  Trapped Atoms

   10.5  Interacting Systems 

\vskip 3mm
{\bf 11. Nonground-State Condensates}

\vskip 3mm

   11.1  Coherent Modes

   11.2  Trap Modulation

   11.3  Interaction Modulation

   11.4  Turbulent Superfluid

   11.5  Heterophase Fluid

\vskip 3mm
{\bf 12. Conclusions} 

\newpage

\section{Principal Theoretical Problems}

In recent years, the topic of Bose-Einstein condensation has been 
attracting very high attention. There have been published the 
books [1,2] and a number of review articles (e.g. [3-12]). This 
great attention is mainly due to a series of beautiful experiments 
with trapped atoms, accomplished in many laboratories of different 
countries and promising a variety of interesting applications. The 
interpretation of experiments requires the development of theory. 
It is well known that there is nothing more practical than a good 
theory. Only a correct theory allows for the proper understanding 
of experiments and can suggest appropriate and realistic technical 
applications.

The theory of real systems with Bose-Einstein condensate was advanced 
by Bogolubov [13-16] who considered {\it uniform weakly nonideal 
low-temperature} Bose gas. Extensions to {\it nonuniform 
zero-temperature weakly interacting} gas are due to Gross [17-19], 
Ginzburg and Pitaevskii [20], and Pitaevskii [21]. This approach has 
been the main tool for describing Bose-condensed systems, since the 
majority of initial experiments with trapped atoms had been 
accomplished with weakly interacting Bose gases at low temperatures, 
using the techniques of cooling and trapping [22].  

Since London [23], it is assumed that superfluidity in $^4$He is 
accompanied by Bose-Einstein condensation, although detecting the 
condensate fraction in helium is a rather difficult experimental task. 
The existence in liquid helium of Bose-Einstein condensate with zero 
momentum has not been directly proved, without model assumptions, 
though the majority of experiments are in agreement with the existence 
of condensate fraction of about $10 \%$ [24]. The possibility that 
superfluidity is accompanied by mid-range atomic correlations [25-27] 
or that it is due to the appearance in superfluid helium of a condensate 
with a finite modulus of momentum [28-31] have also been discussed. 
In his works on superfluid helium, Landau [32] has never assumed the 
condensate existence. This is why the direct observation of Bose-Einstein 
condensation of trapped atoms has become so important and intensively 
studied phenomenon [1-12].      

The trapped Bose gases are dilute and can be cooled down to very low 
temperatures. Usually, they also are weakly interacting. Thus, cold 
trapped atomic gases have become the ideal object for the application
of the Bogolubov theory [13-16].  

However, by employing the Feshbach resonance techniques [33,34] it is 
possible to vary atomic interactions, making them arbitrarily strong.
In addition, the properties of trapped gases at nonzero temperature
have also to be properly described. But the Bogolubov approximation,
designed for weakly interacting low-temperature systems, cannot be 
applied for Bose systems at finite interactions and temperature.   

Attempts to use the Hartree-Fock-Bogolubov approximation resulted 
in the appearance of an unphysical gap in the spectrum [35,36].
While there should be no gap according to the Hugenholtz-Pines 
[37] and Bogolubov [16] theorems. This gap cannot be removed without
loosing the self-consistency of theory, which would lead to the 
distortion of conservation laws and thermodynamic relations [16].
The situation was carefully analyzed by Hohenberg and Martin [38],
who showed that, as soon as the global gauge symmetry, associated 
with the Bose-Einstein condensation, is broken, any theory, in 
the frame of the grand canonical ensemble, becomes either 
nonconserving or acquires a gap in the spectrum. This dramatic 
conclusion is known as the Hohenberg-Martin dilemma of conserving 
versus gapless theories. In this review, it is shown how a correct 
self-consistent theory has to be developed, being both conserving
and gapless, and being valid for finite temperatures and arbitrary 
interactions.  

In the Bogolubov approach, the global gauge symmetry $U(1)$ is 
broken, which yields Bose-Einstein condensation. Hence, this gauge 
symmetry breaking is a sufficient condition for condensation. But 
maybe it is not necessary? Some researchers state that Bose-Einstein 
condensation does not require any symmetry breaking. This delusion 
is explained in the review, where it is emphasized that the gauge 
symmetry breaking is the necessary and sufficient condition for 
Bose-Einstein condensation. 

In recent literature on Bose systems, there often happens a very 
unfortunate mistake, when one omits anomalous averages, arising 
because of the gauge symmetry breaking. But it is straightforward 
to show that this omission is principally wrong from the precise
mathematical point of view. To get an excuse for the unjustified 
omission of anomalous averages, one ascribes such an omission to 
Popov, terming this trick "Popov approximation". Popov, however, 
has never suggested such an incorrect trick, which can be easily 
inferred from his original works [39,40]. 

The general self-consistent theory, presented in the review, is 
based on the Bogolubov shift of field operators, which explicitly
breaks the gauge symmetry. The theory is valid for arbitrary 
interacting Bose systems, whether equilibrium or nonequilibrium, 
uniform or nonuniform, in the presence of any external potentials, 
and at any temperature. External potentials of a special type are 
spatially random potentials. For treating the latter, one often 
uses perturbation theory with respect to disorder. However, it is 
possible to show that such perturbation theory can be misleading, 
yielding wrong results. In this review, a method is described 
that can be used for disorder potentials of any strength.    

One of the most confusing problems, widely discussed in recent 
literature, is the occurrence of thermodynamically anomalous particle 
fluctuations in Bose-condensed systems. In the review, a detailed 
explanation is given that such anomalous fluctuations cannot arise in
any real system, since their presence would make the system unstable, 
thus, precluding its very existence. The appearance of such anomalous 
fluctuations in some theoretical calculations is caused by technical 
mistakes. 

The usual Bose-Einstein condensate corresponds to the accumulation 
of particles on the ground-state level. An important problem, 
considered in the review, is whether it would be admissible to 
create nonground-state condensates. A positive answer is given and 
it is explained how this could be done and what would be the 
features of such condensates.

Throughout the paper, the system of units is employed, where the 
Planck constant $\hbar = 1$ and the Boltzmann constant $k_B =1$.

\section{Criteria of Bose-Einstein Condensation}

\subsection{Einstein Criterion}

Bose-Einstein condensation implies macroscopic accumulation of 
particles on the ground-state level of a Bose system. This means that, 
if the number of condensed particles is $N_0$ and the total number of 
particles in the system is $N$, then Bose-Einstein condensation occurs,
when $N_0$ is proportional to $N$. To formulate this criterion in a 
more precise way, it is necessary to invoke the notion of the 
thermodynamic limit, when the number of particles $N$, as well as the 
system volume $V$, tend to infinity, with their ratio remaining finite:
\be
\label{1}
N \ra \infty \; , \qquad V \ra \infty \; , \qquad \frac{N}{V}
\ra const \;   .
\ee 
Then the Einstein criterion is formulated as the limiting property
\be
\label{2}
\lim_{N\ra\infty} \; \frac{N_0}{N} \; > \; 0 \;   ,
\ee
where the thermodynamic limit (1) is assumed. This is a very general 
criterion that, however, does not hint on how the condensate particle 
number $N_0$ should be found.

\subsection{Yang Criterion}

The Yang criterion [41] is related to the notion of the off-diagonal 
long-range order related to the behavior of reduced density matrices [42].
The first-order reduced density matrix $\rho(\br,\br')$ defines
the limit
\be
\label{3}
\lim_{r\ra\infty} \rho(\br,0) = 
\lim_{N\ra\infty}\; \frac{N_0}{V} \;   ,
\ee
in which $r \equiv |{\bf r}|$. One says that this matrix displays the 
off-diagonal long-range order and Bose-Einstein condensation occurs, 
when
\be
\label{4}
 \lim_{r\ra\infty} \rho(\br,0) > 0 \; .
\ee
The Yang criterion can be useful for uniform systems, but is not suitable 
for confined systems, where the limit of $\rho(\br,0)$, as ${\bf r}\ra\infty$, 
is always zero, while condensation can happen [3,9].

\subsection{Penrose-Onsager Criterion}

Penrose and Onsager [43] showed that the occurrence of condensation is 
reflected in the eigenvalue spectrum of the single-particle density matrix. 
For the latter, the eigenproblem
\be
\label{5}
\int \rho(\br,\br') \vp_k(\br') \; d\br' = n_k \vp_k(\br)
\ee
defines the eigenfunctions $\varphi_k({\bf r})$ and eigenvalues $n_k$, 
labelled by a quantum multi-index $k$. The largest eigenvalue 
\be
\label{6}
N_0 \equiv \sup_k n_k
\ee
gives the number of condensed particles $N_0$. That is, condensation 
occurs, when
\be
\label{7}
\lim_{N\ra\infty}\; \frac{\sup_k n_k}{N} \; > \; 0 \;  .
\ee
This criterion is quite general and can be used for uniform as well as
for nonuniform systems.

\subsection{Order Indices}  

A convenient criterion can be formulated by means of the order indices
for reduced density matrices [44-47]. Order indices can be introduced for 
any operators possessing a norm and trace [48]. Let $\hat{A}$ be such an 
operator. Then the operator order index is defined [48] as
\be
\label{8}
 \om(\hat A) \equiv 
\frac{\log ||\hat A||}{\log|{\rm Tr}\hat A|} \; ,
\ee
where the logarithm can be taken to any base. Considering 
$\hat\rho_1 \equiv [\rho(\br,\br')]$ as a matrix with 
respect to the spatial variables results in the order index
\be
\label{9}
 \om(\hat\rho_1) \equiv 
\frac{\log ||\hat\rho_1||}{\log|{\rm Tr}\rho_1|} \;   .
\ee
Using the expressions
$$
|| \hat\rho_1|| = \sup_k n_k = N_0 \; , \qquad
{\rm Tr}\hat\rho_1 = N \;  ,
$$
yields the order index for the density matrix
\be
\label{10}
\om(\hat\rho_1) = \frac{\log N_0}{\log N} \;  .
\ee
This order index makes it possible to give the classification of 
different types of order: 
\begin{eqnarray}
\label{11}
\begin{array}{ll}
\om(\hat\rho_1) \leq 0 & \qquad (no \; order) \; , \\
0 < \om(\hat\rho_1) < 1 & \qquad (mid-range \; order) \; , \\
\om(\hat\rho_1) = 1 & \qquad (long-range \; order) \;  .
\end{array}
\end{eqnarray}
The latter corresponds to Bose-Einstein condensation, when
\be
\label{12}
 \lim_{N\ra\infty} \om(\hat\rho_1) = 1 \; ,
\ee
in agreement with the previous criteria. Generally, there can
exist Bose systems with mid-range order [45-48]. In such 
systems there is no Bose-Einstein condensate but there happens
a quasi-ordered state that can be called quasicondensate [39].

The order indices are useful in studying confined systems. But for
confined systems, the notion of thermodynamic limit is to be 
generalized. For this purpose, one has to consider {\it extensive}
observable quantities [49,50]. Let $A_N$ be such an observable 
quantity for a system of $N$ particles. The most general form of the 
thermodynamic limit can be given [12,51] as the limiting condition
\be
\label{13}
 N \ra \infty \; , \qquad A_N \ra \infty \; , \qquad
\frac{A_N}{N} \ra const \; .
\ee
Similar conditions with respect to the system ground-state energy 
imply the system {\it thermodynamic stability} [52].

\subsection{Condensate Existence}

The condensation criteria show that Bose-Einstein condensation
imposes the following restriction on the behavior of the 
density-matrix eigenvalues $n_k$. Recall that, by its definition, 
$n_k$ means the particle distribution over the quantum multi-indices 
$k$. According to Eqs. (6) and (7), one has
\be
\label{14}
 \frac{1}{\sup_k n_k} \propto \frac{1}{N} \ra 0 \qquad 
(N\ra\infty) \; .
\ee
If condensation occurs into the state labelled by the multi-index 
$k_0$, so that
$$
 \sup_k n_k = n_{k_0} \; ,
$$
then the {\it condensation condition} [12] is valid:
\be
\label{15}
 \lim_{k\ra k_0} \; \frac{1}{n_k} = 0 \qquad 
(N \ra \infty) \; .
\ee 
Writing $N \ra \infty$, implies, as usual, the thermodynamic limit
in one of the forms, either as in Eq. (1) or as in Eq. (13).

\section{Gauge Symmetry Breaking}

\subsection{Gauge Symmetry}

The global gauge symmetry $U(1)$ for a Hamiltonian $H[\psi]$, which 
is a functional of the field operator $\psi$, means that this 
Hamiltonian is invariant under the gauge transformation
\be
\label{16}
\psi(\br) \ra \psi(\br) e^{i\al}  \;  ,
\ee
where $\alpha$ is a real number. That is,
\be
\label{17}
H[ \psi e^{i\al} ] = H[\psi] \;  .
\ee
Here and in what follows, the time dependence of field operators 
is assumed but is not shown explicitly, when it is not important
and cannot lead to confusion.

The field operator can always be decomposed into an expansion
\be
\label{18}
\psi(\br) = \sum_k a_k \vp_k(\br)
\ee
over an orthonormal complete basis. Though, in general, the basis 
can be arbitrary, for what follows, it is important to choose the 
{\it natural basis}, composed of {\it natural orbitals} [42]. By 
definition, the basis is natural if and only if it is composed of 
the eigenfunctions of the single-particle density matrix, defined 
by the eigenproblem (5). Then the eigenvalues $n_k$ describe the 
particle distribution over the quantum indices $k$.

Bose-Einstein condensation can occur not to any state but only 
into one of the states of the natural basis, that is, into one of 
the natural orbitals. Denoting the related natural orbital by 
$\vp_0(\bf r)$, one can write
\be
\label{19}
\psi(\br) = \psi_0(\br) + \psi_1(\br)   ,
\ee
separating the part corresponding to condensate, 
\be
\label{20}
\psi_0(\br) \equiv a_0 \vp_0(\br) \;  ,
\ee
from the part related to uncondensed particles,
\be
\label{21}
\psi_1(\br) = \sum_{k\neq 0} a_k \vp_k(\br) \;  .
\ee

By construction, the condensate part is orthogonal to that of
uncondensed particles:
\be
\label{22}
\int \psi_0^\dgr(\br) \psi_1(\br)\; d\br = 0 \;  ,
\ee
which follows from the orthogonality of natural orbitals. And by
the definition of the natural orbitals as eigenfunctions of the 
single-particle density matrix, the {\it quantum-number conservation 
condition}
\be
\label{23}
\lgl a_k^\dgr a_p \rgl = \dlt_{kp} \lgl a_k^\dgr a_k \rgl
\ee
is valid. Because of the latter, one has the particular form of the 
{\it quantum conservation condition}
\be
\label{24}
\lgl \psi_0^\dgr(\br) \psi_1(\br) \rgl = 0 \;  .
\ee

The number-operator for condensed particles is
\be
\label{25}
\hat N_0 \equiv \int \psi_0^\dgr(\br) \psi_0(\br)\; d\br = 
a_0^\dgr a_0 \;  .
\ee
And the number-operator for uncondensed particles is
\be
\label{26}
\hat N_1 \equiv \int \psi_1^\dgr(\br) \psi_1(\br)\; d\br =
\sum_{k\neq 0} a_k^\dgr a_k \;  .
\ee
So that the total number-operator reads as
\be
\label{27}
\hat N = \hat N_0 + \hat N_1 \;  .
\ee  

The number of condensed particles is the statistical average
\be
\label{28}
N_0 \equiv \lgl \hat N_0 \rgl = \lgl a_0^\dgr a_0 \rgl \;  .
\ee
According to the condensation criteria, Bose-Einstein condensate
appears when
\be
\label{29}
 \lim_{N\ra\infty} \; \frac{\lgl a_0^\dgr a_0\rgl}{N} > 0 \; .
\ee

Till now, no symmetry breaking has been involved in the consideration.
Because of this, one could naively think that no symmetry breaking is
necessary for treating Bose condensation. However, the above 
consideration is yet nothing but a set of definitions. To understand
whether gauge symmetry breaking is compulsory for treating Bose 
condensation, one has to analyze the properties of the defined quantities.

\subsection{Symmetry Breaking}

There are several ways how the Hamiltonian symmetry could be broken.
The oldest method is by incorporating in the description of the system
an order parameter with a prescribed properties corresponding to a 
thermodynamic phase with the broken symmetry, as is done in mean-field 
approximations [32]. Another traditional way, advanced by Bogolubov 
[15,16], is by adding to the Hamiltonian symmetry-breaking terms, 
getting
\be
\label{30}
H_\ep[\psi] \equiv H[\psi] + \ep \Gm[\psi] \;  ,
\ee
where $\lgl\Gamma[\psi]\rgl_\ep\propto N$ and $\varepsilon$ is a small 
number. The statistical averages, with Hamiltonian (30), are denoted as 
$\lgl\cdots\rgl_\ep$. Upon calculating such an average, one should 
take, first, the thermodynamic limit $N\ra\infty$, after which the limit 
$\varepsilon \ra 0$. The so defined averages are called {\it quasiaverages}.   
It is also possible to combine these two limits in one, prescribing to
$\varepsilon$ a dependence on $N$ and taking the sole thermodynamic limit.
The latter procedure defines {\it thermodynamic quasiaverages} [53]. Other 
methods of symmetry breaking are described in the review [54]. Here, for 
concreteness, the standard way of symmetry breaking by means of the 
Bogolubov quasiaverages will be used.

{\it Spontaneous breaking of gauge symmetry} occurs when
\be
\label{31}
\lim_{\ep\ra 0} \; \lim_{N\ra\infty} \; \frac{1}{N}
\int | \lgl \psi_0(\br) \rgl_\ep |^2 d\br \; > \; 0 \;  .
\ee
This can also be rewritten as
\be
\label{32}
\lim_{\ep\ra 0} \; \lim_{N\ra\infty} \; 
\frac{|\lgl a_0\rgl_\ep|^2}{N} > 0 \; .
\ee
 
By the Cauchy-Schwarz inequality,
\be
\label{33}
|\lgl a_0 \rgl_\ep |^2 \leq \lgl a_0^\dgr a_0 \rgl_\ep
\ee
for any $\varepsilon$. This means that gauge symmetry breaking yields
Bose condensation. 

\vskip 2mm

{\bf Theorem 1}. {\it When gauge symmetry is spontaneously broken, then
there exists Bose-Einstein condensate}.

\vskip 2mm

{\it Proof.} Spontaneous breaking of gauge symmetry corresponds to 
Eq. (32). In view of the Schwarz inequality (33), it follows that
\be
\label{34}
\lim_{\ep\ra 0} \; \lim_{N\ra\infty} \;  
\frac{\lgl a_0^\dgr a_0\rgl_\ep}{N} > 0 \; ,
\ee
which implies Bose-Einstein condensation.

\subsection{Ginibre Theorem}

The Hamiltonian of a Bose system is a functional of the field 
operator $\psi$ that can always be represented as the sum (19) of 
two terms (20) and (21). Thus, Hamiltonian (30) is 
$H_{\varepsilon}[\psi] = H_{\varepsilon}[\psi_0,\psi_1]$. For an 
equilibrium system, this Hamiltonian defines the grand thermodynamic 
potential 
\be
\label{35}  
\Om_\ep \equiv - T \ln {\rm Tr} \; 
\exp \{ - \bt H_\ep[\psi_0,\psi_1] \} \;  ,
\ee
where $T$ is temperature and $\beta \equiv 1/T$. Let us replace the 
operator term $\psi_0$ by a nonoperator quantity $\eta$, getting
\be
\label{36}
 \Om_{\eta\ep} \equiv - T \ln {\rm Tr} \;
\exp \{ - \bt H_\ep[\eta,\psi_1] \} \; ,
\ee
and assuming that this thermodynamic potential is minimized with 
respect to $\eta$, so that
\be
\label{37}
\Om_{\eta\ep} = \inf_x \Om_{x\ep} \;  .
\ee
Ginibre [55] proved the following proposition.

\vskip 2mm

{\bf Theorem 2}. {\it In thermodynamic limit, the thermodynamic 
potentials (35) and (36) coincide}:
\be
\label{38}
\lim_{N\ra\infty} \; \frac{\Om_\ep}{N} =
\lim_{N\ra\infty} \; \frac{\Om_{\eta\ep}}{N} \;  .
\ee

\vskip 2mm

This theorem holds true irrespectively from whether there is Bose 
condensation or not. But if, when minimizing potential (36), one gets
a nonzero $\eta$, then, according to condition (31), there is 
spontaneous gauge symmetry breaking. Hence, because of Theorem 1, Bose
condensation occurs.

\subsection{Bogolubov Theorem}

Let 
\be
\label{39}
C_\ep(\psi_0,\psi_1) \equiv \lgl \ldots \psi_0^\dgr \ldots 
\psi_1^\dgr \ldots \psi_0 \ldots \psi_1 \rgl_\ep
\ee
be a class of correlation functions being the averages, with respect 
to the Hamiltonian $H_{\varepsilon}[\psi_0, \psi_1]$, of the normal 
products of the field operators (20) and (21). And let
\be
\label{40}
C_\ep(\eta,\psi_1) \equiv \lgl \ldots \eta^* \ldots 
\psi_1^\dgr \ldots \eta \ldots \psi_1 \rgl_{\eta\ep}
\ee
be a class of correlation functions being the averages, with respect 
to the Hamiltonian $H_{\varepsilon}[\eta, \psi_1]$, of the normal 
products of the field-operator terms, where the operators $\psi_0$ have
been replaced by a nonoperator quantity $\eta$ that minimizes the 
thermodynamic potential (36). Then the Bogolubov theorem [16] holds.

\vskip 2mm

{\bf Theorem 3}. {\it In thermodynamic limit, the corresponding 
correlation functions from classes (39) and (40) coincide}:
\be
\label{41}
\lim_{N\ra\infty} C_\ep(\psi_0,\psi_1) =
\lim_{N\ra\infty} C_\ep(\eta,\psi_1) \;  .
\ee

\vskip 2mm

As particular consequences from this theorem, it follows that
$$
\lim_{\ep\ra 0} \; \lim_{N\ra\infty} \; \frac{1}{N} \;
\int \lgl \psi_0^\dgr(\br) \psi_0(\br) \rgl_\ep d\br =
\lim_{N\ra\infty} \; \frac{1}{N} \; \int |\eta(\br)|^2 d\br \; ,
$$
\be
\label{42}
\lim_{\ep\ra 0} \; \lim_{N\ra\infty} \lgl \psi_0(\br)
\rgl_\ep = \eta(\br) \;   .
\ee
Invoking the conservation condition (24) yields
$$
\lim_{\ep\ra 0} \; 
\lim_{N\ra\infty} \lgl \psi_1(\br) \rgl_\ep = 0 \; ,
$$
\be
\label{43}
\lim_{\ep\ra 0} \; 
\lim_{N\ra\infty} \lgl \psi(\br) \rgl_\ep = \eta(\br) \;   .
\ee
Hence, if $\eta$ is not zero, the spontaneous gauge symmetry 
breaking takes place. Respectively, Bose condensation occurs. This 
important consequence of the Bogolubov theorem can be formulated 
as the following proposition.

\vskip 2mm

{\bf Theorem 4}. {\it Spontaneous gauge symmetry breaking implies 
Bose-Einstein condensation}:
\be
\label{44}
\lim_{\ep\ra 0} \; \lim_{N\ra\infty} \; 
\frac{|\lgl a_0\rgl_\ep|^2}{N}
= \lim_{\ep\ra 0} \; \lim_{N\ra\infty} \; 
\frac{\lgl a_0^\dgr a_0 \rgl_\ep}{N} \;  .
\ee

\subsection{Roepstorff Theorem}

The above theorems show that spontaneous gauge symmetry breaking 
is a sufficient condition for Bose-Einstein condensation. The fact 
that the former is also the necessary condition for the latter was, 
first, proved by Roepstorff [56] and recently the proof was polished 
in Refs. [57,58].

\vskip 2mm

{\bf Theorem 5}. {\it Bose-Einstein condensation implies spontaneous 
gauge symmetry breaking}:
\be
\label{45}
 \lim_{N\ra\infty} \; \frac{\lgl a_0^\dgr a_0\rgl}{N} 
\leq \lim_{\ep\ra 0} \; \lim_{N\ra\infty} \; 
\frac{|\lgl a_0\rgl_\ep|^2}{N} \; .
\ee

\vskip 2mm

In the left-hand side of inequality (45), the average is taken without
explicitly breaking the gauge symmetry. Combining theorems 4 and 5 
leads to the following conclusion:

\vskip 2mm

{\bf Conclusion}. {\it Spontaneous gauge symmetry breaking is the 
necessary and sufficient condition for Bose-Einstein condensation}.

\section{General Self-Consistent Approach}

\subsection{Representative Ensembles}

A statistical ensemble is a pair $\{\cal{F}, \hat {\rho}\}$ of the space
of microstates $\cal{F}$ and a statistical operator $\hat{\rho}$.
Defining the statistical operator, it is necessary to take into account
all conditions and constraints that uniquely describe the considered 
statistical system. This requirement was emphasized by Gibbs [59,60] and 
Ter Haar [61,62]. Such an ensemble is termed a {representative ensemble}.
The general formulation of the representative ensembles and their 
properties have been given in Refs. [54,63,64].

Constraints, imposed on the system, can be represented as the statistical 
averages of {\it condition operators} $\hat{C}_i$, with $i = 1,2,\ldots$ 
being the index enumerating the condition operators. This gives the set 
of {\it statistical conditions}    
\be
\label{46}
\lgl \hat C_i \rgl = C_i \;  .
\ee
Taking into account the latter defines the {\it grand Hamiltonian}
\be
\label{47}
 H = \hat H + \sum_i \lbd_i \hat C_i \; ,
\ee
in which $\hat{H}$ is the energy operator and $\lambda_i$ are Lagrange 
multipliers guaranteeing the validity of conditions (46).

\subsection{Bogolubov Shift}

The most convenient way of gauge symmetry breaking for Bose systems is
by means of the Bogolubov shift [16] of the field operator, when the 
field operator $\psi$ of a system without condensate is replaced by the 
field operator
\be
\label{48}
\hat\psi(\br) = \eta(\br) + \psi_1(\br) \;  ,
\ee
in which $\eta({\bf r})$ is the condensate wave function and the second 
term is the field operator of uncondensed particles. The latter is a 
Bose operator, with the standard commutation relations
$$
\left [ \psi_1(\br),\psi_1^\dgr(\br) \right ] =
\dlt(\br-\br') \;  .
$$
It is important to remember that the Fock space ${\cal F}(\psi)$, generated
by the operator $\psi$, is orthogonal to the Fock space ${\cal F}(\psi_1)$, 
generated by the operator $\psi_1$, so that after the Bogolubov shift (48)
it is necessary to work in the space ${\cal F}(\psi_1)$. Mathematical details 
can be found in Ref. [65]. 

Similarly to property (22), the condensate wave function is orthogonal to
the field operator of uncondensed particles:
\be
\label{49}
\int \eta^*(\br) \psi_1(\br) \; d\br = 0 \; .
\ee

The quantum-number conservation condition, analogous to Eqs. (24) and (43),
takes the form
\be
\label{50}
\lgl \psi_1(\br) \rgl = 0 \;  .
\ee
Then Eq. (48) yields
\be
\label{51}
 \lgl \hat\psi(\br) \rgl = \eta(\br) \; ,
\ee
which shows that the condensate function plays the role of an order 
parameter.

The condensate function is normalized to the number of condensed particles
\be
\label{52}
N_0 = \int |\eta(\br) |^2 d\br \;  .
\ee
The number of uncondensed particles gives another normalization condition
\be
\label{53}
N_1 = \lgl \hat N_1 \rgl \;  ,
\ee
where the number operator $\hat{N}_1$ is as in Eq. (26). The total number
operator
\be
\label{54}
\hat N \equiv \int \hat\psi^\dgr(\br) \hat\psi(\br) \; d\br = 
N_0 + \hat N_1
\ee
defines the total number of particles
\be
\label{55}
N = \lgl \hat N \rgl = N_0 + N_1 \; .
\ee

In the Bogolubov representation of the field operator (48), the 
condensate function and the field operator of uncondensed particles 
are two independent variables, orthogonal to each other.

\subsection{Grand Hamiltonian}

The general self-consistent theory to be presented in this and in 
the following sections, is based on Refs. [63-71], where all details 
can be found. 

In order to define a representative ensemble, one has to keep in 
mind the normalization conditions (52) and (53). The quantum-number 
conservation condition (50) is another restriction that is necessary 
to take into account. The latter equation can be rewritten in the 
standard form of a statistical condition by introducing the operator
\be
\label{56}
\hat\Lbd  \equiv \int \left [ \lbd(\br) \psi_1^\dgr(\br) +
\lbd^*(\br) \psi_1(\br) \right ] d\br \; , 
\ee
in which $\lambda({\bf r})$ is a complex function that accomplishes 
the role of a Lagrange multiplier guaranteeing the validity of the 
conservation condition (50). For this purpose, it is sufficient [71] 
to choose $\lambda({\bf r})$ such that to kill in the Grand Hamiltonian 
the terms linear in $\psi_1({\bf r})$. The conservation condition (50)
can be represented as
\be
\label{57}
 \lgl \hat\Lbd \rgl = 0 \; .
\ee

Taking into account the given statistical conditions (52), (53), 
and (57) prescribes the form of the grand Hamiltonian
\be
\label{58}
H = \hat H - \mu_0 N_0 - \mu_1 \hat N_1 - \hat\Lbd \;  ,
\ee
in which $\mu_0$ and $\mu_1$ are the related Lagrange multipliers
and $\hat{H} = \hat{H}[\eta,\psi_1]$ is the energy operator. The 
multiplier $\mu_0$ has the meaning of the condensate chemical potential
and $\mu_1$ can be called the chemical potential of uncondensed particles. 

The Hamiltonian average can be represented as
\be
\label{59}
\lgl H \rgl = \lgl \hat H \rgl - \mu N \;  ,
\ee
with $\mu$ being the system chemical potential. Then, from Eq. (58),
it follows that the chemical potential is
\be
\label{60}
\mu = \mu_0 n_0 + \mu_1 n_1 \;  ,
\ee
where the fractions of condensed and uncondensed particles,
\be
\label{61}
n_0 \equiv \frac{N_0}{N} \; , \qquad
n_1 \equiv \frac{N_1}{N}  \;  ,
\ee
are introduced.
  
It is necessary to stress that the number of Lagrange multipliers in
the grand Hamiltonian has to be equal to the number of imposed 
statistical conditions. Only then the statistical ensemble will be
representative. In the other case, the system would not be uniquely 
defined. Here, there are three conditions, the normalization 
conditions (52) and (53) and the conservation condition (57).

It is easy to show that the multipliers $\mu_0$ and $\mu_1$ do not
need to coincide. To this end, let us consider the thermodynamic 
stability condition requiring the extremization of the system free
energy $F = F(T,V,N_0,N_1)$, that is, $\delta F = 0$. This gives
\be
\label{62}
\dlt F = \frac{\prt F}{\prt N_0} \; \dlt N_0 +
\frac{\prt F}{\prt N_1} \; \dlt N_1 = 0 \;  .
\ee
Substituting here
\be
\label{63}
\mu_0 = \frac{\prt F}{\prt N_0} \; , \qquad
\mu_1 = \frac{\prt F}{\prt N_1}
\ee
transforms Eq. (62) to the equation
\be
\label{64}
 \mu_0 \dlt N_0 + \mu_1 \dlt N_1 = 0 \; .
\ee
The total number of particles $N = N_0 + N_1$ is assumed to be fixed,
so that $\delta N = 0$ and $\delta N_0 = - \delta N_1$. Then Eq. (64)
reduces to the relation
\be
\label{65}
(\mu_0 - \mu_1) \dlt N_1 = 0 \;  .
\ee
If $N_1$ were arbitrary, then one would have the equivalence of the 
multipliers $\mu_0$ and $\mu_1$. However, the number of uncondensed 
particles $N_1$ is fixed for each fixed $T, V$, and $N$. That is,
$\delta N_1 = 0$ and Eq. (65) is satisfied for any multipliers. 
Hence the multipliers $\mu_0$ and $\mu_1$ do not have to be equal.

It would be possible to say that $N_0$ is fixed for each given $T,V,N$.
But, clearly, this is the same as to say that $N_1$ is fixed. In any 
case, there always exist two normalization conditions requiring to 
introduce two related Lagrange multipliers. 

The Hamiltonian energy operator is
$$
\hat H = \int \hat\psi^\dgr(\br) \left ( -\; 
\frac{\nabla^2}{2m} + U \right ) \hat\psi(\br)\; d\br \; +
$$
\be
\label{66}
+ \; \frac{1}{2}\; \int \hat\psi^\dgr(\br) \hat\psi^\dgr(\br')
\Phi(\br-\br') \hat\psi(\br') \hat\psi(\br) \; d\br d\br' \;  ,
\ee
where $\Phi (-{\bf r}) = \Phi ({\bf r})$ is a pair interaction potential 
and $U = U({\bf r},t)$ is an external potential that, generally, can 
depend on time $t$. 

Substituting into the grand Hamiltonian (58) the shifted operator (48)
results in the form
\be
\label{67}
H = \sum_{n=0}^4 H^{(n)} \;  ,
\ee
whose terms are classified according to the order of the products of
the field operators $\psi_1$. The zero-order term
$$
H^{(0)} = \int \eta^*(\br) \left ( -\; \frac{\nabla^2}{2m} + U
- \mu_0 \right ) \eta(\br) \; d\br \; +
$$
\be
\label{68}
+ \; \frac{1}{2} \; \int \Phi(\br-\br') | \eta(\br')|^2 
| \eta(\br)|^2 d\br d\br'
\ee
does not contain the operators $\psi_1$. The first-order term 
\be
\label{69}
H^{(1)} = 0
\ee
is zero because of the conservation condition (57). The second-order
term is
$$
H^{(2)} = \int \psi_1^\dgr(\br) \left ( -\; \frac{\nabla^2}{2m} + U
- \mu_1 \right ) \psi_1(\br) \; d\br \; +
$$
$$
+\; \int \Phi(\br-\br') \left [ 
| \eta(\br)|^2 \psi_1^\dgr(\br') \psi_1(\br') +
\eta^*(\br) \eta(\br') \psi_1^\dgr(\br') \psi_1(\br) \; + \right.
$$
\be
\label{70}
\left. + \;
\frac{1}{2}\; \eta^*(\br) \eta^*(\br') \psi_1(\br') \psi_1(\br) +
\frac{1}{2}\; \eta(\br) \eta(\br') \psi_1^\dgr(\br') \psi_1^\dgr(\br) 
\right ] d\br d\br' \; .
\ee
Respectively, one has the third-order term
\be
\label{71}
H^{(3)} = \int \Phi(\br-\br') \left [ 
\eta^*(\br) \psi_1^\dgr(\br') \psi_1(\br')\psi_1(\br) +
\psi_1^\dgr(\br) \psi_1^\dgr(\br') \psi_1(\br') \eta(\br) 
\right ] d\br d\br'
\ee
and the fourth-order term
\be
\label{72}
H^{(4)} = \frac{1}{2} \; \int 
\psi_1^\dgr(\br) \psi_1^\dgr(\br') \Phi(\br-\br') 
\psi_1(\br')\psi_1(\br)\; d\br d\br' \; .
\ee

\subsection{Variational Principle}

In the Heisenberg representation, field operators satisfy the 
Heisenberg equation involving a commutator of the operator with the 
system Hamiltonian. At the same time, in quantum field theory, one 
usually gets the equations for the field operators by extremizing an 
action functional [72,73], which reduces to the variation of the 
Hamiltonian. Conditions, when these two methods are equivalent, are 
clarified in the following propositions.  

\vskip 2mm

{\bf Theorem 6}. {\it Let a field operator $\psi ({\bf r})$ be either
Bose or Fermi operator satisfying, respectively, the commutation or
anticommutation relations
\be
\label{73}
\left [ \psi(\br),\psi^\dgr(\br') \right ]_\mp = 
\dlt(\br-\br') \; , \qquad  
[\psi(\br),\psi(\br') ]_\mp = 0 \; ,
\ee
with the upper sign index being for Bose statistics while the lower, 
for Fermi statistics. Then for the products 
\be
\label{74}
P_{mn} \equiv P_m^+ P_n \; , \qquad 
P_m^+ \equiv \prod_{i=1}^m \psi^\dgr(\br_i) \; , \qquad
P_n \equiv \prod_{i=1}^n \psi(\br_i') \;  ,
\ee   
where $m$ and $n$ are real integers, one has the commutators}
\be
\label{75}
 [ \psi(\br), P_{mn} ] = \frac{\dlt P_{mn}}{\dlt\psi^\dgr(\br)}
+ \left [ (\pm 1)^{m+n} - 1 \right ] P_{mn} \psi(\br) \; .
\ee
 
\vskip 2mm

{\it Proof}. Using the variational derivative
$$
\frac{\dlt\psi^\dgr(\br_i)}{\dlt\psi^\dgr(\br)} =
\dlt(\br-\br_i) \;  ,
$$
it is straightforward to find
$$
\frac{\dlt P_1^+}{\dlt\psi^\dgr(\br)} = \dlt(\br-\br_1) \; ,
\qquad \frac{\dlt P_2^+}{\dlt\psi^\dgr(\br)} = 
\dlt(\br-\br_1) \psi^\dgr(\br_2) \pm 
\dlt(\br-\br_2) \psi^\dgr(\br_1) \; ,
$$
$$
\frac{\dlt P_3^+}{\dlt\psi^\dgr(\br)} = 
\dlt(\br-\br_1) \psi^\dgr(\br_2)\psi^\dgr(\br_3) \pm  
\dlt(\br-\br_2) \psi^\dgr(\br_1) \psi^\dgr(\br_3) +
\dlt(\br-\br_3) \psi^\dgr(\br_1) \psi^\dgr(\br_2) \; ,
$$
and so on. By induction, it follows that
\be
\label{76}
\frac{\dlt P_m^+}{\dlt\psi^\dgr(\br)} =
\sum_{j=1}^m \; (\pm 1)^{j+1} \dlt(\br-\br_j) 
\prod_{i(\neq j)}^m \psi^\dgr(\br_i) \; .
\ee

Using the commutator
$$
\left [ \psi(\br), \psi^\dgr(\br') \right ] =
\dlt(\br-\br') + ( \pm 1 - 1) \psi^\dgr(\br') \psi(\br) \;  ,
$$
we derive
$$
\left [ \psi(\br), P_1^+ \right ] = \dlt(\br-\br_1) +
(\pm 1 - 1) P_1^+ \psi(\br) \; ,
$$
$$
\left [ \psi(\br), P_2^+ \right ] = 
\dlt(\br-\br_1)\psi^\dgr(\br_2) \pm
\dlt(\br-\br_2)\psi^\dgr(\br_1) \; ,
$$
$$
\left [ \psi(\br), P_3^+ \right ] = 
\dlt(\br-\br_1)\psi^\dgr(\br_2) \psi^\dgr(\br_3) \pm
\dlt(\br-\br_2)\psi^\dgr(\br_1)\psi^\dgr(\br_3) +
$$
$$
+ \dlt(\br-\br_3)\psi^\dgr(\br_1)\psi^\dgr(\br_2 ) +
(\pm 1 -1) P_3^+ \psi(\br) \;  ,
$$
and so on. From here, using Eq. (76), by induction, we get
\be
\label{77}
 \left [ \psi(\br), P_m^+ \right ] = 
\frac{\dlt P_m^+}{\dlt\psi^\dgr(\br)} +
\left [ (\pm 1)^m -1 \right ] P_m^+ \psi(\br) \; .
\ee

Also, it is easy to check that
\be
\label{78}
[\psi(\br),P_n] = 
\left [ (\pm 1)^n -1 \right ] P_n \psi(\br) \;  .
\ee

Then, taking into account that, for any three operators 
$\hat{A}, \hat{B}, \hat{C}$, the equality
$$
\left [ \hat A,\hat B \hat C \right ] = 
\left [ \hat A,\hat B  \right ] \hat C +
 \hat B  \left [ \hat A,\hat C  \right ]
$$
is valid, we have
$$
\left [ \psi(\br),P_{mn} \right ] = 
\left [ \psi(\br),P_m^+ \right ] P_n +
P_m^+ [ \psi(\br), P_n] \;  .
$$
Substituting here Eqs. (77) and (78) gives the required Eq. (75). 

\vskip 2mm

{\bf Theorem 7}. {\it Let $\hat{F}[P_{mn}]$ be a linear functional
of the products defined in Eq. (74). And let the linear combination
\be
\label{79}
\hat F = \sum_{mn} c_{mn} \hat F [ P_{mn} ]
\ee
contain only such functionals for which, in the case of Bose 
statistics, $m$ and $n$ are arbitrary while, for the case of Fermi 
statistics, $m+n$ is even. Then
\be
\label{80}
\left [ \psi(\br), \hat F \right ] =
\frac{\dlt \hat F}{\dlt \psi^\dgr(\br)} \;  .
\ee
}
\vskip 2mm 

{\it Proof}. The proof is straightforward, following immediately from 
Eq. (75).

\vskip 2mm

The latter theorem shows that for a large class of functionals the 
commutator with the field operator is equivalent to the variational
derivative. The operators of observable quantities are in this class,
as well as Hamiltonians. This is because for Fermi systems, the field 
operators enter the observables always in pairs, which is necessary 
for spin conservation. This is why the Heisenberg equations for the 
field operators can be written in two equivalent ways, in the form 
of a commutator, as in the left-hand side of Eq. (80), or in the form 
of a variational derivative, as in the right-hand side of that equation.
Note that the standard form of many phenomenological evolution 
equations also involves variational derivatives [74,75].

\subsection{Evolution Equations}

With the grand Hamiltonian (58), the evolution equations for the field 
variables $\eta$ and $\psi_1$ read as
\be
\label{81}
 i\; \frac{\prt}{\prt t} \; \eta(\br,t) =
\frac{\dlt H}{\dlt\eta^*(\br,t)} \; ,
\ee
for the condensate function, and as
\be
\label{82}
 i\; \frac{\prt}{\prt t} \; \psi_1(\br,t) =
\frac{\dlt H}{\dlt\psi_1^\dgr(\br,t)} \;  ,
\ee
for the field operator of uncondensed particles. Recall that, in view 
of Theorem 7,
$$
\frac{\dlt H}{\dlt\psi_1^\dgr(\br,t)} = [ \psi_i(\br,t), H] \;  .
$$

Invoking expression (67) of the grand Hamiltonian gives the equation
$$
i\; \frac{\prt}{\prt t} \; \eta(\br,t) = \left (
- \; \frac{\nabla^2}{2m} + U - \mu_0 \right ) \eta(\br,t) \; +
$$
\be
\label{83}
 + \; \int \Phi(\br-\br') \left [ \hat X_0(\br,\br') +
\hat X(\br,\br') \right ] d\br' \; ,
\ee
in which the notations are introduced:
$$
\hat X_0(\br,\br') \equiv \eta^*(\br') \eta(\br') \eta(\br) \; ,
$$
$$
\hat X(\br,\br') \equiv \psi_1^\dgr(\br') \psi_1(\br')\eta(\br)
+ \psi_1^\dgr(\br')\eta(\br')\psi_1(\br) +
$$
\be
\label{84}
 + \eta^*(\br') \psi_1(\br') \psi_1(\br) +
\psi_1^\dgr(\br') \psi_1(\br') \psi_1(\br) \; .
\ee
In these expressions, for brevity, the explicit dependence on time
is not shown.

Equation (82) yields the equation for the field operator of 
uncondensed particles:
$$
i\; \frac{\prt}{\prt t} \; \psi_1(\br,t) = \left ( -\;
\frac{\nabla^2}{2m} + U - \mu_1 \right ) \psi_1(\br,t) \; +
$$
\be
\label{85}
 + \; \int \Phi(\br-\br') \left [ \hat X_1(\br,\br') +
\hat X(\br,\br') \right ] d\br' \;  ,
\ee
where
\be
\label{86}
\hat X_1(\br,\br') \equiv 
\eta^*(\br') \eta(\br') \psi_1(\br) +
\eta^*(\br')\psi_1(\br')\eta(\br) + 
\psi_1^\dgr(\br')\eta(\br')\eta(\br) \;  .
\ee

An equation for the condensate function follows from averaging 
Eq. (81), with the standard notation for a statistical average of 
an operator $\hat{A}$ as
$$
 \lgl \hat A(t) \rgl \equiv {\rm Tr}\hat\rho(0) \hat A(t) \; ,
$$
where $\hat{\rho}(0)$ is the statistical operator at the initial 
time $t=0$. So that the condensate-function equation is
\be
\label{87}
i\; \frac{\prt}{\prt t} \; \eta(\br,t) =  
\left \lgl \frac{\dlt H}{\dlt\eta^*(\br,t)} \right \rgl \; .
\ee
Averaging the right-hand side of Eq. (83), we shall need the 
notations for the single-particle density matrix
\be
\label{88}
 \rho_1(\br,\br') \equiv \lgl \psi_1^\dgr(\br') \psi_1(\br) \rgl  
\ee
and the anomalous density matrix
\be
\label{89}
\sgm_1(\br,\br') \equiv \lgl \psi_1(\br') \psi_1(\br) \rgl \;  .
\ee
The density of condensed particles is
\be
\label{90}
 \rho_0(\br) \equiv | \eta(\br) |^2  \; ,
\ee
while the density of uncondensed particles is
\be
\label{91}
\rho_1(\br) \equiv \rho_1(\br,\br) = 
\lgl \psi_1^\dgr(\br) \psi_1(\br) \rgl \;  .
\ee
The diagonal element of the anomalous density matrix,
\be
\label{92}
\sgm_1(\br) \equiv \sgm_1(\br,\br) = 
\lgl \psi_1(\br) \psi_1(\br) \rgl \; ,
\ee
defines the density of pair-correlated particles as $|\sgm_1(\br)|$.
The total density of particles in the system is the sum
\be
\label{93}
 \rho(\br) = \rho_0(\br) + \rho_1(\br) \; .
\ee
Also, we shall need the notation for the anomalous triple correlator
\be
\label{94}
\xi(\br,\br') \equiv 
\lgl \psi_1^\dgr(\br') \psi_1(\br') \psi_1(\br) \rgl \;  .
\ee
Employing these notations gives
$$
\hat X_0(\br,\br') = \rho_0(\br') \eta(\br) \; ,
$$
$$
 \lgl \hat X(\br,\br') \rgl = \rho_1(\br')\eta(\br) +
\rho_1(\br,\br')\eta(\br') + \sgm_1(\br,\br')\eta^*(\br') + 
\xi(\br,\br') \; .
$$
Finally, Eqs. (83) and (87) result in the equation for the condensate 
function  
$$
i\; \frac{\prt}{\prt t} \; \eta(\br,t) = \left ( -\;
\frac{\nabla^2}{2m} + U -\mu_0 \right ) \eta(\br,t) \; +
$$
\be
\label{95}
+ \; \int \Phi(\br-\br') \left [ \rho(\br') \eta(\br) +
\rho_1(\br,\br')\eta(\br') + \sgm_1(\br,\br')\eta^*(\br') +
\xi(\br,\br') \right ] d\br' \;  .
\ee

Equations for the densities can be obtained from the above equations, 
with introducing the condensate density of current
\be
\label{96}
\bj_0(\br,t) \equiv -\; \frac{i}{2m} \left [ \eta^*(\br)
\nabla\eta(\br) - \eta(\br) \nabla\eta^*(\br) \right ]
\ee
and the current density of uncondensed particles
\be
\label{97}
 \bj_1(\br,t) \equiv - \; 
\frac{i}{2m} \left \lgl \psi_1^\dgr(\br) \nabla\psi_1(\br) -
\left [ \nabla\psi_1^\dgr(\br) \right ] 
\psi_1(\br) \right \rgl \;   .
\ee
And let us also define the source term
\be
\label{98}
 \Gm(\br,t) \equiv i \int \Phi(\br-\br') \left [
\Xi^*(\br,\br') -\Xi(\br,\br') \right ] d\br' \; ,
\ee
with the anomalous correlation function
$$
\Xi(\br,\br') \equiv \eta^*(\br) \left [
\eta^*(\br') \sgm_1(\br,\br') +
\xi(\br,\br') \right ] \;   .
$$

Then we get the continuity equations for the condensate,
\be
\label{99}
\frac{\prt}{\prt t} \; \rho_0(\br,t) + \nabla \cdot \bj_0(\br,t)
=  \Gm(\br,t) \; ,
\ee
and for uncondensed particles,
\be
\label{100}
 \frac{\prt}{\prt t} \; \rho_1(\br,t) + 
\nabla \cdot \bj_1(\br,t) = -\Gm(\br,t) \; .
\ee
The total density (93) satisfies the continuity equation
\be
\label{101}
 \frac{\prt}{\prt t} \; \rho(\br,t) + \nabla \cdot \bj(\br,t)
= 0 \;  ,
\ee
with the total density of current
\be
\label{102}
 \bj(\br,t) = \bj_0(\br,t) + \bj_1(\br,t) \;  .
\ee

For the anomalous diagonal average (92), we find the equation
\be
\label{103}
i \; \frac{\prt}{\prt t} \; \sgm_1(\br,t) =
2 K(\br,t) + 2( U - \mu_1 ) \sgm_1(\br,t) +
2 \int \Phi(\br-\br') S(\br,\br',t) \; d\br' \;   ,
\ee
where the average anomalous kinetic-energy density is defined as
\be
\label{104}
K(\br,t) = - \; \frac{1}{2} \left \lgl 
\frac{\nabla^2\psi_1(\br)}{2m}\; \psi_1(\br) +
\psi_1(\br) \; \frac{\nabla^2\psi_1(\br)}{2m} \right\rgl  
\ee
and where we use the notation
$$
S(\br,\br',t) = \eta(\br) \eta(\br') \rho_1(\br,\br') +
\eta^*(\br') \eta(\br) \sgm_1(\br,\br') + 
\eta^*(\br')\eta(\br')\sgm_1(\br) +
$$
$$
+ \eta(\br) \xi(\br,\br') + 
\eta(\br')\lgl \psi_1^\dgr(\br') \psi_1(\br) \psi_1(\br) \rgl +
$$
\be
\label{105}
+ 
\eta^*(\br')\lgl \psi_1(\br') \psi_1(\br) \psi_1(\br) \rgl +
\lgl \psi_1^\dgr(\br') \psi_1(\br') \psi_1(\br) \psi_1(\br) \rgl 
+ [ \eta^2(\br)  + \sgm_1(\br) ] \dlt(\br-\br') \;  .
\ee

\section{Superfluidity in Quantum Systems}

\subsection{Superfluid Fraction}

One of the most important features of Bose-condensed systems is 
superfluidity. Therefore it is necessary to have a general definition
for calculating the superfluid fraction. Probably, the most general 
such a definition is by identifying the superfluid fraction as the 
fraction of particles nontrivially responding to a velocity boost.  

The systems Hamiltonian $H = H[\hat{\psi}]$ is a functional of the 
field operator $\hat{\psi}$. The operator of momentum is
\be
\label{106}
\hat\bP \equiv \int 
\hat\psi^\dgr(\br) \hat\bp \hat\psi(\br) \; d\br \;  ,
\ee
where $\hat{\bf p} \equiv -i \nabla$. 

Boosting the system with a velocity $\bv$ leads to the Galilean 
transformation of the field operators in the laboratory frame
\be
\label{107}
\hat\psi_v(\br,t) = \hat\psi(\br-\bv t) \; \exp \left \{
i \left ( m \bv \cdot \br \; - \; \frac{mv^2}{2} \; t 
\right ) \right \} \;  ,
\ee
expressed through the field operators $\hat{\psi}$ in the frame 
accompanying the moving system. Then the operator of momentum in 
the frame at rest,
\be
\label{108}
\hat\bP_v \equiv \int \psi^\dgr_v(\br) \hat\bp \hat\psi_v(\br)\;
d\br \;  ,
\ee
transforms into
\be
\label{109}
\hat\bP_v = \int \hat\psi^\dgr(\br) (\bp + m\bv )
\hat\psi(\br) \; d\br = \hat\bP + m\bv \hat N \; .
\ee

Since
$$
\frac{(\hat\bp+m\bv)^2}{2m} =
\frac{\hat\bp^2}{2m} + \bv \cdot \hat\bp + 
\frac{mv^2}{2} \;  ,
$$
the Hamiltonian $H_v = H[\hat{\psi}_v]$ for the moving 
system becomes
\be
\label{110}
 H_v = H + \int \hat\psi^\dgr(\br) \left ( \bv \cdot \hat\bp +
\frac{mv^2}{2} \right ) \hat\psi(\br) \; d\br \; .
\ee
 
The {\it generalized superfluid fraction} is defined through 
the ratio
\be
\label{111}
 n_s(\bv) \equiv 
\frac{\frac{\prt}{\prt\bv} \cdot \lgl \hat\bP_v\rgl_v}
{ \left \lgl \frac{\prt}{\prt\bv} \cdot \hat\bP_v\right\rgl_v} \; ,
\ee
in which the statistical averages $<\cdots>_v$ are determined for
the moving system with the Hamiltonian $H_v$, given in Eq. (110).
This definition is valid for any system, including nonequilibrium 
and nonuniform systems of arbitrary statistics.

One usually defines the superfluid fraction for a system at rest,
which gives
\be
\label{112}
n_s \equiv \lim_{v\ra 0} n_s(\bv) \;   .
\ee 

For equilibrium systems, the statistical averages are given by
the expressions
\be
\label{113}
\lgl \hat A \rgl_v \equiv \frac{{\rm Tr}\hat A\exp(-\bt H_v)}
{{\rm Tr}\exp(-\bt H_v)} \;  ,
\ee
for the moving system, and by
\be
\label{114}
\lgl \hat A \rgl \equiv 
\frac{{\rm Tr}\hat A e^{-\bt H} }{{\rm Tr} e^{-\bt H}} =
\lim_{v\ra 0} \lgl \hat A \rgl_v  \;  ,
\ee
for the system at rest.   

In the case of equilibrium systems, the derivatives over parameters
can be calculated according to the formulas of Ref. [76]. Thus, we 
have
\be
\label{115}
\frac{\prt}{\prt\bv} \cdot \lgl \hat\bP_v \rgl_v =
\left \lgl \frac{\prt}{\prt\bv} \cdot \hat\bP_v \right \rgl_v
- \bt {\rm cov} \left ( \hat\bP_v,\frac{\prt H_v}{\prt\bv}
\right ) \;  ,
\ee
where the covariance of any two operators, $\hat{A}$ and $\hat{B}$, 
is
$$
{\rm cov} \left ( \hat A,\hat B\right ) \equiv \frac{1}{2} \lgl
\hat A \hat B + \hat B \hat A \rgl_v - 
\lgl \hat A\rgl_v \lgl \hat B\rgl_v \;   .
$$

>From Eqs. (109) and (110), one has
$$
\frac{\prt}{\prt\bv} \cdot \hat\bP_v = 3m\hat N \; ,
\qquad \frac{\prt H_v}{\prt\bv} = \hat\bP_v \;  .
$$
Consequently, fraction (111) becomes
\be
\label{116}
n_s(\bv) = 1 \; - \; \frac{\Dlt^2(\hat\bP_v)}{3mNT} \;  ,
\ee
where the notation for an operator dispersion
$$
\Dlt^2(\hat A_v) \equiv \lgl \hat A_v^2 \rgl_v - 
\lgl \hat A_v \rgl^2_v   
$$
is used. Therefore, for fraction (112), Eq. (116) yields
\be
\label{117}
n_s = 1 \; - \; \frac{\Dlt^2(\hat\bP)}{3mNT} \;  ,
\ee
with the dispersion given as
$$
\Dlt^2(\hat A) \equiv \lgl \hat A^2 \rgl -
\lgl \hat A \rgl^2 \;  .
$$

The quantity
\be
\label{118}
Q \equiv \frac{\Dlt^2(\hat\bP)}{2mN}
\ee
describes the heat dissipated in the considered quantum system. While
the dissipated heat in the classical case reads as
\be
\label{119} 
Q_0 \equiv \frac{3}{2} \; T \;  .
\ee
Hence, the superfluid fraction (117) can be represented by the expression
\be
\label{120}
n_s = 1 \; - \; \frac{Q}{Q_0} \;  .
\ee
For an immovable system, the average momentum $<\hat{\bf P}>$ is zero. 
Then
$$
\Dlt^2(\hat\bP) = \lgl \hat\bP^2 \rgl \qquad
\left ( \lgl \hat\bP \rgl = 0 \right ) \;  .
$$
And the dissipated heat reduces to
\be
\label{121}
Q = \frac{\lgl\hat\bP^2\rgl}{2mN} \;  .
\ee

\subsection{Moment of Inertia}

Another way of defining the superfluid fraction is through the system 
response to rotation. The latter is connected with the angular momentum 
operator
\be
\label{122}
\hat{\bf L}  \equiv 
\int \hat\psi^\dgr(\br) (\br \times \hat\bp) 
\hat\psi(\br) \; d\br \; .
\ee
When the system is rotated with the angular velocity $\vec{\omega}$, the 
related linear velocity is 
\be
\label{123}
\bv_\om \equiv \vec{\om} \times \br \;   .
\ee
Then, in the laboratory frame, the angular momentum operator takes the form
\be
\label{124}
\hat{\bf L}_\om = \int \hat\psi^\dgr(\br) \left [ \br \times
(\hat\bp + m\bv_\om ) \right ] \hat\psi(\br) \; d\br \;  .
\ee
This, using the equality
$$
 \br \times (\vec{\om} \times \br ) = r^2 \vec{\om} -
(\vec{\om}\cdot\br ) \br \; ,
$$
gives
\be
\label{125}
 \hat{\bf L}_\om = \hat{\bf L} + m \int \hat\psi^\dgr(\br)\left [
r^2\vec{\om} - (\vec{\om} \cdot\br ) \br \right ] \hat\psi(\br)\;
d\br \;  .
\ee

The energy Hamiltonian of an immovable system can be written as the sum
\be
\label{126}
\hat H = \hat K + \hat V
\ee
of the kinetic energy operator 
\be
\label{127}
 \hat K \equiv \int \hat\psi^\dgr(\br) \; 
\frac{\hat\bp^2}{2m} \; \hat\psi(\br) \; d\br
\ee
and the potential energy part $\hat{V}$, respectively. 

Under rotation, the potential energy part does not change, but only the 
kinetic part varies, so that the energy Hamiltonian of a rotating system, 
in the laboratory frame, becomes
\be
\label{128}
 \hat H_\om = \hat K_\om + \hat V \; ,
\ee
with the same potential energy operator $\hat{V}$. The kinetic energy 
operator, in the laboratory frame, can be represented [77,78] by the 
formula
\be
\label{129}
 \hat K_\om = \int \hat\psi^\dgr(\br) \;
\frac{(\hat\bp+m\bv_\om)^2}{2m} \;  \hat\psi(\br) \; d\br \; .
\ee
In the rotating frame, where the system is at rest, the kinetic 
energy operator can be obtained from Eq. (129) with replacing 
$\vec{\omega}$ by $-\vec{\omega}$ and, respectively, replacing 
${\bf v}_\omega$ by $-{\bf v}_\omega$. Using the relations
$$
(\vec{\om} \times \br )^2 = \om^2 r^2 - 
(\vec{\om} \cdot \br)^2 \; , \qquad   
(\vec{\om} \times \br ) \cdot \hat\bp = 
\vec{\om} \cdot (\br \times \hat\bp) 
$$
allows us to represent the kinetic energy operator (129) as
\be
\label{130}
 \hat K_\om = \hat K + \vec{\om} \cdot \hat{\bf L} +
\frac{m}{2} \int \hat \psi(\br) \left [ \om^2 r^2 -
(\vec{\om}\cdot\br )^2 \right ] \hat\psi(\br) \; d\br \; .
\ee 
Thus, the energy Hamiltonian (128), in the laboratory frame, takes 
the form
\be
\label{131}
\hat H_\om = \hat H + \vec{\om} \cdot \hat{\bf L}  +
\frac{m}{2} \int \hat \psi(\br) \left [ \om^2 r^2 -
(\vec{\om}\cdot\br )^2 \right ] \hat\psi(\br) \; d\br \; .
\ee

Rotating systems are characterized by the inertia tensor
\be
\label{132} 
 \hat I_{\al\bt} \equiv \frac{\prt \hat L_\om^\al}{\prt\om_\bt}  
\ee
that, in view of Eq. (125), reads as
\be
\label{133}
\hat I_{\al\bt} = m \int \hat\psi^\dgr(\br) \left (
r^2 \dlt_{\al\bt} - r_\al r_\bt 
\right ) \hat\psi(\br)\; d\br \;  .
\ee

If one chooses the axis z in the direction of the angular velocity,
so that 
\be
\label{134}
\vec{\om} = \om \bfe_z \;  ,
\ee
then the angular momentum (125) is given by the expression
\be
\label{135}
 \hat L_\om^z = \hat L_z + \om \hat I_{zz} \; ,
\ee
with the inertia tensor
\be
\label{136}
\hat I_{zz} = m \int \hat\psi^\dgr(\br) \left ( x^2 + y^2 
\right ) \hat\psi(\br) \; d\br \;  ,
\ee
where the relation $r^2 - z^2 = x^2 + y^2$ is used. The energy 
Hamiltonian (128), characterizing the system energy in the laboratory 
frame, can be represented as
\be
\label{137}
\hat H_\om = \hat H + \om \hat L_z +
\frac{\om^2}{2} \; \hat I_{zz} \;  ,
\ee
with $\hat H$ from Eq. (126).

The generalized superfluid fraction is defined as
\be
\label{138}
n_s(\om) \equiv 
\frac{\frac{\prt}{\prt\om} \lgl \hat L_\om^z\rgl_\om}
{\left\lgl \frac{\prt}{\prt\om} \; \hat L_\om^z \right\rgl_\om} \;  .
\ee

For an equilibrium system, we can again employ the formulas of 
differentiation over parameters [76], leading to the derivative
\be
\label{139}
\frac{\prt}{\prt\om} \; \lgl \hat L_\om^z \rgl_\om =
\left \lgl \frac{\prt}{\prt\om}\; \hat L_\om^z \right \rgl_\om -
\bt {\rm cov} \left ( \hat L_\om^z, \frac{\prt\hat H_\om}{\prt\om}
\right ) \;  .
\ee
Substituting here
$$
\frac{\prt\hat L_\om^z}{\prt\om} = \hat I_{zz} \; , \qquad
\frac{\prt\hat H_\om}{\prt\om} = \hat L_\om^z \;  ,
$$
we come to the expression
\be
\label{140}
n_s(\om) = 1 \; - \; 
\frac{\Dlt^2(\hat L_\om^z )}{T\lgl \hat I_{zz} \rgl_\om} \; .
\ee

Considering the superfluid fraction in the nonrotating limit,
\be
\label{141}
n_s \equiv \lim_{\om\ra 0} n_s(\om) \;  ,
\ee
and using the notation
\be
\label{142}
 I_{zz} \equiv \lim_{\om\ra 0} \lgl \hat I_{zz} \rgl_\om =
m \int \left ( x^2 + y^2 \right ) \rho(\br) \; d\br \; ,
\ee
we obtain the superfluid fraction in the form
\be
\label{143}
 n_s = 1 \; - \; \frac{\Dlt^2(\hat L_z)}{T I_{zz}} \; .
\ee
The dispersion of $\hat{L}_z$ is calculated with the Hamiltonian
for a nonrotating system.

Introducing the notation
\be
\label{144}
I_{eff} \equiv \bt \Dlt^2(\hat L_z)   
\ee
allows us to represent the superfluid fraction (143) as
\be
\label{145}
 n_s = 1 \; - \; \frac{I_{eff}}{I_{zz}} \; .
\ee
For a nonrotating system, one has
$$
 \Dlt^2(\hat L_z) = \lgl \hat L_z^2 \rgl \qquad 
( \lgl \hat L_z \rgl = 0 ) \; .
$$
Hence $I_{eff} = \beta <\hat{L}_z^2>$.

\subsection{Equivalence of Definitions}

The definitions of the superfluid fraction, considered in Sec. 5.1 
and in Sec. 5.2, are equivalent with each other. To show this, one
can take a cylindrical annulus of radius $R$, width $\delta$, and 
length $L$, such that $\delta \ll R$. The volume of this annulus is 
$V \simeq 2 \pi R L \delta$. Then the classical inertia tensor (142) 
is $I_{zz} \simeq m N R^2$. The angular momentum (122) can be written 
as
\be
\label{146}
\hat L_z = \int \hat\psi^\dgr(\br) \left ( - i\; 
\frac{\prt}{\prt\vp} \right ) \hat\psi(\br) \; d\br \; ,
\ee
where $\vp$ is the angle of the cylindrical system of coordinates.

For the annulus of large radius $R$, making the round along the 
annulus circumference, one has the path element 
$\delta l = R \delta \vp$. Therefore the angular momentum (146) can 
be represented as 
\be
\label{147}
 \hat L_z = R \hat P_l \; ,
\ee
being proportional to the momentum
\be
\label{148}
\hat P_l \equiv \int \hat\psi^\dgr(\br) \left ( -i \;
\frac{\prt}{\prt l} \right ) \hat\psi(\br) \; d\br \;  .
\ee
Then the superfluid fraction (143) becomes
\be
\label{149}
n_s = 1 \; - \; \frac{\Dlt^2(\hat P_l)}{mNT} \;  .
\ee
The same formula follows from the consideration of Sec. 5.1, 
if one takes the velocity boost along the annulus circumference.

\subsection{Local Superfluidity}

In some cases, it is important to know the spatial distribution of the 
superfluid fraction that would be given by the spatial dependence 
$n_s({\bf r})$. This can be necessary, when one considers equilibrium 
nonuniform systems or systems in local equilibrium [79,80].    

To describe local superfluidity, we can consider the momentum density
\be
\label{150}
\hat\bP(\br) \equiv \hat\psi^\dgr(\br) \hat\bp\hat\psi(\br) \;  .
\ee
Following Sec. 5.1, we introduce a velocity boost, which leads to the 
momentum density
\be
\label{151}
\hat\bP_v(\br) \equiv \hat\psi^\dgr(\br) (\hat\bp + m\bv ) 
\hat\psi(\br)
\ee
in the laboratory frame. The local superfluid fraction is defined as
\be
\label{152}
 n_s(\br) \equiv \lim_{v\ra 0} \;
\frac{ \frac{\prt}{\prt\bv}\cdot\lgl\hat\bP_v(\br)\rgl_v }
{ \left\lgl \frac{\prt}{\prt\bv}\cdot\hat\bP_v(\br)\right\rgl_v } \;  .
\ee
Because of form (151), one has
$$
\frac{\prt}{\prt\bv} \cdot \hat\bP_v(\br) = 
3m \hat\psi^\dgr(\br) \hat\psi(\br) \;  .
$$
Then the local superfluid fraction (152) reduces to
\be
\label{153}
n_s(\br) = 1 \; - \; 
\frac{{\rm cov}(\hat\bP(\br),\hat\bP)}{3m\rho(\br) T} \;  .
\ee
Owing to the relation
\be
\label{154}
 \rho_s(\br) = n_s(\br) \rho(\br) \; ,
\ee
we get the local superfluid density
\be
\label{155}
 \rho_s(\br) = \rho(\br) \; - \; 
\frac{{\rm cov}(\hat\bP(\br),\hat\bP)}{3m T} \; .
\ee
Integrating the above equation over ${\bf r}$ and considering the 
average fraction
$$
 n_s = \frac{1}{N} \int  \rho_s(\br) \; d\br 
$$
would bring us back to formula (117).

\subsection{Superfluidity and Condensation}

Usually, Bose-Einstein condensation is accompanied by superfluidity. 
However, there is no straightforward relation between these phenomena
and the related fractions [3,12]. Thus, in two-dimensional systems
at finite temperature, there is no Bose condensation, but there can 
exist superfluidity. And in spatially random systems, there can happen 
local Bose condensation without superfluidity. 

The relation between Bose condensation and superfluidity depends on the 
type of the effective particle spectrum and system dimensionality. To 
illustrate this, let us consider a $d$-dimensional Bose gas with an 
effective particle spectrum
\be
\label{156}
 \om_k = A k^n - \mu \; ,
\ee
where $A$ and $n$ are positive parameters and $\bk$ is $d$-dimensional
momentum. For the $d$-dimensional case, the superfluid fraction (117) 
takes the form
\be
\label{157}
n_s = 1 \; - \; \frac{\lgl \hat\bP^2 \rgl}{N m T d} \;  .
\ee
The integration over the $d$-dimensional momenta involves the relation
$$
\frac{d\bk}{(2\pi)^d} \; \ra \; 
\frac{2k^{d-1}dk}{(4\pi)^{d/2}\Gm(d/2)} \;  ,
$$
in which $\Gamma(x)$ is the gamma function.
 
For the condensation temperature, we find
\be
\label{158}
 T_c = A \left [ 
\frac{(4\pi)^{d/2}\Gm(d/2)n\rho}{2\Gm(d/n)\zeta(d/n)} 
\right ]^{n/d} \; ,
\ee
where $\zeta(x)$ is the Riemann zeta function. The latter, can be 
represented in several forms:
\be
\label{159}
\zeta(x) = \sum_{j=1}^\infty \; \frac{1}{j^x} =
\frac{1}{\Gm(x)} \int_0^\infty \; 
\frac{t^{x-1}}{e^t-1}\; dt \;  ,
\ee
when ${\rm Re} x > 1$, and
\be
\label{160}
 \zeta(x) = \frac{1}{(1-2^{1-x})\Gm(x)} \int_0^\infty\;
\frac{u^{x-1}}{e^u+1}\; du \; ,
\ee
if ${\rm Re} x > 0$.
 
Taking into account that $\Gamma(x)>0$ for $x>0$ and $\zeta(x)<0$
in the interval $0 < x < 1$ tells us that there is no condensation 
for $d < n$. When $d = n$, then $T_c = 0$. And $T_c>0$ for $d>n$. 

For $d > n$, the condensate fraction below $T_c$ is given by the 
expression
\be
\label{161}
 n_0 = 1 - \left ( \frac{T}{T_c} \right )^{d/n} \qquad
(T \leq T_c) \; ,
\ee
while the superfluid fraction, under $\mu = 0$, is
\be
\label{162}
 n_s = 1 - B\zeta\left ( \frac{d+2-n}{n} 
\right ) T^{(d+2-n)/n} \; ,
\ee
where
\be
\label{163}
 B \equiv \frac{2(d+2-n)\Gm\left(\frac{d+2-n}{n}\right )}
{(4\pi)^{d/2}\Gm\left(\frac{d}{2}\right)A^{(d+2)/n}m\rho n^2d} \; .
\ee
If there is no condensate, then $\mu$ is defined by the equation
\be
\label{164}
\rho = \frac{2\Gm(d/n)g_{d/n}(z) T^{d/n}}
{(4\pi)^{d/2}\Gm(d/2)nA^{d/n}}   ,
\ee
in which $z \equiv e^{\beta\mu}$ is fugacity and
$$
g_n(z) \equiv \frac{1}{\Gm(n)} \int_0^\infty\;
\frac{zu^{n-1}}{e^u-z}\; du
$$
is the Bose function. The superfluid fraction, in the absence 
of condensate, is 
\be
\label{165}
n_s = 1 - Bg_{(d+2-n)/n}(z) T^{(d+2-n)/n}   .
\ee 

Generally speaking, Bose-Einstein condensation is neither necessary 
nor sufficient for superfluidity. These phenomena are connected with
different system features. Bose condensation implies the appearance
of coherence in the system, while superfluidity is related to the 
presence of sufficiently strong pair correlations. Thus, there can 
occur four possibilities, depending on the values of the condensate 
and superfluid fractions: \\
(i) {\it incoherent normal fluid}
$$
   n_0 = 0 \; , \qquad n_s = 0 ;
$$
(ii) {\it coherent normal fluid}
$$
   n_0 > 0 \; , \qquad n_s = 0 ;
$$
(iii) {\it incoherent superfluid}
$$
   n_0 = 0 \; , \qquad n_s > 0 ;
$$
(iv) {\it coherent superfluid}
$$
   n_0 > 0 \; , \qquad n_s > 0 .
$$
In this classification, we do not take into account that the system
can form a solid [12].

\section{Equilibrium Uniform Systems}

\subsection{Information Functional}

The definition of statistical averages involves the use of a statistical 
operator. The form of the latter, in the case of an equilibrium system, 
can be found from the {\it principle of minimal information}. This 
principle requires that, composing an information functional, one has 
to take into account all conditions and constraints that uniquely define 
the considered system [81]. Only then the corresponding statistical 
ensemble will be representative and will correctly describe the system. 
In the other case, if not all necessary constraints have been taken into 
account, so that the system is not uniquely described, the ensemble is not 
representative and cannot correctly characterize the system. In such a 
case, one confronts different problems, for instance, the occurrence of 
thermodynamic instability or nonequivalence of ensembles. However, all 
those problems are caused by the use of nonrepresentative ensembles and
have nothing to do with physics. A detailed discussion of these problems
can be found in Ref. [63]. The construction of representative ensembles 
for Bose-condensed systems is given in Refs. [64,71].

A statistical operator $\hat{\rho}$ of an equilibrium system should be 
the minimizer of the Shannon information $\hat{\rho} \ln \hat{\rho}$, 
under given statistical conditions. The first evident condition is the
normalization
\be
\label{166}  
\lgl \hat 1 \rgl \equiv {\rm Tr}\hat\rho = 1 \; ,
\ee
with $\hat{1}$ being the unity operator. Then, one defines the internal 
energy $E$ through the average
\be
\label{167}
\lgl \hat H \rgl \equiv {\rm Tr}\hat\rho \hat H = E \; .
\ee
The normalization condition (52) for the condensate function can also 
be presented in the standard form of a statistical condition as
\be
\label{168}
\lgl \hat N_0 \rgl \equiv {\rm Tr}\hat \rho \hat N_0 = N_0 \; ,
\ee
where $\hat{N}_0 \equiv N_0 \hat{1}$. Normalization (53), for the number 
of uncondensed particles, can be written as
\be
\label{169}
 \lgl \hat N_1 \rgl \equiv {\rm Tr}\hat \rho \hat N_1 = N_1 \; .
\ee
Finally, the conservation condition (57) reads as
\be
\label{170}
\lgl \hat\Lbd \rgl \equiv {\rm Tr}\hat \rho \hat\Lbd = 0 \; .
\ee

Note that, in general, the conditional operators do not need to be 
necessarily commutative with the energy operator [80]. For instance,
here the operator $\hat{N}_0$ commutes with 
$\hat{H}$, but $\hat{\Lambda}$ does not have to commute with the 
latter. 

It is also worth stressing that the average quantities, involved in
the statistical conditions, do not need to be directly prescribed, 
but they have to be uniquely defined by fixing other thermodynamic 
parameters. Thus, internal energy is not prescribed directly in 
either canonical or grand canonical ensembles, but it is uniquely 
defined through the fixed temperature, the number of particles in 
the system, and volume. Similarly, the number of condensed 
particles may be not directly given, but it is uniquely defined, 
and can be measured, by fixing other thermodynamic parameters, 
temperature, total number of particles, and volume. For confined 
systems, instead of volume, the external potential is given.

The information functional, under the above conditions, takes the 
form
$$
I[ \hat\rho ] =  {\rm Tr}\hat \rho \ln \hat\rho + 
\lbd_0({\rm Tr}\hat\rho -1 ) + \bt ({\rm Tr}\hat\rho \hat H - E) -
$$
\be
\label{171}
- \bt\mu_0 ({\rm Tr}\hat\rho \hat N_0 - N_0) -
 \bt\mu_1 ({\rm Tr}\hat\rho \hat N_1 - N_1) -
\bt {\rm Tr}\hat\rho \hat\Lbd \; ,
\ee
in which the corresponding Lagrange multipliers are introduced. 
Minimizing this functional with respect to $\hat{\rho}$ yields 
the statistical operator
\be
\label{172}
 \hat\rho = \frac{e^{-\bt H} }{{\rm Tr}e^{-\bt H} } \;  ,
\ee
with the same grand Hamiltonian (58).

\subsection{Momentum Representation}

For a uniform system, it is convenient to pass to the momentum 
representation by means of the Fourier transformation with plane waves.
This is because the plane waves are the natural orbitals for a uniform 
system, which implies that they are the eigenfunctions of the density 
matrix in the sense of eigenproblem (5).

The field operator of uncondensed particles transforms as
\be
\label{173}
\psi_1(\br) = \frac{1}{\sqrt{V}} \; 
\sum_{k\neq 0} a_k e^{i\bk\cdot\br} \; , \qquad 
a_k = \frac{1}{\sqrt{V}} \int 
\psi_1(\br) e^{-i\bk\cdot\br} d\br \; .
\ee
We assume that the pair interaction potential is Fourier transformable,
\be
\label{174}
\Phi(\br) = \frac{1}{V} \sum_k \Phi_k e^{i\bk\cdot\br} \; , \qquad
\Phi_k = \int \Phi(\br) e^{-i\bk\cdot\br} d\br \;  .
\ee

The condensate function $\eta({\bf r})$ for a uniform system, is a 
constant $\eta$, such that
\be
\label{175}
\rho_0(\br) = |\eta|^2 = \rho_0 \;  .
\ee

These transformations are substituted into the grand Hamiltonian (67).
Then the zero-order term (68) becomes
\be
\label{176}
H^{(0)} =\left ( \frac{1}{2}\; \rho_0 \Phi_0 -
\mu_0 \right ) N_0 \;  .
\ee
The first-order term $H_1$ is automatically zero, as in Eq. (69). The 
second-order term (70) reads as
$$
H^{(2)} = \sum_{k\neq 0} \left [ \frac{k^2}{2m} + \rho_0
(\Phi_0 + \Phi_k ) -\mu_1 \right ] a_k^\dgr a_k \; +
$$
\be
\label{177}
+\; \frac{1}{2} \sum_{k\neq 0} \rho_0 \Phi_k \left (
a_k^\dgr a_{-k}^\dgr + a_{-k} a_k \right ) \;  .
\ee
The third-order term (71) is    
\be
\label{178}
H^{(3)} = \sqrt{\frac{\rho_0}{V} } \; {\sum_{kp}}' \Phi_p 
\left ( a_k^\dgr a_{k+p} a_{-p} + a_{-p}^\dgr a_{k+p}^\dgr a_k 
\right ) \;  ,
\ee
where in the sum 
$$
 \bk \neq 0 \; , \qquad \bp \neq 0 \; , \qquad 
\bk + \bp \neq 0 \; .
$$
The fourth-order term (72) takes the form
\be
\label{179}
 H^{(4)} = \frac{1}{2V} \sum_q\; {\sum_{kp}}' 
\Phi_q a_k^\dgr a_p^\dgr a_{p+q} a_{k-q} \;  ,
\ee
where
$$
\bk \neq 0 \; , \qquad \bp \neq 0 \; , \qquad 
\bp + \bq \neq 0 \; ,\qquad \bk - \bq \neq 0 \;   .
$$

\subsection{Condensate Function}

In the case of an equilibrium system, the condensate function does not
depend on time,
\be
\label{180}
\frac{\prt}{\prt t} \; \eta(\br,t) = 0 \;  .
\ee
Therefore, Eq. (95) reduces to the eigenvalue problem
$$
\left [ -\; \frac{\nabla^2}{2m} + U(\br) \right ] \eta(\br) \; +
$$
\be
\label{181}
+ \; \int \Phi(\br-\br') [ \rho(\br') \eta(\br) +
\rho_1(\br,\br') \eta(\br') + \sgm_1(\br,\br') \eta^*(\br')
+ \xi(\br,\br') ] d\br' = \mu_0 \eta(\br) \;  .
\ee

A uniform system presupposes the absence of a nonuniform external 
potential. Hence, one can set $U = 0$. The average densities $\rho_0$ and 
$\rho_1$ are constant. The total particle density is
\be
\label{182}
 \rho = \rho(\br) = \rho_0 + \rho_1 \; .
\ee
Then Eq. (181) gives
\be
\label{183}
 \mu_0 = \rho \Phi_0 + \int \Phi(\br) \left [ \rho_1(\br,0) +
\sgm_1(\br,0) + \frac{\xi(\br,0)}{\sqrt{\rho_0}} \right ] d\br \; .
\ee

The normal density matrix is written as
\be
\label{184}
 \rho_1(\br,\br') = \frac{1}{V} \sum_{k\neq 0}
n_k e^{i\bk\cdot(\br-\br')} \; ,
\ee
where
\be
\label{185}
 n_k \equiv \lgl a_k^\dgr a_k \rgl \; .
\ee
And the anomalous average
\be
\label{186}
\sgm_1(\br,\br') = \frac{1}{V} \sum_{k\neq 0} 
\sgm_k e^{i\bk\cdot(\br-\br')}
\ee
is expressed through
\be
\label{187}
 \sgm_k \equiv \lgl a_k a_{-k} \rgl \; .
\ee
The triple anomalous correlator (94) can be represented as
\be
\label{188}
 \xi(\br,\br') = \frac{1}{V} 
\sum_{k\neq 0} \xi_k e^{i\bk\cdot(\br-\br')} \; ,
\ee
with
\be
\label{189}
\xi_k = \frac{1}{\sqrt{V} } \sum_{p\neq 0} 
\lgl a_k a_p a_{-k-p} \rgl  \;  .
\ee
 
The diagonal element of Eq. (184) gives the density of 
uncondensed particles 
\be
\label{190}
 \rho_1 = \rho_1(\br,\br) = \frac{1}{V} \sum_{k\neq 0} n_k \; .
\ee
The diagonal element of the anomalous average (186) is
\be
\label{191}
\sgm_1 = \sgm_1(\br,\br) = \frac{1}{V} \sum_{k\neq 0} \sgm_k \;  .
\ee
And the triple correlator (188) leads to
\be
\label{192}
\xi = \xi(\br,\br) = \frac{1}{V} \sum_{k\neq 0} \xi_k \;  .
\ee
The condensate chemical potential (183) can be rewritten in the form
\be
\label{193}
\mu_0 = \rho \Phi_0 + \frac{1}{V} \sum_{k\neq 0} \left (
n_k + \sgm_k + \frac{\xi_k}{\sqrt{\rho_0} } \right ) \Phi_k \; .
\ee

\subsection{Green Functions}

There are several types of Green functions. Here, we shall deal with 
the causal Green functions [81,82] that are called propagators. The 
set $\{{\bf r}_j, t_j\}$  of the spatial variable ${\bf r}_j$ and time 
$t_j$ will be denoted, for brevity, just as $j$. If there are other 
internal variables, they can also be included in the notation $j$.

For a Bose-condensed system, one considers four types of Green 
functions:
$$
G_{11}(12) = - i \lgl \hat T \psi_1(1) \psi_1^\dgr(2) \rgl \; , 
\qquad
G_{12}(12) = - i \lgl \hat T \psi_1(1) \psi_1(2) \rgl \; ,
$$
\be
\label{194}
G_{21}(12) = - i \lgl \hat T \psi_1^\dgr(1) \psi_1^\dgr(2) \rgl \; , 
\qquad
G_{22}(12) = - i \lgl \hat T \psi_1^\dgr(1) \psi_1(2) \rgl \;   ,
\ee
in which $\hat{T}$ is chronological operator. It is convenient [83] to 
introduce the retarded interaction
\be
\label{195}
\Phi(12) \equiv \Phi(\br_1-\br_2) \dlt(t_1 - t_2 + 0 ) \;  .
\ee    
Also, one defines the inverse propagators
$$
G_{11}^{-1}(12) = \left [ i \; \frac{\prt}{\prt t_1} +
\frac{\nabla_1^2}{2m} - U(1) + \mu_1 \right ] \dlt(12) - 
\Sigma_{11}(12) \; ,
$$
$$
G_{12}^{-1}(12) = - \Sigma_{12}(12) \; , \qquad 
G_{21}^{-1}(12) = - \Sigma_{21}(12) \; ,
$$
\be
\label{196}
G_{22}^{-1}(12) = \left [ - i \; \frac{\prt}{\prt t_1} +
\frac{\nabla_1^2}{2m} - U(1) + \mu_1 \right ] \dlt(12) - 
\Sigma_{22}(12) \; ,
\ee
where $\Sigma_{\alpha\beta}(12)$ is self-energy. Using these, one can 
write the equations of motion in the matrix form
$$
G_{11}^{-1} G_{11} + G_{12}^{-1} G_{21} = \hat 1 \; , \qquad
G_{11}^{-1} G_{12} + G_{12}^{-1} G_{22} = 0 \; ,
$$
\be
\label{197}
G_{21}^{-1} G_{11} + G_{22}^{-1} G_{21} = 0 \; , \qquad
G_{21}^{-1} G_{12} + G_{22}^{-1} G_{22} = \hat 1 \;  .
\ee

For a uniform system, when $U=0$, one passes to the Fourier transforms
of the Green functions $G_{\alpha\beta}({\bf k},\omega)$, inverse 
propagators $G_{\alpha\beta}^{-1}({\bf k},\omega)$, and self-energies
$\Sigma_{\alpha\beta}({\bf k},\omega)$. The inverse propagators (196) 
transform into
$$
G_{11}^{-1}(\bk,\om) = \om \; - \; \frac{k^2}{2m} + \mu_1
-\Sigma_{11}(\bk,\om) \; , \qquad 
G_{12}^{-1}(\bk,\om) = - \Sigma_{12}(\bk,\om) \; , 
$$
\be
\label{198}
G_{21}^{-1}(\bk,\om) = - \Sigma_{21}(\bk,\om) \; , \qquad
G_{22}^{-1}(\bk,\om) = - \om \; - \; \frac{k^2}{2m} + \mu_1
-\Sigma_{22}(\bk,\om) \;    .
\ee
The Green functions enjoy the properties
$$
G_{\al\bt}(-\bk,\om) = G_{\al\bt}(\bk,\om) \; , \qquad
G_{11}(\bk,-\om) = G_{22}(\bk,\om) \; , 
$$
\be
\label{199}
G_{12}(\bk,-\om) = G_{21}(\bk,\om) = G_{12}(\bk,\om) \; .
\ee
And the self-energies also share the same properties
$$
\Sigma_{\al\bt}(-\bk,\om) = \Sigma_{\al\bt}(\bk,\om) \; , \qquad
\Sigma_{11}(\bk,-\om) = \Sigma_{22}(\bk,\om) \; ,
$$
\be
\label{200}
\Sigma_{12}(\bk,-\om) = \Sigma_{21}(\bk,\om) = 
\Sigma_{12}(\bk,\om) \; .
\ee

Equations (197) yield
\be
\label{201}
G_{11}(\bk,\om) = \frac{\om+k^2/2m - \mu_1 + \Sigma_{11}(\bk,\om)}
{D(\bk,\om)} \;  , \qquad
G_{12}(\bk,\om) = -\;\frac{\Sigma_{12}(\bk,\om)}{D(\bk,\om)} \;  ,
\ee
with the denominator
\be
\label{202}
 D(\bk,\om) = \Sigma_{12}^2(\bk,\om) - 
G_{11}^{-1}(\bk,\om) G_{22}^{-1}(\bk,\om) \; .
\ee

\subsection{Hugenholtz-Pines Relation}

Hugenholtz and Pines [37], using perturbation theory at zero temperature,
found the relation
\be
\label{203}
 \mu_1 = \Sigma_{11}(0,0) - \Sigma_{12}(0,0) \; .
\ee

The most general proof of this relation, for any temperature, was given 
by Bogolubov [16]. He proved the theorem, according to which
\be
\label{204}
 \left | G_{11}(\bk,0) \right | \geq \frac{mn_0}{2k^2} \;  ,
\ee
where $n_0$ is the condensate fraction, and
\be
\label{205}
\left | G_{11}(\bk,0) - G_{12}(\bk,0) \right | \geq 
\frac{mn_0}{k^2} \;   .
\ee

>From inequality (204), one has
\be
\label{206}
\lim_{k\ra 0} \; \lim_{\om\ra 0} D(\bk,\om) = 0 \;  .
\ee
And from inequality (205), it follows that
\be
\label{207}
 \left | \frac{k^2}{2m} \; - \; \mu_1 + \Sigma_{11}(\bk,0)
- \Sigma_{12}(\bk,0) \right | \leq \frac{k^2}{mn_0} \; .
\ee
The latter inequality leads to the Hugenholtz-Pines relation (203). 

It is important to stress that the expression for $\mu_1$, given by 
Eq. (203), is exact and, generally, it is different from the exact 
value of $\mu_0$ in Eq. (183).  

The Hugenholtz-Pines relation is equivalent to the fact that the 
particle spectrum is gapless, which follows from the following.

The spectrum $\varepsilon_k$  is given by the zeroes of the Green-function 
denominator:
\be
\label{208}
 D(\bk,\ep_k) = 0 \; ,
\ee
which gives the equation
\be
\label{209}
 \ep_k = \frac{1}{2} \left [ \Sigma_{11}(\bk,\ep_k) -
\Sigma_{22}(\bk,\ep_k) \right ] + \sqrt{\om_k^2 - 
\Sigma_{12}^2(\bk,\ep_k) }  \; ,
\ee
where
\be
\label{210}
\om_k \equiv \frac{k^2}{2m} + \frac{1}{2} \left [
\Sigma_{11}(\bk,\ep_k) + \Sigma_{22}(\bk,\ep_k) 
\right ] - \mu_1 \;  .
\ee
In view of condition (206), the limit 
\be
\label{211}
\lim_{k\ra 0} \ep_k = 0
\ee
is valid, that is, the spectrum is gapless.

To find the long-wave spectrum behavior, keeping in mind that the 
spectrum is uniquely defined by Eq. (209), we can use the expansion
\be
\label{212}
 \Sigma_{\al\bt}(k,\ep_k) \simeq \Sigma_{\al\bt}(0,0) +
\Sigma_{\al\bt}' k^2 \;  ,
\ee
in which $k \ra 0$ and
$$
\Sigma_{\al\bt}' \equiv \lim_{k\ra 0} \; 
\frac{\prt}{\prt k^2}\;  \Sigma_{\al\bt} (\bk,\ep_k) \;  .
$$
Then, defining the sound velocity
\be
\label{213}
c \equiv \sqrt{\frac{1}{m^*}\; \Sigma_{12}(0,0) }
\ee
and the effective mass
\be
\label{214}
 m^* \equiv 
\frac{m}{1 + m\left (\Sigma_{11}'+\Sigma_{22}'-2\Sigma_{12}'\right )}  ,
\ee
we get the acoustic spectrum
\be
\label{215}
 \ep_k \simeq ck \qquad (k\ra 0) \; .
\ee

Equation (213), characterizing the general feature of the long-wave 
spectrum, has been obtained without approximations, assuming only the
validity of expansion (212). Therefore, in a Bose-condensed system, the 
anomalous self-energy $\Sigma_{12}(0,0)$ must be nonzero in order to 
define a meaningful nonzero sound velocity. The zero sound velocity 
would mean the system instability. Since expression (213) involves no 
perturbation theory and no approximations, the condition 
$$
\Sigma_{12}(0,0) \neq 0
$$
is general, as soon as expansion (212) is valid.

\section{Hartree-Fock-Bogolubov Approximation}

\subsection{Nonuniform Matter}

To realize practical calculations, it is necessary to resort to some 
approximation. The Bogolubov approximation [13,14] is valid for low 
temperatures and asymptotically weak interactions. The more general 
approximation, that would be valid for all temperatures and any 
interaction strength, is the Hartree-Fock-Bogolubov (HFB) approximation.
Early works [35,36], employing this approximation, confronted the 
inconsistency problem discussed in Sec. 1, because of a gap in the
particle spectrum. This happened as a result of the use of a 
nonrepresentative ensemble. Employing the representative ensemble 
of Sec. 4 yields no gap and no any other problems. The HFB approximation, 
applied in the frame of the self-consistent theory of Sec. 4, is gapless
and conserving [63-71].

The HFB approximation simplifies the general Hamiltonian (67). For 
generality, we consider, first, the nonuniform case. 

The third-order term (71) in the HFB approximation is zero. And in the 
fourth-order term (72), the HFB approximation gives
$$
\psi_1^\dgr(\br) \psi_1^\dgr(\br') \psi_1(\br') \psi_1(\br) =
\rho_1(\br) \psi_1^\dgr(\br') \psi_1(\br') +  
\rho_1(\br') \psi_1^\dgr(\br) \psi_1(\br) +
\rho_1(\br',\br) \psi_1^\dgr(\br') \psi_1(\br) +
$$
$$
+ \rho_1(\br,\br') \psi_1^\dgr(\br) \psi_1(\br') +
\sgm_1(\br,\br') \psi_1^\dgr(\br) \psi_1(\br') +
$$
\be
\label{216}
+ \sgm_1^*(\br',\br) \psi_1^\dgr(\br') \psi_1(\br) -
\rho_1(\br) \rho_1(\br') - | \rho_1(\br,\br')|^2 -
|\sgm_1(\br,\br') |^2 \; .
\ee
In what follows, it is convenient to use the notation for the total 
single-particle density matrix
\be
\label{217}
\rho(\br,\br') \equiv \eta(\br) \eta^*(\br') + 
\rho_1(\br,\br')
\ee
and for the total anomalous average
\be
\label{218}
 \sgm(\br,\br') \equiv \eta(\br) \eta(\br') + \sgm_1(\br,\br')\;  .
\ee
These equations reduce the grand Hamiltonian (67) to the HFB form
$$
H_{HFB} = E_{HFB} + \int \psi_1^\dgr(\br) \left ( -\; 
\frac{\nabla^2}{2m} + U - \mu_1 \right ) \psi_1(\br) \; d\br \; +
$$
$$
+ \; \int \Phi(\br-\br') \left [ 
\rho(\br') \psi_1^\dgr(\br) \psi_1(\br) +
\rho(\br',\br) \psi_1^\dgr(\br') \psi_1(\br) + \right.
$$
\be
\label{219}
+ \left.
\frac{1}{2} \; \sgm(\br,\br') \psi_1^\dgr(\br') \psi_1^\dgr(\br) + 
\frac{1}{2} \; \sgm^*(\br,\br')\psi_1(\br') \psi_1(\br) 
\right ] d\br d\br' \; ,
\ee
in which the nonoperator term is
\be
\label{220}
E_{HFB} = H^{(0)} - \; \frac{1}{2} \int \Phi(\br-\br') 
\left [ \rho_1(\br) \rho_1(\br') + | \rho_1(\br,\br')|^2 +
| \sgm_1(\br,\br')|^2 \right ] d\br d\br' \;  .
\ee

The condensate-function equation (95) becomes
$$
i \; \frac{\prt}{\prt t} \; \eta(\br,t) = \left ( -\;
\frac{\nabla^2}{2m} + U - \mu_0 \right ) \eta(\br) \; +
$$
\be
\label{221}
+ \; \int \Phi(\br-\br') \left [ \rho(\br') \eta(\br) +
\rho_1(\br,\br') \eta(\br') + \sgm_1(\br,\br') \eta^*(\br') 
\right ] d\br ' \;  .
\ee
And the equation of motion (85) for the operator of uncondensed 
particles now reads as
$$
i \; \frac{\prt}{\prt t} \; \psi_1(\br,t) = \left ( -\;
\frac{\nabla^2}{2m} + U - \mu_1 \right ) \psi_1(\br) \; +
$$
\be
\label{222}
 +\; \int \Phi(\br-\br') \left [ \rho(\br') \psi_1(\br) +
\rho(\br,\br') \psi_1(\br') + \sgm(\br,\br')\psi_1^\dgr(\br') 
\right ] d\br ' \;   .
\ee

In the case of an equilibrium system, Eq. (221) reduces to the 
eigenproblem 
$$
\left ( - \; \frac{\nabla^2}{2m} + U \right ) \eta(\br)\; +
$$
\be
\label{223}
  +\;
\int \Phi(\br-\br') \left [ \rho(\br') \eta(\br) + 
\rho_1(\br,\br') \eta(\br') + \sgm_1(\br,\br') \eta^*(\br') 
\right ] d\br' = \mu_0 \eta(\br)  
\ee
defining the condensate function and the the condensate chemical 
potential
$$
\mu_0 = \frac{1}{N_0} \int \eta^*(\br) \left [ -\; 
\frac{\nabla^2}{2m} + U(\br) \right ] \eta(\br) \; d\br \; +
$$
\be
\label{224}
 + \; \frac{1}{N_0} \int \Phi(\br-\br') \left [ 
\rho_0(\br) \rho(\br') + 
\rho_1(\br,\br')\eta^*(\br) \eta(\br') +
\sgm_1(\br,\br')\eta^*(\br)\eta^*(\br') 
\right ] d\br d\br' \; .
\ee

\subsection{Bogolubov Transformations} 

The HFB Hamiltonian (219) is a quadratic form with respect to the 
operators $\psi_1$. As any quadratic form, it can be diagonalized
by means of the Bogolubov canonical transformations, whose general
properties are described in detail in the book [84]. In the present 
case, the Bogolubov transformations read as
\be
\label{225}
  \psi_1(\br) = \sum_k \left [ u_k(\br) b_k + 
v_k^*(\br) b_k^\dgr \right ] \; , \qquad 
b_k = \int \left [ u_k^*(\br) \psi_1(\br) - 
v_k^*(\br) \psi_1^\dgr(\br) \right ] d\br \; .
\ee
Since $\psi_1$ is a Bose operator, it should be:
$$
 \sum_k \left [ u_k(\br) u_k^*(\br') - 
v_k^*(\br) v_k(\br') \right ] = \dlt(\br-\br')\; , 
$$
\be
\label{226}
\sum_k \left [ u_k(\br) v_k^*(\br') - 
v_k^*(\br) u_k(\br') \right ] =  0 \; .
\ee
And, the condition that $b_k$ is also a Bose operator leads to the 
relations
\be
\label{227}
\int \left [ u_k^*(\br) u_p(\br) - v_k^*(\br) v_p(\br) 
\right ] d\br = \dlt_{kp} \; , \qquad 
\int \left [ u_k(\br) v_p(\br) - v_k(\br) u_p(\br) 
\right ] d\br = 0 \; .
\ee
The coefficient functions $u_k$ and $v_k$ are to be defined by the 
requirement of the Hamiltonian diagonalization, under conditions (226) 
and (227).

Let us introduce the notations
$$
\om(\br,\br') \equiv \left [ -\; \frac{\nabla^2}{2m} + U(\br) -
\mu_1 + \int \Phi(\br-\br') \rho(\br')\; d\br' \right ]
\dlt(\br-\br') \; + 
$$
\be
\label{228}
+\; \Phi(\br-\br') \rho(\br,\br')
\ee
and 
\be
\label{229}
 \Dlt(\br,\br') \equiv \Phi(\br-\br') \sgm(\br,\br') \;  .
\ee
Then the Hamiltonian diagonalization leads to the {\it Bogolubov equations}
$$
\int [ \om(\br,\br') u_k(\br') + 
\Dlt(\br,\br') v_k(\br') ] d\br' = \ep_k u_k(\br) \; ,
$$
\be
\label{230}
\int [ \om^*(\br,\br') v_k(\br') + 
\Dlt^*(\br,\br') u_k(\br') ] d\br' = -\ep_k v_k(\br) \; .
\ee
This is the eigenproblem for the Bogolubov functions $u_k$ and $v_k$ and 
the Bogolubov spectrum $\varepsilon_k$.

The resulting diagonal Hamiltonian is
\be
\label{231}
H_B = E_B + \sum_k \ep_k b_k^\dgr b_k \;  ,
\ee
with the nonoperator term
\be
\label{232}
E_B = E_{HFB} - \sum_k \ep_k \int | v_k(\br)|^2 d\br \;  .
\ee
  
The quasiparticles, described by the operators $b_k$, are called 
{\it bogolons}. Their quantum-number distribution is easily calculated, giving
\be
\label{233}
 \pi_k \equiv \lgl b_k^\dgr b_k \rgl = 
\left ( e^{\bt\ep_k} - 1 \right )^{-1} \; ,
\ee
which can also be represented as
\be
\label{234}
 \pi_k = \frac{1}{2} \left [ \coth\left ( \frac{\ep_k}{2T} \right )
- 1 \right ] \; .
\ee 

The normal density matrix (88) takes the form
\be
\label{235}
 \rho_1(\br,\br') = \sum_k [ \pi_k u_k(\br) u_k^*(\br')
+ (1 + \pi_k) v_k^*(\br) v_k(\br') ] \; ,
\ee
while the anomalous average (89) becomes
\be
\label{236}
\sgm_1(\br,\br') = \sum_k [ \pi_k u_k(\br) v_k^*(\br')
+ (1 + \pi_k) v_k^*(\br) u_k(\br') ] \; .
\ee
The density of uncondensed particles (91) is
\be
\label{237}
\rho_1(\br) = \sum_k \left [ \pi_k |u_k(\br)|^2 + 
(1+\pi_k) |v_k(\br)|^2 \right ]
\ee
and the diagonal anomalous average (92) is
\be
\label{238}
\sgm_1(\br) = \sum_k (1 + 2\pi_k) u_k(\br) v_k^*(\br) \;  .
\ee

The grand thermodynamic potential
\be
\label{239}
 \Om \equiv - T \ln {\rm Tr} e^{-\bt H} \; ,
\ee
under Hamiltonian (231), reads as
\be
\label{240}
\Om = E_B + T \sum_k \ln \left ( 1 - e^{-\bt \ep_k} 
\right ) \;  ,
\ee
where the first term, defined in Eq. (232), gives
$$
E_B = -\; \frac{1}{2} \int \Phi(\br-\br') \left [
\rho_0(\br) \rho_0(\br') + \right.
$$
$$
+ \left. 2\rho_0(\br)\rho_1(\br') +
2\eta^*(\br)\eta(\br')\rho_1(\br,\br') + 2\eta^*(\br)\eta^*(\br')
\sgm_1(\br,\br') + \right.
$$
\be
\label{241}
+ \left. \rho_1(\br) \rho_1(\br') + |\rho_1(\br,\br')|^2 +
|\sgm_1(\br,\br') |^2 \right ] d\br d\br' \; 
- \; \sum_k \ep_k \int |v_k(\br)|^2 d\br \;  .
\ee

The above equations are valid for any nonuniform matter, with an arbitrary 
external potential $U({\bf r})$.

\subsection{Uniform Matter}

The previous equations simplify for a uniform case, when there is no 
external potential. Setting $U = 0$, we can use the Fourier 
transformation (173) and follow the way of Sec. 6.

Instead of expressions (228) and (229), we now have
\be
\label{242}
\om_k \equiv \frac{k^2}{2m} + \rho\Phi_0 + \rho_0 \Phi_k +
\frac{1}{V} \sum_{p\neq 0} n_p \Phi_{k+p} - \mu_1
\ee
and
\be
\label{243}
\Dlt_k \equiv \rho_0 \Phi_k + 
\frac{1}{V} \sum_{p\neq 0} \sgm_p \Phi_{k+p} \;  .
\ee
The HFB Hamiltonian (219) reduces to
\be
\label{244}
H_{HFB} = E_{HFB} + \sum_{k\neq 0} \om_k a_k^\dgr a_k +
\frac{1}{2} \sum_{k\neq 0} \Dlt_k \left ( a_k^\dgr a_{-k}^\dgr
+ a_{-k} a_k \right )  \;  ,
\ee
with the nonoperator term
\be
\label{245}
 E_{HFB} = H^{(0)} - \; \frac{1}{2} \; \rho_1^2 \Phi_0 V -\;
\frac{1}{2V} {\sum_{kp}}' \Phi_{k+p} (n_k n_p +
\sgm_k \sgm_p ) \;  ,
\ee
in which ${\bf k} \neq 0, {\bf p} \neq 0$. 

Instead of the Bogolubov canonical transformations (225), one has
\be
\label{246}
a_k = u_k b_k + v_{-k}^* b_{-k}^\dgr \; , \qquad
b_k = u_k^* a_k - v_k^* a_{-k}^\dgr \; .
\ee
And the Bogolubov equations (230) become
\be
\label{247}
 ( \om_k - \ep_k) u_k + \Dlt_k v_k = 0 \; , \qquad
\Dlt_k u_k + (\om_k + \ep_k) v_k = 0 \; .
\ee
The Bogolubov Hamiltonian (231) has the same form, but with
\be
\label{248}
 E_B = E_{HFB} + \frac{1}{2} \sum_{k\neq 0} (\ep_k -\om_k) \; ,
\ee
instead of Eq. (232). 

The coefficient functions $u_k$ and $v_k$ are defined by the Bogolubov 
equations (247), under conditions
\be
\label{249}
 | u_k|^2 - | v_{-k}|^2 = 1 \; , \qquad
u_k v_k^* - v_{-k}^* u_{-k} = 0 \;  ,
\ee
replacing conditions (226) and (227). These functions, due to the 
system uniformity and isotropy, are real and symmetric with respect 
to the momentum inversion ${\bf k} \ra -{\bf k}$. As a result, one has
$$
u_k^2 - v_k^2 = 1 \; , \qquad 
u_k^2 + v_k^2 = \frac{\om_k}{\ep_k} \; ,\qquad 
u_k v_k = -\; \frac{\Dlt_k}{2\ep_k} \; ,
$$
\be
\label{250}
 u_k^2 = \frac{\om_k + \ep_k}{2\ep_k} \; , \qquad
 v_k^2 = \frac{\om_k - \ep_k}{2\ep_k} \; .
\ee

The Bogolubov spectrum becomes
\be
\label{251}
 \ep_k = \sqrt{ \om_k^2 - \Dlt_k^2 } \;  .
\ee
As is known from Sec. 6, the spectrum has to be gapless, which gives
\be
\label{252}
\mu_1 = \rho \Phi_0 + 
\frac{1}{V} \sum_{k\neq 0} (n_k - \sgm_k) \Phi_k \;  .
\ee
This is different from the condensate chemical potential (224) that is
\be
\label{253}
\mu_0 = \rho \Phi_0 + 
\frac{1}{V} \sum_{k\neq 0} (n_k + \sgm_k) \Phi_k \;  .
\ee

With $\mu_1$ from Eq. (252), expression (242) is
\be
\label{254}
 \om_k = \frac{k^2}{2m} + \rho_0 \Phi_k + \frac{1}{V} 
\sum_{p\neq 0} (n_p \Phi_{k+p} - n_p \Phi_p + \sgm_p \Phi_p ) \;   .
\ee
In the long-wave limit, the Bogolubov spectrum (251) is of acoustic form (215),
with the sound velocity
\be
\label{255}
 c = \sqrt{ \frac{\Dlt}{m^*} } \; ,
\ee
in which
\be
\label{256}
\Dlt \equiv \lim_{k\ra 0} \Dlt_k = \rho_0 \Phi_0 +
\frac{1}{V} \sum_{p\neq 0} \sgm_p \Phi_p
\ee
and the effective mass is
\be
\label{257}
 m^* \equiv 
\frac{m}{1 + \frac{2m}{V}\sum_{p\neq 0}(n_p-\sgm_p)\Phi_p'} \; ,
\ee
where
$$
\Phi_p' \equiv \frac{\prt}{\prt p^2}\; \Phi_p \;  .
$$   

>From Eqs. (255) and (256), we have
\be
\label{258}
 \Dlt \equiv m^* c^2 = \rho_0 \Phi_0 +
\frac{1}{V} \sum_{p\neq 0} \sgm_p \Phi_p \; .
\ee
Hence, expression (254) can be written as
\be
\label{259}
\om_k = m^* c^2 + \frac{k^2}{2m} + \rho_0 ( \Phi_k - \Phi_0)
+ \frac{1}{V} \sum_{p\neq 0} n_p ( \Phi_{k+p} - \Phi_p) \;  .
\ee

Comparing Eqs. (213) and (255) yields
\be
\label{260}
 \Sigma_{12}(0,0) = \rho_0 \Phi_0 + 
\frac{1}{V} \sum_{p\neq 0} \sgm_p \Phi_p \; .
\ee
And from the Hugenholtz-Pines relation (203), with $\mu_1$ from 
Eq. (252), we get 
\be
\label{261}
\Sigma_{11}(0,0) = (\rho+ \rho_0) \Phi_0 + 
\frac{1}{V} \sum_{p\neq 0} n_p \Phi_p \;  .
\ee
Of course, the same Eqs. (260) and (261) can be derived directly from 
the Green function equations. 

The condensate chemical potential (253) can be written as
\be
\label{262}
 \mu_0 = \Sigma_{11}(0,0) + \Sigma_{12}(0,0) - 
2\rho_0 \Phi_0 \; .
\ee
The difference between Eqs. (252) and (253) takes the form
\be
\label{263}
 \mu_0 - \mu_1 = 2 \left [ \Sigma_{12}(0,0) - 
\rho_0 \Phi_0 \right ] \;  ,
\ee
which again tells us that these chemical potentials are different. 
They coincide only in the Bogolubov approximation [13,14], when 
$\Sigma_{12}(0,0)$ equals $\rho_0 \Phi_0$. Then $\mu_0$ and $\mu_1$ 
both are also equal to $\rho_0 \Phi_0$ and, hence, to each other. 

The momentum distribution (185) is
\be
\label{264}
 n_k = \frac{\om_k}{2\ep_k} \; 
\coth \left ( \frac{\ep_k}{2T} \right ) -\; \frac{1}{2} \; ,
\ee
while the anomalous average (187) reads as
\be
\label{265}
 \sgm_k = -\; \frac{\Dlt_k}{2\ep_k} \; 
\coth \left ( \frac{\ep_k}{2T} \right ) \;  .
\ee

The grand potential (239) enjoys the same form (240), but with
$$
E_B = -\; \frac{V}{2} \int \Phi(\br) \left [ \rho^2 + 
2\rho_0 \rho_1(\br,0) + 2\rho_0 \sgm_1(\br,0) + \right.
$$
\be
\label{266}
 + \left. | \rho_1(\br,0)|^2 + | \sgm_1(\br,0) |^2 
\right ] d\br \; + \; \frac{1}{2} \sum_k (\ep_k - \om_k) \;  ,
\ee
which can be transformed to
$$
E_B = - \; \frac{N}{2} \; \rho \Phi_0 - 
\rho_0 \sum_p (n_p + \sgm_p) \Phi_p -
$$
\be
\label{267}
 - \; \frac{1}{2V} \sum_{kp} ( n_k n_p + \sgm_k \sgm_p ) 
\Phi_{k+p} \; + \; \frac{1}{2} \sum_k (\ep_k - \om_k) \; .
\ee

\subsection{Local-Density Approximation}

When there exists an external potential $U({\bf r})$ and the system 
is nonuniform, one can use the equations from Sec. 7.2. It is also 
possible to resort to the local-density approximation [1-3]. The 
local-density, or semi-classical, approximation [85,86] is applicable 
when the external potential is sufficiently smooth, such that
\be
\label{268}
  \left | \frac{l_0}{U_0} \; \frac{\prt U(\br)}{\prt\br}
\right | \ll 1 \; ,
\ee
where $U_0$ and $l_0$ are the characteristic depth and length of the
potential, respectively.    

In this approximation, one looks for the solutions of the Bogolubov 
equations (230), represented as
\be
\label{269}
u_k(\br) = u(\bk,\br) \; 
\frac{e^{i\bk\cdot\br}}{\sqrt{V}} \;  , \qquad
v_k(\br) = v(\bk,\br) \;\frac{e^{i\bk\cdot\br}}{\sqrt{V}} \; ,
\ee
where the functions $u({\bf k}, {\bf r})$ and $v({\bf k}, {\bf r})$
are assumed to be slowly varying as compared to the exponentials, so 
that
\be
\label{270}
 |\nabla u(\bk,\br) | \ll k | u(\bk,\br) | \; , \qquad 
|\nabla v(\bk,\br) | \ll k | v(\bk,\br) | \; .
\ee
Then, using the notations
\be
\label{271}
 \om(\bk,\br) \equiv \frac{k^2}{2m} + U(\br) + 
2\Phi_0 \rho(\br) - \mu_1(\br) 
\ee
and
\be
\label{272}
\Dlt(\br) \equiv [ \rho_0(\br) + \sgm_1(\br) ] \Phi_0 \;  ,
\ee
one reduces the Bogolubov equations (230) to the form 
$$
[ \om(\bk,\br) - \ep(\bk,\br) ] u(\bk,\br) +
\Dlt(\br) v(\bk,\br) = 0 \; ,
$$
\be
\label{273}
 \Dlt^*(\br) u(\bk,\br) + [ \om^*(\bk,\br) +
\ep(\bk,\br) ] v(\bk,\br) = 0 \; ,
\ee
in which
\be
\label{274}
\Phi_0 \equiv \int \Phi(\br) \; d\br \;  .
\ee

The following procedure is analogous to the uniform case. For the 
coefficient functions, we have
$$
u^2(\bk,\br) - v^2(\bk,\br) = 1 \; , \qquad
u^2(\bk,\br) + v^2(\bk,\br) = 
\frac{\om(\bk,\br)}{\ep(\bk,\br)} \; ,
$$
$$ 
u(\bk,\br) v(\bk,\br) = -\; 
\frac{\Dlt(\br)}{2\ep(\bk,\br)} \; ,
$$
\be
\label{275}
u^2(\bk,\br) = \frac{\om(\bk,\br)+\ep(\bk,\br)}{2\ep(\bk,\br)} \; , 
\qquad 
v^2(\bk,\br) = \frac{\om(\bk,\br)-\ep(\bk,\br)}{2\ep(\bk,\br)} \; .
\ee
The local Bogolubov spectrum is
\be
\label{276}
 \ep(\bk,\br) = \sqrt{\om^2(\bk,\br) - \Dlt^2(\br) } \; .
\ee
From the requirement that the spectrum be gapless,
\be
\label{277}
\lim_{k\ra 0} \ep(\bk,\br) = 0 \;  ,
\ee
we find
\be
\label{278}
\mu_1(\br) = U(\br) + [ \rho_0(\br) + 2\rho_1(\br) -
\sgm_1(\br) ] \Phi_0 \;  .
\ee
Denoting
\be
\label{279} 
 \Dlt(\br) \equiv mc^2(\br) \; ,
\ee
from Eq. (272), we get
\be
\label{280}
  mc^2(\br) = [\rho_0(\br) + \sgm_1(\br) ] \Phi_0 \; .
\ee
Then Eq. (271) becomes
\be
\label{281}
 \om(\bk,\br) = mc^2(\br) + \frac{k^2}{2m} \; .
\ee
The local Bogolubov spectrum (276) takes the form
\be
\label{282}
 \ep(\bk,\br) = 
\sqrt{ c^2(\br) k^2 + \left ( \frac{k^2}{2m} \right )^2 } \;  .
\ee
This shows that $c({\bf r})$ is the local sound velocity. 

With spectrum (282), the bogolon momentum distribution (234) reads as
\be
\label{283}
 \pi(\bk,\br) = \frac{1}{2} \left [ \coth\left\{
\frac{\ep(\bk,\br)}{2T} \right \} - 1 \right ] \; .
\ee
In view of the system isotropy, the symmetry properties
\be
\label{284}
 \ep(-\bk,\br) = \ep(\bk,\br) \; , \qquad 
\pi(-\bk,\br) = \pi(\bk,\br)   
\ee
are valid.

The single-particle density matrix (235) now transforms into
\be
\label{285}
 \rho_1(\br,\br') = \frac{1}{V} 
\sum_k n(\bk,\br) e^{i\bk\cdot(\br-\br')} \; ,
\ee
while the anomalous average (236) becomes
\be
\label{286}
\sgm_1(\br,\br') = \frac{1}{V} 
\sum_k \sgm(\bk,\br) e^{i\bk\cdot(\br-\br')} \; .
\ee
Here the particle local momentum distribution, replacing Eq. (264), is
\be
\label{287}
n(\bk,\br) = \frac{\om(\bk,\br)}{2\ep(\bk,\br)} \;
\coth \left [ \frac{\ep(\bk,\br)}{2T} \right ] - \; \frac{1}{2}   
\ee
and, instead of the anomalous average (265), one has
\be
\label{288}
 \sgm(\bk,\br) = -\; \frac{mc^2(\br)}{2\ep(\bk,\br)} \;
\coth \left [ \frac{\ep(\bk,\br)}{2T} \right ]  .
\ee
The density of uncondensed particles (237) gives
\be
\label{289}
 \rho_1(\br) = \frac{1}{V} \sum_k n(\bk,\br)  
\ee  
and the anomalous average (238) is
\be
\label{290}
 \sgm_1(\br) = \frac{1}{V} \sum_k \sgm(\bk,\br) \; .
\ee

The grand potential (239) reads as
\be
\label{291}
\Om = E_B + T \int \ln \left [ 1 - \exp \left\{ -\bt \ep(\bk,\br)
\right \} \right ] \frac{d\bk}{(2\pi)^3} \; d\br \; .
\ee
Here the first term, after the dimensional regularization of the 
expression
\be
\label{292}
\int [ \ep(\bk,\br) -\om(\bk,\br) ] \frac{d\bk}{(2\pi)^3} =
\frac{16m^4}{15\pi^2} \; c^5(\br) \;  ,
\ee
takes the form
$$
E_B = -\; \frac{\Phi_0}{2} \int \left [ \rho^2(\br) + 
2\rho_0(\br) \rho_1(\br) + 2\rho_0(\br) \sgm_1(\br)
+ \rho_1^2(\br) + \sgm_1^2(\br) \right ] d\br \; +
$$
\be
\label{293}
 + \; \frac{8m^4}{15\pi^2} \int c^5(\br) d\br \; .
\ee

When the system is constrained inside a fixed volume $V$, then the grand 
potential $\Omega = - PV$ defines the system pressure $P = - \Omega/V$,
irrespectively of whether the system is uniform or not. But, when a 
nonuniform system is confined inside a trapping potential that does 
not have rigid boundaries constraining the system inside a given volume,
then the system pressure cannot be defined as $-\Omega/V$. It is possible,
being based on the generalized definition of thermodynamic limit (13), to 
introduce an effective volume and effective pressure. However, these 
quantities are different for different potentials and, moreover, they are 
not uniquely defined even for a given potential, hence, they would have
no physical meaning.

What is well defined for any nonuniform system is the local pressure 
$p({\bf r})$ that enters the grand potential through the equality
\be
\label{294}
 \Om = - \int p(\br) \; d\br \; .
\ee
For the grand potential (291), the local pressure is
$$
p(\br) = -T \int \ln \left [ 1 - \exp\{ - \bt \ep(\bk,\br) \}
\right ] \frac{d\bk}{(2\pi)^3} \; +
$$
\be
\label{295}
+\; \frac{\Phi_0}{2} \left [ \rho^2(\br) + 
2\rho_0(\br) \rho_1(\br) + 2\rho_0(\br) \sgm_1(\br) +
\rho_1^2(\br) + \sgm_1^2(\br) \right ]\; - \;
\frac{8m^4}{15\pi^2} \; c^5(\br) \; .
\ee

Equation (295) can be represented as the sum
$$
p(\br) = p_0(\br) + p_T(\br) \; ,
$$
in which
$$
p_0(\br) = \left [ \rho^2(\br) - \rho_0^2(\br) \right ] \Phi_0 \;
+ \; \frac{m^2c^4(\br)}{2\Phi_0} \; - \; 
\frac{8m^4}{15\pi^2} \; c^5(\br)
$$
and 
$$
p_T(\br) = - T \int \ln [ 1 - \exp\{ - \bt \ep(\bk,\br) \} ]\;
\frac{d\bk}{(2\pi)^3} \; .
$$
The latter term, when temperature decreases, tends to zero as
$$
p_T(\br) \simeq \frac{T^4}{2\pi^2c^3(\br)} \qquad (T \ra 0) \;  .
$$

For asymptotically weak interactions, when $\Phi_0 \ra 0$, Eq. (280),
defining the local sound velocity, reduces to
$$
mc^2(\br) \simeq \rho_0(\br) \Phi_0 \; .  
$$
In that case, the local pressure (295) simplifies to
$$
p(\br) = \frac{1}{2} \; \rho^2(\br) \Phi_0 - 
T \int \ln [ 1 - \exp \{ -\bt \ep(\bk,\br) \} ] \; 
\frac{d\bk}{(2\pi)^3} \;  ,
$$
with the local Bogolubov spectrum
$$
\ep(\bk,\br) = \sqrt{\rho_0(\br) \Phi_0 \; \frac{k^2}{m} + 
\left ( \frac{k^2}{2m} \right )^2 } \;  .
$$

Such local thermodynamic quantities are common for nonuniform systems, 
both equilibrium [87] and quasiequilibrium [88,89].

\subsection{Particle Densities}

In the local-density approximation, it is straightforward to find the 
densities of particles. Thus, the condensate density is
\be
\label{296}
  \rho_0(\br) = | \eta(\br) |^2 \; .
\ee
For an equilibrium system, the condensate function is real. Equation (223) 
for the condensate function, in the local-density approximation, becomes
\be
\label{297}
 \left [ -\; \frac{\nabla^2}{2m} + U(\br) \right ] \eta(\br)
+ \Phi_0 [ \rho_0(\br) + 2\rho_1(\br) + \sgm_1(\br) ] \eta(\br) 
= \mu_0\eta(\br) \; .
\ee

The simplest way of solving this equation is by means of the Thomas-Fermi 
approximation, when one neglects the spatial derivative, which yields
\be
\label{298}
\rho_{TF}(\br) = \frac{\mu_0-U(\br)}{\Phi_0} \; - \;
2\rho_1(\br) - \sgm_1(\br) \; .
\ee
In the case of cylindrical symmetry, one can introduce the Thomas-Fermi
volume $V_{TF} = \pi R^2 L$, with the Thomas-Fermi radius $R$ and the  
Thomas-Fermi length $L$ defined by the equations
$$
\mu_0 = U(R,0) + \Phi_0 [ 2\rho_1(R,0) + \sgm_1(R,0) ] \; ,
$$
\be
\label{299} 
\mu_0 = U\left ( 0,\frac{L}{2} \right ) + 
\Phi_0 \left [ 2\rho_1\left ( 0,\frac{L}{2} \right ) + 
\sgm_1 \left ( 0, \frac{L}{2} \right ) \right ] \;   .
\ee
In the Thomas-Fermi approximation, the condensate density is nonzero 
only inside the Thomas-Fermi volume, where
\be
\label{300}
\rho_0(\br) = \rho_{TF}(\br) \Theta(R-r) \Theta\left (
\frac{L}{2} \; - |z| \right ) \;  ,
\ee
with $\Theta(\cdot)$ being the unit step function. Of course, more 
correctly, the condensate function should be calculated by directly 
solving Eq. (297).  
 
The density of uncondensed particles (289) can be written as
\be
\label{301}
\rho_1(\br) = \frac{1}{2} \int \left [ 
\frac{\om(\bk,\br)}{\ep(\bk,\br)} \; - \; 1 \right ] 
\frac{d\bk}{(2\pi)^3} \; 
+ \; \frac{1}{2} \int \frac{\om(\bk,\br)}{\ep(\bk,\br)} 
\left \{ \coth \left [ \frac{\ep(\bk,\br)}{2T} \right ] -1 
\right \} \frac{d\bk}{(2\pi)^3} \;  .
\ee
And the anomalous average (290) is
\be
\label{302}
 \sgm_1(\br) = -\; \frac{1}{2} \int 
\frac{mc^2(\br)}{\ep(\bk,\br)} \;
\coth \left [ \frac{\ep(\bk,\br)}{2T} \right ]
\frac{d\bk}{(2\pi)^3} \;  .
\ee

At zero temperature, the anomalous average becomes
\be
\label{303}
\sgm_0(\br) = -\; \frac{1}{2} 
\int \frac{mc^2(\br)}{\ep(\bk,\br)} \; \frac{d\bk}{(2\pi)^3} \;  .
\ee
This integral diverges. It can be regularized invoking the dimensional 
regularization that is well defined for asymptotically weak 
interactions [4]. Employing the dimensional regularization for finite 
interactions requires that the limiting condition
\be
\label{304}
\sgm_0(\br) \ra 0 \qquad (\rho_0 \ra 0)   
\ee
be satisfied [12,66,67,69-71]. This condition takes into account that
the anomalous averages and Bose condensate always exist together, both 
being due to the common reason of gauge symmetry breaking. As soon as
the condensate density is nonzero, the anomalous average is also nonzero.
And, conversely, when the condensate density becomes zero, the anomalous 
averages have also to disappear. 

Another limiting condition is
\be
\label{305}
 \sgm_0(\br) \ra 0 \qquad (\Phi_0 \ra 0) \; .
\ee
This condition takes into account that the anomalous average nullifies 
for the ideal Bose gas [12,66,67,69-71]. 

Under conditions (304) and (305), the dimensional regularization gives
$$
\int \frac{1}{\ep(\bk,\br)} \; \frac{d\bk}{(2\pi)^3} =
-\; \frac{2m}{\pi^2} \sqrt{m\Phi_0\rho_0(\br) } \; .
$$
Then Eq. (303) reduces to
\be
\label{306}
 \sgm_0(\br) = \frac{m^2c^2(\br)}{\pi^2} \;
\sqrt{m\Phi_0\rho_0(\br) } \; .
\ee  

Thus, at temperatures outside the critical region, the anomalous average 
(302) can be represented in the form
\be
\label{307}
 \sgm_1(\br) = \sgm_0 - \; \frac{1}{2} \int 
\frac{mc^2(\br)}{\ep(\bk,\br)}
\left \{ \coth\left [ \frac{\ep(\bk,\br)}{2T}\right ] -1 \right \}\;
\frac{d\bk}{(2\pi)^3} \; .
\ee  
This form can also be used even in the critical region, provided that 
interactions are weak. Strictly speaking, form (307) is valid when one 
of the following conditions holds true:
\be
\label{308}
 \frac{T}{T_c} \ll 1 \; , \qquad 
\frac{\rho\Phi_0}{T_c} \ll 1 \; ,
\ee
where $T_c$ is the critical temperature.

In the vicinity of the transition point $T_c$, where $c({\bf r}) \ra 0$,
the anomalous average (302) behaves as
\be
\label{309}
 \sgm_1(\br) \simeq -\; \frac{m^2 T}{2\pi} \; c(\br) \qquad
(T\ra T_c) \; .
\ee
This behavior guarantees that the Bose condensation transition is of 
second order for any interaction strength [12,69-71].  

For the convenience of calculations, the density (301) of uncondensed 
particles can be transformed into
\be
\label{310}
 \rho_1(\br) = \frac{m^3c^3(\br)}{3\pi^2} \left \{ 1 +
\frac{3}{2\sqrt{2}} \int_0^\infty \left ( 
\sqrt{1+x^2}-1 \right )^{1/2}  \left ( \coth \left [
\frac{mc^2(\br)}{2T} \; x \right ] - 1 \right ) dx \right \}
\ee
and the anomalous average (307), into 
\be
\label{311}
\sgm_1(\br) = \sgm_0(\br) - \; 
\frac{m^3c^3(\br)}{2\sqrt{2}\pi^2} \int_0^\infty
\frac{\left ( \sqrt{1+x^2}-1\right)^{1/2}}{\sqrt{1+x^2}}
\left ( \coth \left [
\frac{mc^2(\br)}{2T} \; x \right ] - 1 \right ) dx \;  .
\ee
The sound velocity here is defined by Eq. (280).

The local superfluid density has been introduced in Eq. (155). In the 
local-density approximation, for an equilibrium system, we have
\be
\label{312}
 \rho_s(\br) = \rho(\br) - \; \frac{2Q(\br)}{3T} \; ,
\ee
with the local dissipated heat
\be
\label{313}
 Q(\br) = \int \frac{k^2}{2m} \left [ n(\bk,\br) +
n^2(\bk,\br) - \sgm^2(\bk,\br) \right ] 
\frac{d\bk}{(2\pi)^3} \; .
\ee
In view of Eqs. (287) and (288), this yields
$$
 Q(\br) = \frac{1}{(4\pi)^2m} \int_0^\infty
\frac{k^4 dk}{\sinh^2[\ep(\bk,\br)/2T]} \; ,
$$
which can be transformed into
\be
\label{314}
 Q(\br) = \frac{m^4c^5(\br)}{(2\pi)^2\sqrt{2}} \int_0^\infty
\frac{\left ( \sqrt{1+x^2}-1 \right )^{3/2} xdx }
{\sqrt{1+x^2}\sinh^2[mc^2(\br)x/2T]} \;  .
\ee  

It is necessary to stress the importance of taking account of the 
anomalous average. If in Eq. (313), one would omit this anomalous 
average, then the dissipated heat would be infinite, hence the 
superfluid density would not exist at all. But, taking the anomalous 
average into account renders the dissipated heat (314) a well defined 
finite quantity. The fact that the anomalous average is crucially 
important for describing superfluidity should be apparent remembering
that $|\sigma_1({\bf r})|$ is the density of pair-correlated particles.
These pair correlations are, actually, responsible for the existence 
of superfluidity as such. Therefore, when there are no pair 
correlations, there is no supefluidity.  

Having all particle densities defined makes it possible to study their 
spatial distributions and to calculate the average condensate, $n_0$, 
and superfluid, $n_s$, fractions, as well as the fraction $n_1$ of 
uncondensed particles, given by the equations
$$
n_0 = \frac{1}{N} \int \rho_0(\br) \; d\br \; , \qquad
n_s = \frac{1}{N} \int \rho_s(\br) \; d\br \; , 
$$
\be
\label{315}
n_1 = \frac{1}{N} \int \rho_1(\br) \; d\br \; , \qquad 
n_0 + n_1 = 1 \;  .
\ee

\section{Local Interaction Potential}

\subsection{Grand Hamiltonian}

Till now, the consideration, for generality, has been accomplished
for any type of the symmetric interaction potential 
$\Phi(-{\bf r}) = \Phi({\bf r})$, with the sole restriction that this 
potential be integrable, such that integral (274), defining $\Phi_0$, 
be finite. 

When particles interact with each other through a potential, whose 
effective interaction radius $r_0$ is much shorter than the mean 
interparticle distance $a$, then this potential can be represented
in the local form
\be
\label{316}
\Phi(\br) = \Phi_0 \dlt(\br) \; , \qquad
\Phi_0 \equiv 4\pi \; \frac{a_s}{m} \;  ,
\ee
in which the interaction strength $\Phi_0$ is expressed through s-wave 
scattering length $a_s$ and mass $m$. For uniform systems, the potential 
is called stable [90] when $\Phi_0$ is positive. For trapped atoms, a 
finite system can be stable also for negative interactions [1-3,91].   

The grand Hamiltonian (67), for the local interaction potential (316),
contains the following terms. The zero-order term (68) reads as
\be
\label{317}
H^{(0)} = \int \eta^*(\br) \left ( -\; 
\frac{\nabla^2}{2m} + U - \mu_0 \right ) \eta(\br) \; d\br \; + \;
\frac{\Phi_0}{2} \int | \eta(\br)|^4 d\br \; .
\ee
The first-order term, as always, is zero. The second-order term (70) is
$$
H^{(2)} = \int \psi_1^\dgr(\br) \left ( -\; 
\frac{\nabla^2}{2m} + U - \mu_1 \right ) \psi_1(\br) \; d\br \; +
$$
\be
\label{318}
 + \; \Phi_0 \int \left [ 
2 | \eta(\br)|^2 \psi_1^\dgr(\br) \psi_1(\br) + 
\frac{1}{2} ( \eta^*(\br) )^2 \psi_1(\br) \psi(\br) + 
\frac{1}{2} ( \eta(\br))^2 \psi_1^\dgr(\br) \psi_1^\dgr(\br) 
\right ] d\br \; .
\ee 
The third-order term (71) becomes
\be
\label{319}
 H^{(3)} = \Phi_0 \int \left [ 
\eta^*(\br) \psi_1^\dgr(\br) \psi_1(\br) \psi_1(\br) +
\psi_1^\dgr(\br) \psi_1^\dgr(\br) \psi_1(\br) \eta(\br) 
\right ] d \br \; .
\ee
And the fourth-order term (72) reduces to
\be
\label{320}
 H^{(4)} = \frac{\Phi_0}{2} \int 
\psi_1^\dgr(\br) \psi_1^\dgr(\br) \psi_1(\br) \psi_1(\br) \; 
d\br \; .
\ee

\subsection{Evolution Equations}

Evolution equations, derived in Sec. 4.5, simplify for the local 
potential (316). The same notations (90) to (93) can be used. But,
instead of (94), we define
\be
\label{321}
\xi(\br) \equiv 
\lgl \psi_1^\dgr(\br) \psi_1(\br) \psi_1(\br) \rgl \; .
\ee
In addition, we shall employ the notation
\be
\label{322}
\xi_1(\br) \equiv 
\lgl \psi_1(\br) \psi_1(\br) \psi_1(\br) \rgl \;  .
\ee

Equation (95) for the condensate function yields
$$
i \; \frac{\prt}{\prt t} \; \eta(\br) = \left ( -\;
\frac{\nabla^2}{2m} + U -\mu_0 \right ) \eta(\br ) \; +
$$
\be
\label{323}
+\; \Phi_0 \left [ \rho_0(\br) \eta(\br) + 
2\rho_1(\br) \eta(\br) + \sgm_1(\br) \eta^*(\br) + 
\xi(\br) \right ] \; .
\ee
The continuity equations (99) to (101) have the same form, but with
the source term
\be
\label{324}
 \Gm(\br,t) = i \Phi_0 \left [ \Xi^*(\br) - 
\Xi(\br) \right ] \; ,
\ee
with the anomalous correlation function
$$
 \Xi(\br) = \eta^*(\br) [ \eta^*(\br) \sgm_1(\br) +
\xi(\br) ] \; .
$$ 
Equation (85) for the operator of uncondensed particles changes to
\be
\label{325}
i\; \frac{\prt}{\prt t} \; \psi_1(\br,t) = \left ( -\;
\frac{\nabla^2}{2m} + U - \mu_1 \right ) \psi_1(\br,t) +
 \Phi_0 \left [ \hat X_1(\br,\br) + 
\hat X(\br,\br) \right ] \; .
\ee

Equation (103) for the anomalous average becomes
$$
i\; \frac{\prt}{\prt t} \; \sgm_1(\br,t) = 2K(\br) +
2 ( U - \mu_1 ) \sgm_1(\br) \; +
$$
$$
 + \; 2 \Phi_0 \left [ 
\eta^2(\br) \rho_1(\br) + 2\rho_0(\br) \sgm_1(\br) + 
2\eta(\br)\xi(\br) + \eta^*(\br)\xi_1(\br) + 
\lgl \psi_1^\dgr(\br) \psi_1(\br) \psi_1(\br) \psi_1(\br) \rgl 
\right ] \; +
$$
\be
\label{326}
 + \; 2 \left [ \eta^2(\br) + \sgm_1(\br) \right ] \Phi(0) \; .
\ee
The quantity $\Phi(0)$, under the local potential (316), is not defined
and requires to be specified by additional constraints. 

A straightforward formal way of giving some meaning to this quantity would 
be by remembering that the delta potential (316) is the limiting form of 
a potential with a finite interaction range $r_0$, such that $r_0 \ll a$. 
For instance, potential (316) could be treated as the limiting form of the 
potential
\be
\label{327}
\Phi(\br) = A \exp \left ( -\; \frac{3r^2}{2r_0^2} 
\right ) \;  ,
\ee
where $r_0 \ra 0$, so that the integral 
$$
\Phi_0 \equiv \int \Phi(\br) \; d\br
$$
is fixed as in Eq. (316). The interaction radius is defined as
\be
\label{328}
 r_0^2 \equiv \frac{1}{\Phi_0} \int r^2 \Phi(\br) \; d\br \; .
\ee
These requirements give
$$
 A = \left ( \frac{3}{2\pi} \right )^{3/2} 
\frac{\Phi_0}{r_0^3} \; .
$$
Then, for potential (327), the quantity $\Phi(0)$ should be defined as
\be
\label{329}
 \Phi(0) = A = 3 \sqrt{\frac{6}{\pi} } \; 
\frac{a_s}{mr_0^3} \; .
\ee
 
However, we have to always remember that the local interaction potential
(316) is an {\it effective} potential modeling particle interactions for 
the processes occurring at the interparticle distance much larger than the 
interaction radius. In order to characterize the processes at short 
distance, one has to use a different effective potential that takes into 
account particle correlations [5,81,83]. The latter, in particular, show
that two particles cannot exist at the same spatial point. This is equivalent 
to saying that $\Phi(0)$ must be set to zero.  

The four-operator term can be simplified as 
\be
\label{330}
\lgl \psi_1^\dgr(\br) \psi_1(\br) \psi_1(\br) \psi_1(\br) \rgl
 = 3 \rho_1(\br) \sgm_1(\br) \;  ,
\ee
while the three-operator terms are left untouched. 

Then the evolution equation (103) for the anomalous average leads to
$$
i\; \frac{\prt}{\prt t}\; \sgm_1(\br) = 2K(\br) + 
2 ( U - \mu_1 ) \sgm_1(\br) \; +
$$
\be
\label{331}
+\; 2 \Phi_0 \left [ \eta^2(\br) \rho_1(\br) +
2\rho_0(\br) \sgm_1(\br) + 3\rho_1(\br)\sgm_1(\br) +
2\eta(\br) \xi(\br) + \eta^*(\br) \xi_1(\br) \right ] \; ,
\ee
with the anomalous kinetic-energy density $K({\bf r})$ given by Eq. (104) and with 
$\Phi(0)$ set to zero.

The derived evolution equations can be used for studying the initiation 
of Bose-Einstein condensation and also the decoherence processes in  
systems with spontaneous symmetry breaking. For finite systems, with $N$
degrees of freedom, coherence can persist [92-95] during the time not 
longer than that of order $N/T$.

\subsection{Equilibrium Systems}

Considering equilibrium systems, we follow Secs. 6 and 7, substituting 
in the corresponding equations the local potential (316). The grand 
Hamiltonian (219), in the HFB approximation, reads as
$$
H_{HFB} = E_{HFB} + \int \psi_1^\dgr(\br) \left ( -\; 
\frac{\nabla^2}{2m} + U - \mu_1 \right ) \psi_1(\br) \; d\br \; +
$$
\be
\label{332}
+ \; \Phi_0 \int \left [ 
2\rho(\br) \psi_1^\dgr(\br) \psi_1(\br)
+ \frac{1}{2} \; \sgm(\br) \psi_1^\dgr(\br) \psi_1^\dgr(\br)
+ \frac{1}{2} \; \sgm^*(\br) \psi_1(\br) \psi_1(\br) 
\right ] d\br \;  ,
\ee
where the nonoperator term is
\be
\label{333}
 E_{HFB} = H^{(0)} -\; \frac{\Phi_0}{2} \int \left [
2\rho_1^2(\br) + \sgm_1^2(\br) \right ] d\br \; ,
\ee
and the notation
\be
\label{334}
\rho(\br) \equiv \rho_0(\br) + \rho_1(\br) \; , 
\qquad \sgm(\br) \equiv \eta^2(\br) + \sgm_1(\br)
\ee
is used. The condensate-function equation (223) becomes
\be
\label{335}
\left ( -\; \frac{\nabla^2}{2m} + U \right ) \eta(\br)
+ \Phi_0 \left [ \rho_0(\br) \eta(\br) + 
2\rho_1(\br) \eta(\br) + \sgm_1(\br) \eta^*(\br) + 
\xi(\br) \right ] = \mu_0 \eta(\br) \; .
\ee
For the condensate chemical potential (224), we have
$$
\mu_0 = \frac{1}{N_0} \int \eta^*(\br) \left ( -\;
\frac{\nabla^2}{2m} + U \right ) \eta(\br)\; d\br \; +
$$
\be
\label{336}
+ \; \frac{\Phi_0}{N_0} \int \left [ \rho_0^2(\br) +
2\rho_0(\br) \rho_1(\br) + \left ( \eta^*(\br) \right )^2
\sgm_1(\br) \right ] d\br \;  .
\ee

Employing the Bogolubov transformations (225) yields the Bogolubov 
equations
\be
\label{337}
 \hat\om(\br) u_k(\br) + \Dlt(\br) v_k(\br) = 
\ep_k u_k(\br) \; , \qquad
\hat\om(\br) v_k(\br) + \Dlt^*(\br) u_k(\br) = 
- \ep_k v_k(\br) \;
\ee
replacing Eqs. (230), with the operator
\be
\label{338}
 \hat\om(\br) \equiv -\; \frac{\nabla^2}{2m} + U(\br) -
\mu_1 + 2\Phi_0 \rho(\br) \; ,
\ee
instead of Eq. (228), and with
\be
\label{339}
 \Dlt(\br) \equiv \Phi_0 \left [ \eta^2(\br) + \sgm_1(\br)
\right ]  \; ,
\ee
instead of Eq. (229). The chemical potential $\mu_1$ is defined by 
the requirement that the spectrum $\varepsilon_k$ be gapless. 

The grand potential has the same form (240), with
$$
E_B = -\; \frac{\Phi_0}{2} \int \left [ \rho_0^2(\br) +
4\rho_0(\br) \rho_1(\br) + 2 \rho_1^2(\br) +
2\left (\eta^*(\br) \right )^2 \sgm_1(\br) + \sgm_1^2(\br) 
\right ] d \br \; -
$$
\be
\label{340}
- \; \sum_k \int \ep_k | v_k(\br) |^2 d\br \;  ,
\ee
in agreement with Eq. (241).

\subsection{Uniform Systems}

Resorting to the Fourier transformation (173) gives the following 
terms of the grand Hamiltonian (67). The zero-order term (317) is
\be
\label{341}
 H^{(0)} = \left ( \frac{1}{2}\; \rho^2 \Phi_0 -
\mu_0 \rho_0 \right ) V \; .
\ee
The first-order term is, as always, zero. The second-order term 
(318) reads as
\be
\label{342}
H^{(2)} = \sum_{k\neq 0} \left ( \frac{k^2}{2m} + 2\rho_0\Phi_0
-\mu_1 \right ) a_k^\dgr a_k \; 
+ \; \frac{1}{2} \; \rho_0 \Phi_0 \sum_{k\neq 0} \left (
a_k^\dgr a_{-k}^\dgr + a_{-k} a_k \right ) \;  .
\ee
The third-order term (319) yields
\be
\label{343}
 H^{(3)} = \sqrt{\frac{\rho_0}{V}} \; \Phi_0 \;
{\sum_{kp}}' \left ( a_k^\dgr a_{k+p} a_{-p} +
a_{-p}^\dgr a_{k+p}^\dgr a_k \right ) \; ,
\ee
where
$$
 \bk \neq 0 \; , \qquad \bk + \bp \neq 0 \; , 
\qquad \bp \neq 0 \; .
$$
And the fourth-order term (320) becomes
\be
\label{344}
 H^{(4)} = \frac{\Phi_0}{2V} \sum_q \; {\sum_{kp}}'
a_k^\dgr a_p^\dgr a_{p+q} a_{k-q} \; ,
\ee
in which
$$
\bk \neq 0 \; , \qquad \bp \neq 0 \; , \qquad 
\bp + \bq \neq 0 \; , \qquad \bk - \bq \neq 0 \;   .
$$
Equation (183) defines the condensate chemical potential
\be
\label{345}
 \mu_0 = \rho \Phi_0 + \frac{\Phi_0}{V} \sum_{k\neq 0} 
\left ( n_k + \sgm_k + \frac{\xi_k}{\sqrt{\rho_0} } 
\right ) \;.
\ee
  
The grand Hamiltonian (332) in the HFB approximation transforms to
\be
\label{346}
 H_{HFB} = E_{HFB} + \sum_{k\neq 0} \om_k a_k^\dgr a_k +
\frac{\Dlt}{2} \sum_{k\neq 0} \left (
a_k^\dgr a_{-k}^\dgr + a_{-k} a_k \right ) \;  ,
\ee
where the nonoperator term (333) is given by the expression 
\be
\label{347}
 \frac{E_{HFB}}{V} = \frac{\Phi_0}{2} \left (
\rho_0^2 - 2 \rho_1^2 + \sgm_1^2 \right ) -
\mu_0 \rho_0  \; .
\ee
Also, here the notation
\be
\label{348}
\om_k = \frac{k^2}{2m} - \mu_1 + 2\rho \Phi_0
\ee
is used, and, instead of Eq. (339), we have
\be
\label{349}
\Dlt = \Phi_0 ( \rho_0 + \sgm_1 ) \;  .
\ee
The diagonalization of Hamiltonian (346) is done by the Bogolubov 
canonical transformations (246), resulting in the Hamiltonian of 
the Bogolubov form (231).

The condensate chemical potential (253), or (336), reads as
\be
\label{350}
\mu_0 = \Phi_0 ( \rho_0 + 2\rho_1 + \sgm_1 ) \;    .
\ee
And the chemical potential (252) becomes
\be
\label{351}
 \mu_1 = \Phi_0 ( \rho_0 + 2\rho_1 - \sgm_1 ) \;   .
\ee
Using the latter in Eq. (348) gives
\be
\label{352}
 \om_k = \frac{k^2}{2m} + \Phi_0(\rho_0 + \sgm_1 )  .
\ee

The long-wave spectrum is acoustic,
\be
\label{353}
 \ep_k \simeq ck \qquad (k \ra 0 ) \; ,
\ee
with the sound velocity
\be
\label{354}
c \equiv \sqrt{ \frac{\Dlt}{m} } \; ,
\ee
which is defined by the equation
\be
\label{355}
 mc^2 = \Phi_0 (\rho_0 + \sgm_1 ) \; .
\ee
Combining Eqs. (352) and (355) yields
\be
\label{356}
\om_k = mc^2 + \frac{k^2}{2m} \;  .
\ee

The solution to the Bogolubov equations (247) results in the 
Bogolubov spectrum
\be
\label{357}
\ep_k = \sqrt{ (ck)^2 + 
\left ( \frac{k^2}{2m}\right )^2 } \;  .
\ee
 
Calculating Eq. (340), we resort to the dimensional regularization 
giving
$$
\int ( \ep_k - \om_k ) \; \frac{d\bk}{(2\pi)^3} =
\frac{16m^4 c^5}{15\pi^2} \; .
$$
Then Eq. (340) reduces to
\be
\label{358}
 \frac{E_B}{V} = -\; \frac{\Phi_0}{2} \left [
\rho^2 + 2\rho_0 (\rho_1 + \sgm_1 ) +
\rho_1^2 + \sgm_1^2 \right ] + 
\frac{8m^4 c^5}{15\pi^2} \;  .
\ee

The system pressure can be expressed through the grand potential (240),
which gives
\be
\label{359}
 p \equiv -\; \frac{\Om}{V} = -\; \frac{E_B}{V} - T \int
\ln \left ( 1 - e^{-\bt\ep_k} \right )
\frac{d\bk}{(2\pi)^3} \; .
\ee
The integral in Eq. (359) corresponds to thermal pressure. It can be 
calculated by transforming it to the form
$$
\int \ln \left ( 1 - e^{-\bt\ep_k} \right ) 
\frac{d\bk}{(2\pi)^3}
= \frac{(mc)^3}{2\sqrt{2}\pi^2} \int_0^\infty \ln \left [
1 - \exp \left ( -\; \frac{mc^2}{T}\; x \right ) \right ] \;
\frac{\left ( \sqrt{1+x^2}-1\right )^{1/2} x dx}{\sqrt{1+x^2}}\; .
$$
At low temperatures, such that $T \ll m c^2$, one can expand the 
integral as
$$
\int \ln \left ( 1 - e^{-\bt\ep_k} \right )
\frac{d\bk}{(2\pi)^3} \simeq 
- \frac{(mc)^3}{2\pi^2} \left [ \left ( 
\frac{T}{mc^2}\right )^3 - \frac{15}{2} \left ( 
\frac{T}{mc^2} \right )^5 \right ] \;  .
$$
Therefore, the zero-temperature pressure is
\be
\label{360}
p = - \; \frac{E_B}{V} \qquad ( T = 0 ) \;  .
\ee

The internal energy is given by the expression 
\be
\label{361}
 E \equiv \lgl H \rgl + \mu N \;   ,
\ee 
in which the average of the grand Hamiltonian is as in Eq. (59) and 
the system chemical potential is defined in Eq. (60). At zero 
temperature, the internal energy (361) yields the ground-state energy
\be
\label{362}
 E_0 = E_B + \mu N \qquad (T = 0) \; ,
\ee
where we take into account that
$$
\lgl H \rgl = \lgl H_B \rgl = E_B \qquad (T = 0 ) \;  .
$$
Then the ground-state energy is given by the equation
\be
\label{363}
 \frac{E_0}{N} = \frac{2\pi a_s}{m\rho} \left ( \rho^2 +
\rho_1^2 - 2\rho_1 \sgm_1 - \sgm_1^2 \right ) +
\frac{8m^4c^5}{15\pi^2\rho} \;  .
\ee

For convenience, let us introduce the dimensionless ground-state energy
\be
\label{364}
e_0 \equiv \frac{2m E}{N\rho^{3/2}}   
\ee
and the dimensionless {\it gas parameter}
\be
\label{365}
 \gm \equiv \rho^{1/3} a_s \;  .
\ee
The ground-state energy (364) at weak interactions, when $\gamma \ll 1$, 
allows [69,70] for the expansion
\be
\label{366}
 e_0 \simeq 4\pi \gm + \frac{512}{15} \; \sqrt{\pi} \;\gm^{5/2}
+ \frac{512}{9} \; \gm^4 \; .
\ee
The first two terms here reproduce the Lee-Huang-Yang result [96-98].
When particle interactions are strong, so that $\gamma \gg 1$, then [69]
one has 
\be
\label{367}
 e_0 \simeq 8\pi\gm + \frac{6}{5} \left ( 9\pi^4 \right )^{1/3}
- \; \frac{3}{4} \left ( 3\pi^5 \right )^{1/3} \frac{1}{\gm} +
\frac{1}{64} \left ( 3\pi^8 \right )^{1/3} \frac{1}{\gm^4} \;  .
\ee

\subsection{Atomic Fractions}

For the local interaction potential (316), it is straightforward to 
calculate all atomic densities and fractions. The condensate density
\be
\label{368}
\rho_0 = \rho -\rho_1
\ee
is expressed through the density of uncondensed particles
\be
\label{369}
 \rho_1 = \int n_k \; \frac{d\bk}{(2\pi)^3} =
\int \left [ \frac{\om_k}{2\ep_k} \; 
\coth \left ( \frac{\ep_k}{2T} \right )  - 
\; \frac{1}{2} \right ] \frac{d\bk}{(2\pi)^3}
\ee
that can be rewritten as
\be
\label{370}
\rho_1 = \frac{1}{2} \int \left ( \frac{\om_k}{\ep_k} - 1 
\right )
\frac{d\bk}{(2\pi)^3} \; + \; \int \frac{\om_k}{2\ep_k} 
\left [  \coth \left ( \frac{\ep_k}{2T} \right ) -1 \right ] 
\frac{d\bk}{(2\pi)^3} \; .
\ee
The anomalous average
\be
\label{371}
\sgm_1 = \int \sgm_k \; \frac{d\bk}{(2\pi)^3} = -
\int \frac{mc^2}{2\ep_k} \; 
\coth \left ( \frac{\ep_k}{2T} \right )
\frac{d\bk}{(2\pi)^3}
\ee
can be treated as in Sec. 7, by separating the term
\be
\label{372}
 \sgm_0 \equiv -\int \frac{mc^2}{2\ep_k}\; 
\frac{d\bk}{(2\pi)^3} \; ,
\ee
which gives
\be
\label{373}
\sgm_1 = \sgm_0 - \int \frac{mc^2}{2\ep_k} \left [
\coth \left ( \frac{\ep_k}{2T} \right ) -1 \right ]
\frac{d\bk}{(2\pi)^3} \; .
\ee
   
Expression (372) diverges, but can be regularized invoking the 
dimensional regularization, as in Sec. 7, resulting in
$$
 \int \frac{1}{\ep_K}\; \frac{d\bk}{(2\pi)^3} =
-\; \frac{2m}{\pi^2} \; \sqrt{m\rho_0\Phi_0 } \; .
$$
Then Eq. (372) becomes
\be
\label{374}
 \sgm_0 = \left ( 
\frac{mc}{\pi} \right )^2 \sqrt{m\rho_0\Phi_0 } \;   .
\ee
The anomalous average (373), with the separated term (374), is valid
if at least one of conditions (308) is satisfied. 

In the vicinity of the condensation temperature $T_c$, it is necessary 
to use the anomalous average in the form
\be
\label{375}
 \sgm_1 \simeq -\; \frac{mc^2 T}{2\pi}  \qquad 
( T \ra T_c) \; ,
\ee    
which follows from Eq. (371) and guarantees the second-order phase 
transition.

The density of uncondensed particles (370) can be represented as
\be
\label{376}
 \rho_1 = \frac{m^3c^3}{3\pi^2} \left \{ 1 +
\frac{3}{2\sqrt{2}} \int_0^\infty 
\left ( \sqrt{1+x^2} -1 \right )^{1/2}
\left [ \coth \left ( \frac{mc^2}{2T}\; x \right ) -1 
\right ] dx \right \} \; ,
\ee
and the anomalous average (373), as
\be
\label{377}
\sgm_1 = \sgm_0 - \; \frac{m^3c^3}{2\sqrt{2}\pi^2} \int_0^\infty
\frac{\left ( \sqrt{1+x^2}-1\right )^{1/2}}{\sqrt{1+x^2} } \left [
\coth \left ( \frac{mc^2}{2T}\; x \right ) -1 \right ] dx \;  .
\ee

The superfluid fraction
\be
\label{378}
n_s = 1 - \; \frac{2Q}{3T}
\ee
is expressed through the dissipated heat
\be
\label{379}
Q = \frac{\Dlt^2(\hat\bP)}{2mN} =
\frac{\lgl\hat\bP^2\rgl}{2mN} \;  .
\ee
In the HFB approximation, we get
\be
\label{380}
 Q = \frac{1}{\rho} \int \frac{k^2}{2m} \left ( n_k + n_k^2
-\sgm_k^2 \right ) \frac{d\bk}{(2\pi)^3} \;  ,
\ee
which reduces to
\be
\label{381}
Q = \frac{1}{(4\pi)^2m\rho} \int_0^\infty
\frac{k^4d\bk}{\sinh^2(\ep_k/2T)} \;  .
\ee
The superfluid fraction (378) leads to the superfluid density
\be
\label{382}
 \rho_s = \rho -\; 
\frac{(mc)^5}{6\sqrt{2}\pi^2 mT} \int_0^\infty 
\frac{\left ( \sqrt{1+x^2}-1\right )^{3/2}xdx}
{\sqrt{1+x^2}\sinh^2(mc^2x/2T)} \; .
\ee

The particle densities are related to particle fractions
\be
\label{383}
 n_0 \equiv \frac{\rho_0}{\rho} = 1 - n_1 \; , \qquad 
n_1 \equiv \frac{\rho_1}{\rho} \; , \qquad 
n_s \equiv \frac{\rho_s}{\rho} \; .
\ee 
Also, let us define the dimensionless anomalous average
\be
\label{384}
 \sgm \equiv \frac{\sgm_1}{\rho}  
\ee
and the dimensionless sound velocity
\be
\label{385}
 s \equiv \frac{mc}{\rho^{1/3} } \; .
\ee

At zero temperature, we have
$$
n_0 = 1 - \; \frac{s^3}{3\pi^2} \; , \qquad 
n_1 = \frac{s^3}{3\pi^2} \; , \qquad n_s = 1 \; ,
$$
\be
\label{386}
\sgm = \frac{2s^2}{\pi^{3/2}} \; \sqrt{\gm n_0} \qquad 
(T = 0) \;  .
\ee
When interactions are weak, such that $\gamma \ll 1$, then the 
following expansions are valid for the condensate fraction
\be
\label{387}
 n_0 \simeq 1 - \; \frac{8}{3\sqrt{\pi}} \; \gm^{3/2} - \;
\frac{64}{3\pi} \; \gm^3 - \; 
\frac{640}{9\pi^{3/2}} \; \gm^{9/2} \; ,
\ee
sound velocity
\be
\label{388}
 s \simeq 2 \sqrt{\pi}\; \; \gm^{1/2} + 
\frac{16}{3} \; \gm^2 + \frac{32}{9\sqrt{\pi}} \; \gm^{7/2} - \; 
\frac{3904}{27\pi} \; \gm^5 \; ,
\ee
and the anomalous average
\be
\label{389}
\sgm \simeq \frac{8}{\sqrt{\pi}} \; \gm^{3/2} +
\frac{32}{\pi} \; \gm^3 - \; 
\frac{64}{\pi^{3/2}} \; \gm^{9/2} \;   .
\ee
As is seen, the anomalous average is three times larger than the normal 
fraction of uncondensed particles:
\be
\label{390}
\frac{\sgm}{n_1} \simeq 3 \qquad ( \gm \ll 1 ) \;  .
\ee
This emphasizes again that the anomalous average in no way can be 
neglected for a Bose-condensed system.

For strong interactions, when $\gamma \gg 1$, we find the following 
expansions for the condensate fraction
\be
\label{391}
 n_0 \simeq \frac{\pi}{64} \; \frac{1}{\gm^3} - \;
\frac{1}{512} \left ( \frac{\pi^5}{9} 
\right )^{1/3} \frac{1}{\gm^5} \; ,
\ee
sound velocity
\be
\label{392}
s \simeq \left ( 3\pi^2 \right )^{1/3} - \; \frac{1}{64}
\left ( \frac{\pi^5}{9} \right )^{1/3} \; \frac{1}{\gm^3}\;
 + \; \frac{1}{1536} 
\left ( \frac{\pi^7}{3} \right )^{1/3} \frac{1}{\gm^5} \; ,
\ee
and the anomalous average
\be
\label{393}
 \sgm \simeq \frac{(9\pi)^{1/3}}{4} \; \frac{1}{\gm} - \;
\frac{\pi}{64} \; \frac{1}{\gm^3} - \; \frac{1}{128} 
\left ( \frac{\pi^4}{3} \right )^{1/3} \frac{1}{\gm^4} + 
\frac{1}{512} \left ( \frac{\pi^5}{9} \right )^{1/3} \frac{1}{\gm^5} \; .
\ee
Though now the anomalous average is smaller than $n_1$, but it is much 
larger than the condensate fraction:
\be
\label{394}
\frac{\sgm}{n_0} \simeq 15.516 \gm^2 \qquad (\gm \gg 1) \;  .
\ee
Thus, the anomalous average is always of crucial importance and can never 
be neglected.

At low temperatures, such that
\be
\label{395}
 \frac{T}{mc^2} \ll 1 \; ,
\ee
we find [69-71] the fraction of uncondensed particles
\be
\label{396}
 n_1 \simeq \frac{(mc)^3}{3\pi^2\rho} +
\frac{(mc)^3}{12\rho} \left ( \frac{T}{mc^2} \right )^2 \; ,
\ee
anomalous average
\be
\label{397}
  \sgm \simeq \frac{\sgm_0}{\rho} - \frac{(mc)^3}{12\rho}
\left ( \frac{T}{mc^2} \right )^2 \; ,
\ee
condensate fraction
\be
\label{398}
 n_0 \simeq 1 - \; \frac{(mc)^3}{3\pi^2\rho} -\;
\frac{(mc)^3}{12\rho} \; \left ( \frac{T}{mc^2} \right )^2 \;  ,
\ee
and the superfluid fraction
\be
\label{399}
n_s \simeq 1 - \; \frac{2\pi^2(mc)^3}{45\rho} 
\left ( \frac{T}{mc^2} \right )^4 \;   .
\ee
Notice that the temperature corrections for $\sigma$ are the same as 
for $n_1$ and $n_0$.

Bose-Einstein condensation happens at the temperature 
\be
\label{400}
 T_c = \frac{2\pi}{m} \left [ \frac{\rho}{\zeta(3/2)} 
\right ]^{2/3} \;  .
\ee
In the critical region, where $T \ra T_c$, so that
\be
\label{401}
  \frac{mc^2}{T} \ll 1 \; ,
\ee 
we find [69-71] the expansions
$$
n_1 \simeq \left ( \frac{T}{T_c} \right )^{3/2} +
\frac{(mc)^3}{3\pi^2\rho} \; , \qquad
\sgm \simeq - \; \frac{m^2 c T}{2\pi\rho} \; , \qquad
n_0 \simeq 1 - \left ( \frac{T}{T_c} \right )^{3/2} -\;
\frac{(mc)^3}{3\pi^2\rho} \; ,
$$
\be
\label{402}
n_s \simeq 1 - \left ( \frac{T}{T_c} \right )^{3/2} + \;
\frac{\zeta(1/2)}{\zeta(3/2)} 
\left ( \frac{T}{T_c} \right )^{1/2} \; 
\frac{mc^2}{T_c} \; .
\ee
The superfluid fraction disappears together with the condensate 
fraction.

Passing to dimensionless quantities, it is convenient to consider the 
temperature deviation
\be
\label{403}
\tau \equiv 1 - \; \frac{T}{T_c} \qquad ( T \leq T_c )
\ee
from the dimensionless transition temperature
\be
\label{404}
t_c \equiv \frac{mT_c}{\rho^{1/3}} =
\frac{2\pi}{[\zeta(3/2)]^{2/3}} = 3.312498 \; .
\ee
Then we obtain
$$
s \simeq \frac{3\pi}{t_c}\; \tau + \frac{9\pi}{4t_c} \left (
1 - \; \frac{2\pi}{\gm t_c^2} \right ) \tau^2 \; , \qquad
n_1 \simeq 1 - \; \frac{3}{2} \; \tau + 
\frac{3}{8} \; \tau^2 \; ,
$$
$$
\sgm \simeq -\; \frac{3}{2} \; \tau + \frac{3}{8} \left (
1 + \frac{6\pi}{\gm t_c^2} \right ) \tau^2 \; , \qquad
n_0 \simeq \frac{3}{2} \; \tau - \; \frac{3}{8} \; \tau^2 \; ,
$$
\be
\label{405}
 n_s \simeq \frac{3}{2} \; \tau - \; \frac{3}{8} \left (
1 + \frac{132.411}{t_c^3} \right ) \tau^2 \; .
\ee

As is evident, though the anomalous average tends to zero, as 
$T \ra T_c$, but it is of the same order as the condensate fraction, 
hence, again the anomalous average cannot be omitted. If it were 
neglected, the transition would become of first order, which is 
principally incorrect [67,71]. While accurately taking account of 
the anomalous average renders the Bose-Einstein condensation the 
correct second-order transition, as is obvious from expansions (405).

\section{Disordered Bose Systems}

\subsection{Random Potentials}

The properties of Bose systems can be essentially changed by imposing 
external spatially random potentials. In this section, the theory is 
presented for the case when such random potentials are imposed on a 
uniform system. There is vast literature studying Bose systems inside 
randomly perturbed periodic lattices (see the review article [12] and 
recent works [99-101], where further references can be found). But this 
is a different problem that is not touched in the present section. In 
several papers (e.g. [102-106]) the influence of weak disorder on a 
uniform Bose-condensed system has been studied. Strong disorder can be 
treated by means of numerical Monte Carlo simulations [107]. In the 
present section, the analytical theory is described, which is valid for 
arbitrarily strong disorder. The consideration below is based on Refs. 
[108-110].   

The system is described by the grand Hamiltonian
\be
\label{406}
 H = \hat H - \mu_0 N_0 - \mu_1 \hat N_1 - \hat\Lbd \; ,
\ee
with the energy Hamiltonian
\be
\label{407}
\hat H = \int \hat\psi^\dgr(\br) \left [ -\; 
\frac{\nabla^2}{2m} + U(\br) + \xi(\br) \right 
] \hat\psi(\br) \; d\br \; + \; \frac{\Phi_0}{2} \int 
\hat\psi^\dgr(\br) \hat\psi^\dgr(\br) 
\hat\psi(\br) \hat\psi(\br) \; d\br
\ee
containing a random external potential $\xi({\bf r})$. Other notations
are the same as in the previous sections. 

The random potential, without the loss of generality, can be treated as
zero-centered, such that
\be
\label{408}
 \lgl \lgl \xi(\br) \rgl \rgl = 0 \; .
\ee 
The double brackets imply the related stochastic averaging [111]. 
The random-potential correlations are characterized by the correlation 
function
\be
\label{409}
 \lgl \lgl \xi(\br) \xi(\br') \rgl \rgl = R(\br-\br') \; .
\ee
One can use the Fourier transformations
$$
\xi(\br) = \frac{1}{\sqrt{V}} 
\sum_k \xi_k e^{i\bk\cdot\br} \; ,
\qquad
\xi_k = \frac{1}{\sqrt{V}} 
\int \xi(\br) e^{-i\bk\cdot\br} d\br \; ,
$$
\be
\label{410}
R(\br) = \frac{1}{V} \sum_k R_k e^{i\bk\cdot\br} \; ,
\qquad
R_k = \int R(\br) e^{-i\bk\cdot\br} d\br \;   .
\ee
Then Eq. (409) yields the correlators
\be
\label{411}
 \lgl \lgl \xi_k^* \xi_p \rgl \rgl = \dlt_{kp} R_k \; , 
\qquad
\lgl \lgl \xi_k \xi_p \rgl \rgl = \dlt_{-kp} R_k \; .
\ee

The quantum statistical averaging, involving a Hamiltonian $H$, 
for an operator $\hat {A}$, is denoted as  
\be
\label{412}
\lgl \hat A \rgl_H \equiv 
\frac{{\rm Tr}\hat Ae^{-\bt H}}{{\rm Tr} e^{-\bt H}} \;   .
\ee
The total averaging, including both the quantum and stochastic averagings,
is denoted as 
\be
\label{413}
 \lgl \hat A \rgl \equiv 
\lgl \lgl \left ( \lgl \hat A \rgl_H \right ) \rgl \rgl \;  .
\ee

The grand thermodynamic potential is given by the expression
\be
\label{414}
 \Om \equiv - T \lgl \lgl \ln {\rm Tr} e^{-\bt H} \rgl \rgl \; ,
\ee
corresponding to the frozen disorder.   

In addition to the particle densities, considered in the previous sections,
for a random system, it is necessary to introduce one more density. This is
the {\it glassy density} [108]
\be
\label{415}
 \rho_G \equiv \frac{1}{V} \int 
\lgl \lgl | \lgl \psi_1(\br) \rgl_H |^2 \rgl \rgl d\br \; .
\ee
With the Fourier transform
$$
\psi_1(\br) = \frac{1}{\sqrt{V}} 
\sum_{k\neq 0} a_k e^{i\bk\cdot\br} \;  ,
$$  
we come to
\be
\label{416}
\rho_G = \frac{1}{V} \sum_{k\neq 0} \lgl \lgl | \al_k|^2 \rgl \rgl \;  ,
\ee
where
\be
\label{417}
 \al_k \equiv \lgl a_k \rgl_H \; .
\ee
Because of condition (50), we have
\be
\label{418}
 \lgl \lgl \al_k \rgl \rgl = \lgl a_k \rgl = 0 \; .
\ee
However, quantity (417) is not zero. The {\it glassy fraction} is given 
by
\be
\label{419}
n_G \equiv \frac{\rho_G}{\rho} = \frac{1}{N} \int
\lgl \lgl | \lgl \psi_1(\br) \rgl_H |^2 \rgl \rgl d\br \;  ,
\ee
which can be represented as
\be
\label{420}
n_G = \frac{1}{N} \sum_{k\neq 0} 
\lgl \lgl | \al_k|^2 \rgl \rgl \;  .
\ee

To better illustrate the idea of the approach we aim at developing, 
let us set $U = 0$. This will simplify the consideration. Then the 
grand Hamiltonian writes as
\be
\label{421}
 H = \sum_{n=0}^4 H^{(n)} \; + \; H_\xi \; ,
\ee
where the first sum consists of terms (341) to (344), while the last 
part
\be
\label{422}
H_\xi = \rho_0 \xi_0 \sqrt{V} + \sqrt{\rho_0}\; \sum_{k\neq 0}
\left ( a_k^\dgr \xi_k + \xi_k^* a_k \right ) + \frac{1}{\sqrt{V}}
\sum_{kp(\neq 0)} a_k^\dgr a_p \xi_{k-p}
\ee
is due to the presence of the random potential.

\subsection{Stochastic Decoupling}

The sum in Hamiltonian (421) can be treated in the standard way by 
resorting to the HFB approximation, as in the previous sections. But 
the part (422), characterizing the action on particles of the random 
potential, has to be treated with caution. If one would apply to this 
part the simple HFB-type approximation 
$$
 a_k^\dgr a_p \xi_{k-p} \ra \lgl a_k^\dgr a_p \rgl \xi_{k-p}
+ a_k^\dgr a_p \lgl \xi_{k-p} \rgl - 
\lgl a_k^\dgr a_p \rgl \lgl \xi_{k-p} \rgl \; ,
$$
then the influence of this part, because of Eq. (408), would reduce 
to the trivial mean-field form  
$$
 \frac{1}{N} \sum_{kp(\neq 0)} a_k^\dgr a_p \xi_{k-p} \ra 
\rho_1 \xi_0 \sqrt{V} \;  ,
$$
containing no nontrivial information on the action of the random 
potential on particles. 

In order not to loose the information on the influence of the 
random potential, we employ the idea of {\it stochastic decoupling} 
that has been used earlier for taking into account stochastic 
effects in different systems, such as resonant atoms [112-115] 
and spin assemblies [116-121]. 

In the present case, the idea is that the simplification of the 
third-order expression in the last term of Eq. (422) should include 
only the quantum statistical averaging, but not the stochastic 
averaging, thus retaining undisturbed stochastic correlations. This 
idea can be represented in several equivalent ways. We can write
\be
\label{423}
\lgl a_k^\dgr a_p \xi_{k-p} \rgl = 
\lgl \lgl \al_k^* \al_p \xi_{k-p} \rgl \rgl \;  ,
\ee
which is equivalent to 
\be
\label{424}
 \lgl a_k^\dgr a_p \xi_{k-p} \rgl_H = 
\al_k^* \al_p \xi_{k-p} \; .
\ee
In turn, the latter is equivalent to the decoupling
\be
\label{425}
a_k^\dgr a_p = a_k^\dgr \al_p + \al_k^* a_p - 
\al_k^* \al_p \;  .
\ee

Then, we introduce [108-110] the {\it nonuniform canonical transformation}
\be
\label{426}
a_k = u_k b_k + v_{-k}^* b_{-k}^\dgr + w_k \vp_k \;  ,
\ee
whose coefficient functions are defined so that to diagonalize the grand 
Hamiltonian (421) in terms of the operators $b_k$. The latter are treated 
as quantum variables with the condition
\be
\label{427}
 \lgl b_k \rgl_H = 0 \; .
\ee
The variables $\varphi_k$ represent stochastic fields. In view of Eqs. (426)
and (427), we have
\be
\label{428}
 \al_k \equiv \lgl a_k \rgl_H = w_k \vp_k \; .
\ee

Diagonalizing Hamiltonian (421) results in the relations
$$
u_k^2 = \frac{\om_k+\ep_k}{2\ep_k} \; , 
\qquad v_k^2 = \frac{\om_k-\ep_k}{2\ep_k} \; ,
$$
\be
\label{429}
 u_k v_k = -\; \frac{mc^2}{2\ep_k} \; , 
\qquad w_k = -\; \frac{1}{2\om_k+mc^2} \;   ,
\ee
in which
\be
\label{430}
 \om_k = \frac{k^2}{2m} + mc^2 \; .
\ee
The Bogolubov spectrum 
\be
\label{431}
  \ep_k =\sqrt{(ck)^2 + \left ( \frac{k^2}{2m} \right )^2 } 
\ee
has the standard form, with the sound velocity defined by the equation
\be
\label{432}
 mc^2 = \left ( n_0 + \frac{\sgm_1}{\rho} 
\right ) \rho \Phi_0 \; .
\ee
The stochastic field satisfies the Fredholm equation
\be
\label{433}
\vp_k = \sqrt{\rho_0}\; \xi_k - \; \frac{1}{\sqrt{V}}
\sum_p \frac{\xi_{k-p}\vp_p}{\om_p+mc^2} \;  .
\ee

Hamiltonian (421) acquires the diagonal form
\be
\label{434}
 H = E_B + \sum_k \ep_k b_k^\dgr b_k \; +\; H_{ran} \;  ,
\ee
where the last term
\be
\label{435}
H_{ran} \equiv \vp_0 \; \sqrt{N_0}
\ee
characterizes the explicit influence of the random potential on the system 
energy.

For the particle momentum distribution (185), we get
\be
\label{436}
 n_k = \frac{\om_k}{2\ep_k} \;
\coth \left ( \frac{\ep_k}{2T} \right ) - \; \frac{1}{2} 
+ \lgl \lgl | \al_k|^2 \rgl \rgl \; ,
\ee
and for the anomalous average (187),
\be
\label{437}
 \sgm_k = - \; \frac{mc^2}{2\ep_k} \; 
\coth \left ( \frac{\ep_k}{2T} \right )  + 
\lgl \lgl | \al_k|^2 \rgl \rgl \; .
\ee
Expressions (436) and (437) possess, as compared with Eqs. (264) and (265),
additional terms caused by the random potential. From Eqs. (428) and (429), 
it follows that
\be
\label{438}
 \lgl \lgl | \al_k|^2 \rgl \rgl = 
\frac{\lgl\lgl | \vp_k|^2 \rgl\rgl}{(\om_k+mc^2)^2} \;  .
\ee

The partial chemical potentials (350) and (351) are of the same form
\be
\label{439}
\mu_0 = ( \rho + \rho_1 + \sgm_1 ) \Phi_0 \; , \qquad
\mu_1 = (\rho + \rho_1 -\sgm_1 ) \Phi_0 \;  .
\ee
But the quantities
$$
\rho_1 = \frac{1}{V} \sum_{k\neq 0} n_k \; \qquad
\sgm_1 = \frac{1}{V} \sum_{k\neq 0} \sgm_k
$$
entering them are now different. The density of uncondensed particles
becomes the sum of two terms,
\be
\label{440}
\rho_1 = \rho_N + \rho_G \;  .
\ee
The first term is the normal density
\be
\label{441}
\rho_N = \frac{1}{2} \int \left [ \frac{\om_k}{\ep_k} \;
\coth\left ( \frac{\ep_k}{2T}\right ) - 1 \right ] 
\frac{d\bk}{(2\pi)^3}  
\ee
that, as earlier, is due to finite temperature and interactions, while 
the second term,
\be 
\label{442}
 \rho_G = \int \frac{\lgl\lgl|\vp_k|^2\rgl\rgl}{(\om_k+mc^2)^2}
\frac{d\bk}{(2\pi)^3} \;  ,
\ee
is the glassy density produced by the random potential. 

The anomalous average is also the sum of two terms,
\be
\label{443}
\sgm_1 = \sgm_N + \rho_G \;  .
\ee
The first term is
\be
\label{444}
 \sgm_N = -\; \frac{1}{2} \int \frac{mc^2}{\ep_k}\; 
\coth \left ( \frac{\ep_k}{2T} 
\right ) \frac{d\bk}{(2\pi)^3} \; ,
\ee
while the second term, caused by the presence of the random potential,
coincides with the glassy density (442). 

The partial chemical potentials (439), with expressions (440) and (443),
become
\be
\label{445}
 \mu_0 = ( \rho + \rho_N + \sgm_N + 2\rho_G ) \Phi_0 \; ,
\qquad \mu_1 = (\rho + \rho_N - \sgm_N ) \Phi_0 \; ,
\ee
essentially differing from each other.

The superfluid density (378) requires the knowledge of the dissipated 
heat (379). The latter also reduces to the two-term sum
\be
\label{446}
 Q = Q_N + Q_G \; .
\ee
The first term is analogous to Eq. (380) giving
\be
\label{447}
Q_N = \frac{1}{8m\rho^2} \int 
\frac{k^2}{\sinh^2(\ep_k/2T)} \;
\frac{d\bk}{(2\pi)^3} \; .
\ee
And the second term
\be
\label{448}
Q_G = \frac{1}{2m\rho} \int 
\frac{k^2\lgl\lgl|\vp_k|^2\rgl\rgl}{\ep_k(\om_k+mc^2)} \;
\coth \left ( \frac{\ep_k}{2T}\right ) \;
\frac{d\bk}{(2\pi)^3}
\ee
is the heat dissipated by the glassy fraction. Thus, the superfluid 
density (378) takes the form
\be
\label{449}
n_s = 1 - \; \frac{2Q_N}{3T} \; -\; \frac{2Q_G}{3T} \;  .
\ee

\subsection{Perturbation-Theory Failure}

Considering the case of weak disorder, it is tempting to resort to 
perturbation theory with respect to disorder strength. In doing this,
one has to keep in mind that such a perturbation theory can fail. 
Therefore, the results derived by means of perturbation theory may be
not reliable. To illustrate this, let us consider the average energy
\be
\label{450}
 E_{ran} \equiv \lgl H_{ran} \rgl = 
\lgl\lgl \vp_0 \rgl\rgl \sqrt{N_0} \; ,
\ee
related to the random term (435). 

Assuming that disorder is weak, one could think that Eq. (433) could be 
treated perturbatively, by means of the iteration procedure starting 
with
$$
\vp_k^{(0)} = \sqrt{\rho_0}\; \xi_k \; .
$$
The first iteration gives
$$
 \vp_k^{(1)} = \sqrt{\rho_0}\; \xi_k - \sqrt{\frac{\rho_0}{V}}\; 
\sum_p \frac{\xi_{k-p}\xi_p}{\om_p+mc^2} \; .
$$
Using this in Eq. (450) yields
$$
 E_{ran}^{(1)} = - \rho_0 \sum_p 
\frac{\lgl\lgl|\xi_p|^2\rgl\rgl}{\om_p+mc^2} \; .
$$
In view of correlators (411), one gets
\be
\label{451}
  E_{ran}^{(1)} = - \int \frac{N_0R_p}{\om_p+mc^2} \;
\frac{d\bp}{(2\pi)^3} \; .
\ee
That is, the direct influence of the random potential would lead to the 
decrease of the system energy. It is exactly this expression (451) that 
has been obtained by several authors (see, e.g., [122]) employing 
perturbation theory.

However, from Eqs. (418) and (428), involving no perturbation theory,
it is seen that
$$
 \lgl \lgl \al_k \rgl \rgl = 0 \; , \qquad 
\lgl\lgl \vp_k \rgl\rgl = 0 \; .
$$
Consequently, the random energy (450) is exactly zero:
\be
\label{452}
E_{ran} \equiv \lgl H_{ran} \rgl = 0 \;  .
\ee 

Also, using perturbation theory in calculating sound velocity, some 
authors (e.g. [122]) find that the speed of sound would increase due 
to the random potential. Contrary to this, in our theory [108-110], 
the sound velocity decreases, which looks more natural and is in 
agreement with other [105] calculations. Really, it looks to be clear 
that the occurrence of an additional random potential should lead to 
additional scattering and, hence, to the decrease of sound velocity. 
Some other contradictions resulting from the use of perturbation 
theory are illustrated in Refs. [108,110].   
 
Note that all consideration above is valid for any type of disorder 
characterized by the corresponding correlation function (409).

\subsection{Local Correlations}

To proceed further, let us consider local correlations, described by 
the delta-correlated disorder, when the correlation function (409) is
\be
\label{453}
 R(\br) = R_0 \dlt(\br) \; .
\ee
Then Eq. (411) gives
\be
\label{454}
 \lgl\lgl \xi_k^* \xi_p \rgl\rgl = \dlt_{kp} R_0 \; .
\ee
The solution to the Fredholm equation (433) can be well approximated 
[108] by
\be
\label{455}
 \vp_k = \frac{\sqrt{\rho_0} \xi_k}
{1 + \frac{1}{\sqrt{V}} \sum_p \frac{\xi_p}{\om_p+mc^2}} \; .
\ee

It is convenient to introduce [108] the {\it disorder parameter}
\be
\label{456}
\zeta \equiv \frac{a}{l_{loc}} = \frac{1}{\rho^{1/3}l_{loc}} \; ,  
\ee
being the ratio of the mean interparticle distance versus the 
{\it localization length}
\be
\label{457}
 l_{loc} \equiv \frac{4\pi}{7m^2 R_0} \; .
\ee
Employing the self-similar approximation theory [123-132] results in
\be
\label{458}
 \lgl\lgl | \vp_k|^2 \rgl\rgl = 
\frac{\rho_0 R_0s^{3/7}}{(s-\zeta)^{3/7}} \; ,
\ee
where $s$ is the dimensionless sound velocity (385).

The local disorder, with the delta correlation (453), allows for more
straightforward calculations. At the same time, it gives good 
understanding of the influence of disorder on the system even for the 
general case of nonlocal disorder. If the random potential $\xi(\br)$
is characterized by a finite strength $V_R$, with the correlation 
function (409) having a finite correlation length $l_R$, then, to pass 
to that case, one should make the replacement
\be
\label{459}
 R_0 = V_R^2\; l_R^3 \; ,
\ee 
which follows directly from the definition of correlator (409). As a result, 
the localization length (457) becomes
\be
\label{460}
l_{loc} = \frac{4\pi}{7m^2V_R^2\;l_R^3} \;  .
\ee

The particle fractions
\be
\label{461}
n_0 \equiv \frac{\rho_0}{\rho} \; , \qquad 
n_N \equiv \frac{\rho_N}{\rho} \; , \qquad
n_G \equiv \frac{\rho_G}{\rho}
\ee
satisfy the normalization
\be
\label{462}
 n_0 + n_N + n_G = 1 \; .
\ee

Let us define the dimensionless anomalous averages
\be
\label{463}
\sgm \equiv \frac{\sgm_N}{\rho} \; , \qquad
\frac{\sgm_1}{\rho} = \sgm + n_G \;  .
\ee
Passing to dimensionless quantities, let us use the gas parameter 
$\gamma$, defined in Eq. (365), dimensionless sound velocity (385), 
and dimensionless temperature
\be
\label{464}
 t \equiv \frac{mT}{\rho^{2/3} } \;  .
\ee
Then, the equation for the sound velocity (432) reads as
\be
\label{465}
 s^2 = 4\pi\gm ( 1 - n_G + \sgm ) \; .
\ee
  
For the normal fraction of uncondensed particles, we have
\be
\label{466}
n_N = \frac{s^3}{3\pi^2} \left \{ 1 +
\frac{3}{2\sqrt{2}} \int_0^\infty \left ( \sqrt{1+x^2} - 1
\right )^{1/2} \left [ 
\coth \left ( \frac{s^2 x}{2t} \right ) - 1 \right ] 
dx \right \} \; .
\ee
The glassy fraction is
\be
\label{467}
 n_G = 
\frac{( 1- n_N )\zeta}{\zeta+7s^{4/7}(s-\zeta)^{3/7}} \; .
\ee
And the superfluid fraction becomes
\be
\label{468}
 n_s = 1 - \; \frac{4}{3} \; n_G - \; 
\frac{s^5}{6\sqrt{2}\pi^2 t} \int_0^\infty
\frac{\left ( \sqrt{1+x^2}-1\right )^{3/2}xdx}
{\sqrt{1+x^2}\sinh^2(s^2x/2t)} \;  .
\ee

The Bose-Einstein condensation temperature $T_c$, in the presence of 
disorder, decreases linearly with the increasing disorder strength, 
as compared to the transition temperature $T^0_c$, given by Eq. (400), 
for the system without disorder. The relative transition temperature
decrease, for $\zeta < 1$, follows [108] the law
\be
\label{469}
 \dlt T_c \equiv \frac{T_c - T_c^0}{T_c^0} = -\; 
\frac{2\zeta}{9\pi} \;  .
\ee
This is in agreement with Monte Carlo simulations [107].

Depending on the relation between the localization length $l_{loc}$
and the {\it coherence length}
\be
\label{470}
 l_{coh} \sim \frac{\int r|\rho(\br,0)|d\br}{\int|\rho(\br,0)|d\br} \; ,
\ee
there can exist three different phases [110]. 

{\it Superfluid phase} exists, when the localization length is larger 
than the coherence length:
\be
\label{471}
l_{loc} > l_{coh} \;  .
\ee
In this case, the disorder is yet weak and cannot destroy the system 
coherence.

{\it Bose glass} can occur, when the localization length becomes shorter
than the coherence length, but yet larger than the mean interparticle 
distance:
\be
\label{472}
 a < l_{loc} < l_{coh} \;  .
\ee
Then a kind of granular condensate can exist, being localized in 
different spatial regions that are separated from each other by the 
normal nonsuperfluid phase.

{\it Normal glass} appears, when the localization length is shorter
that the mean interparticle distance:
\be
\label{473}
l_{loc} < a \;   .
\ee
Therefore no coherence between particles can arise, all of them being 
localized in separate regions of deep random wells.

\subsection{Bose Glass}

The peculiar phase of the Bose glass is the random mixture of 
Bose-condensed droplets, localized in different spatial regions that 
are separated from each other by the normal phase. In the Bose-condensed 
regions, the gauge symmetry is locally broken, while in the regions of 
the normal phase, the gauge symmetry is preserved. All these regions 
are randomly distributed in space and it is even possible that they 
chaotically change their spatial locations. Also, they are not 
necessarily compact and may be ramified having fractal geometry [133,134].

Such a randomly mixed system is a particular case of heterophase systems,
whose examples are ubiquitous in condensed matter physics. In this respect,
it is possible to mention paramagnets with local magnetic ordering 
revealing spin waves [135-139], many ferroelectrics [140-143] and
superconductors [144-149], colossal magnetoresistant materials [150-152],
and some other systems reviewed in Refs. [54,153-157].  

The typical features of these heterophase materials are: (i) the embryos 
of one phase inside another are {\it mesoscopic}, their characteristic 
sizes being much larger than the mean interparticle distance but shorter 
than the system length; (ii) the spatial distribution of the embryos, as 
well as their shapes are {\it random}; (iii) the system, as a whole, is 
{\it quasiequilibrium}, being either stable, or at least metastable, with 
the lifetime essentially longer than the local equilibration time. 

Such materials, with randomly distributed mesoscopic embryos of one phase 
inside another should be distinguished from systems composed of large 
stationary domains and from Gibbs mixtures of coexisting macroscopic 
phases [158]. For the equilibrium macroscopic phases, coexisting with each 
other, one has to consider the interfacial free energy [159]. The notion 
of interfacial free energy arises when one considers {\it uniform} 
macroscopic phases, while the mesoscopic heterophase inclusions are 
{\it nonuniform}. For quasiequilibrium mesoscopic embryos, the interfacial 
regions are not well defined, being often ramified and nonequilibrium. 
Quasiequilibrium embryos of competing phases are also different from 
nonequilibrium nuclei arising in kinetic phase transitions [160,161].

A general approach to treating such random heterophase mixtures has been
advanced and developed in Refs. [162-171], and summarized in the review 
articles [54,156,157]. Here, this approach is applied for describing the 
Bose glass.       

Assume that we aim at describing the heterophase mixture of normal 
uncondensed phase and Bose-condensed phase. The condensed phase exists in 
the form of mesoscopic embryos surrounded by the normal uncondensed phase. 
The effective total volume, occupied by each phase is $V_\nu$, with the 
index $\nu = 1,2$ enumerating the phases. The {\it geometric weight} of 
each phase is 
\be
\label{474}
 w_\nu \equiv \frac{V_\nu}{V} \qquad (V = V_1 + V_2 ) \; ,
\ee
where $V$ is the system volume. This weight enjoys the standard 
probability properties
\be
\label{475}
w_1 + w_2 = 1 \; , \qquad 0 \leq w_\nu\leq 1 \;  ,
\ee
because of which it can be termed the {\it geometric probability}.

The nonuniform mixture of phases, consisting of mesoscopic embryos, requires
the use of two types of averages, the statistical averaging over the particle 
degrees of freedom and the configuration averaging over all admissible 
random spatial phase configurations. Accomplishing the configuration averaging 
over phase configurations [54,156,157] results in the appearance of 
the {\it renormalized Hamiltonian}
\be
\label{476}
 \widetilde H = H_1 \bigoplus H_2 \; ,
\ee
consisting of two terms corresponding to each of the phases. This Hamiltonian
is defined on the {\it mixture space}
\be
\label{477}
\cM = \cH_1 \bigotimes \cH_2 \; .
\ee
By its mathematical structure, this space is the tensor product of the 
weighted Hilbert spaces [54,156,157]. It can be treated as a particular  
case of a fiber bundle [172]. 

The effective statistical operator for the random mixture becomes
\be
\label{478}
 \hat\rho = \hat\rho_1 \bigotimes \hat\rho_2 \;  ,
\ee
with $\hat{\rho}_\nu$ being the effective statistical operators for each
of the phases.

Thus, after the configuration averaging, we come to the statistical 
ensemble $\{\hat{\rho}, \cal{M}\}$ that is the collection of the partial 
phase ensembles $\{\hat{\rho}_\nu, \cal{H}_\nu\}$. The partial ensembles
can be called the reduced, or restricted, ensembles, since they are defined
on the restricted spaces of microscopic states, typical of the 
corresponding phase [173-175]. All details are given in the reviews 
[54,156,157].

The Hamiltonians $H_\nu$ are the effective phase Hamiltonians. For the 
Bose-condensed phase,
\be
\label{479}
H_1 = \hat H_1 - \mu_0 N_0 - \mu_1 \hat N_1 - \hat\Lbd \;  ,
\ee
while for the normal uncondensed phase,
\be
\label{480}
 H_2 = \hat H_2 - \mu_2 \hat N_2 \; ,
\ee
where $\hat{N}_2$ is the number operator for the normal phase.

The energy Hamiltonian for the Bose-condensed phase reads as
$$
\hat H_1 = w_1 \int \hat\psi^\dgr(\br) \left [ -\;
\frac{\nabla^2}{2m} + U(\br) + \xi(\br) 
\right ] \hat\psi(\br) \; d\br \; +
$$
\be
\label{481}
 + \; \frac{w_1^2}{2} \int 
\hat\psi^\dgr(\br)\hat\psi^\dgr(\br')
\Phi(\br-\br') \hat\psi(\br') \hat\psi(\br) \; 
d\br d\br' \;,
\ee
with the Bogolubov-shifted field operator
$$
\hat\psi(\br) = \eta(\br) + \psi_1(\br) \; .
$$
For the normal phase, the energy Hamiltonian is
$$
\hat H_2 = w_2 \int \psi_2^\dgr(\br) \left [ -\;
\frac{\nabla^2}{2m} + U(\br) + \xi(\br) \right ] 
\psi_2(\br) \; d\br \; +
$$
\be
\label{482}
 + \; \frac{w_2^2}{2} \int 
\psi_2^\dgr(\br)\psi_2^\dgr(\br')
\Phi(\br-\br') \psi_2(\br') \psi_2(\br) \; 
d\br d\br' \;   .
\ee
The interaction potential here is written in the general form. But, 
in particular, it can take the local form (316).

In the broken-symmetry phase, we have, as earlier, the densities of 
condensed and uncondensed particles, respectively,
\be
\label{483}
 \rho_0(\br) = | \eta(\br)|^2 \; , \qquad
\rho_1(\br) = \lgl \psi_1^\dgr(\br) \psi_1(\br) \rgl \; .
\ee
And in the normal phase, there is only the density of normal particles
\be
\label{484}
 \rho_2(\br) = \lgl \psi_2^\dgr(\br) \psi_2(\br) \rgl \; .
\ee

The numbers of particles in the whole heterophase system are written 
as follows. The number of condensed particles is
\be
\label{485}
 N_0 = w_1 \int \rho_0(\br) \; d\br \; .
\ee
Here and below, the integration is over the whole system. The number 
of uncondensed particles is given by the average
\be
\label{486}
 N_1 = \lgl \hat N_1 \rgl =
w_1 \int \rho_1(\br) \; d\br  
\ee
of the number operator
\be
\label{487}
 \hat N_1 = w_1 \int 
\psi_1^\dgr(\br) \psi_1(\br) \; d\br \; .
\ee
The number of normal particles is the average
\be
\label{488}
N_2 = \lgl \hat N_2 \rgl = w_2 \int \rho_2(\br) \; d\br   
\ee
of the number operator
\be
\label{489}
 \hat N_2 = w_2 
\int \psi_2^\dgr(\br) \psi_2(\br) \; d\br \; .
\ee
The total number of particles in the system is the sum
\be
\label{490}
N = N_0 + N_1 + N_2 \;  .
\ee
The related fractions of condensed, $n_0$, uncondensed, $n_1$, and normal, 
$n_2$, particles satisfy the normalization
\be
\label{491}
 n_0 + n_1 + n_2 = 1 \; .
\ee

The system chemical potential is
\be
\label{492}
 \mu = \mu_0 n_0 + \mu_1 n_1 + \mu_2 n_2 \; .
\ee
>From the condition of equilibrium, it follows [12] that
\be
\label{493}
 \mu_2 = \frac{\mu_0n_0 + \mu_1 n_1}{n_0 + n_1} = \mu \; .
\ee

The grand thermodynamic potential is defined as in the previous sections,
\be
\label{494}
\Om = - T \lgl\lgl \ln {\rm Tr} e^{-\bt\widetilde H} \rgl\rgl \; ,
\ee
with the double brackets implying the stochastic averaging over the 
random external potential $\xi({\bf r})$. The geometric weights $w_\nu$ 
are defined to be the minimizers of the grand potential (494), under the 
normalization condition (475). The latter can be taken into account 
explicitly by introducing the notation
\be
\label{495}
 w_1 \equiv w \; , \qquad w_2 \equiv 1 - w \; .
\ee
Then the minimization of the grand potential (494) implies
\be
\label{496}
\frac{\prt\Om}{\prt w} = 0 \; , \qquad
\frac{\prt^2\Om}{\prt w^2} > 0 \;  .
\ee
The first condition gives the equation 
\be
\label{497}
 \lgl\lgl \left ( 
\lgl \frac{\prt\widetilde H}{\prt w} \rgl_{\widetilde H} 
\right ) \rgl\rgl = 
\lgl \frac{\prt\widetilde H}{\prt w} \rgl = 0 
\ee
for the weight $w$, while the second, the stability condition
\be
\label{498}
 \left ( \lgl \frac{\prt^2\widetilde H}{\prt w^2} \rgl 
\right ) \; > \; \bt \lgl \left (
\frac{\prt\widetilde H}{\prt w} \right )^2 \rgl \;.
\ee
Since the right-hand side of inequality (498) is non-negative, the 
sufficient stability condition is
\be
\label{499}
\lgl \frac{\prt^2\widetilde H}{\prt w^2} \rgl > 0 \;   .
\ee
 
Let us use the notation for the single-particle terms of the 
condensed phase,
$$
K_1 \equiv \int \left \lgl \hat\psi^\dgr(\br) \left [ -\;
\frac{\nabla^2}{2m} + U(\br) + \xi(\br) \right ] 
\hat\psi(\br) \right \rgl d\br \; -
$$
\be
\label{500}
-\; \mu_0 \int \rho_0(\br) \; d\br \; - \; 
\mu_1 \int \rho_1(\br) \; d\br \;  ,
\ee
and the normal phase,
\be
\label{501}
K_2 \equiv \int \left \lgl \psi_2^\dgr(\br) \left [ -\;
\frac{\nabla^2}{2m} + U(\br) + \xi(\br) \right ] 
\psi_2(\br) \right \rgl d\br \; - \; 
\mu_2 \int \rho_2(\br) \; d\br \;  ,
\ee
respectively. Similarly, we can define the interaction terms for the 
condensed phase,
\be
\label{502}
\Phi_1 \equiv 
\int \lgl \hat\psi^\dgr(\br) \hat\psi^\dgr(\br')
\Phi(\br-\br') \hat\psi(\br') \hat\psi(\br) \rgl 
d\br d\br ' \;  ,
\ee
and the normal phase,
\be
\label{503}
\Phi_2 \equiv 
\int \lgl \psi_2^\dgr(\br) \psi_2^\dgr(\br')
\Phi(\br-\br') \psi_2(\br') \psi_2(\br) \rgl 
d\br d\br ' \; .
\ee
In the above expressions (500) and (502), it is assumed that the linear 
in $\psi_1$ terms are omitted, being cancelled by the term $\hat{\Lambda}$ 
in Hamiltonian (479).  

Then Eq. (497) yields the equation for the geometric weight of the 
condensed phase
\be
\label{504}
 w = \frac{\Phi_2 + K_2 - K_1}{\Phi_1 + \Phi_2 }  
\ee
and the stability condition (499) results in the inequality
\be
\label{505}
 \Phi_1 + \Phi_2 > 0 \; .
\ee
The latter condition shows that the heterophase mixture can exist only
for particles with repulsive interactions.

\section{Particle Fluctuations and Stability}

\subsection{Stability Conditions}

Fluctuations of observable quantities in statistical systems are 
characterized by the dispersions of self-adjoint operators 
corresponding to observables. Let $\hat{A}$ be the operator of an 
observable quantity given by the statistical average $\lgl\hat{A}\rgl$ 
of this operator. The fluctuations of this observable quantity are 
quantified by the dispersion
\be
\label{506}
\Dlt^2(\hat A) \equiv \lgl {\hat A}^2 \rgl -
\lgl \hat A \rgl^2 \;   .
\ee

The fluctuations of an observable quantity are called 
{\it thermodynamically normal} when the related dispersion is 
proportional to $N^\alpha$, with $\alpha$ not larger than one. 
And the fluctuations are termed {\it thermodynamically anomalous}
if the corresponding dispersion is proportional to $N^\alpha$, 
with $\alpha$ larger than one. In recent literature on Bose systems 
there has appeared a number of articles claiming the occurrence 
of thermodynamically anomalous fluctuations of the particle number 
in Bose systems everywhere below the transition temperature. 

In the papers [63,176-178] and reviews [5,9,12], it has been 
explained that the occurrence of such anomalous fluctuations 
contradicts the basic principles of statistical physics and that 
their appearance in some theoretical works is due merely to 
incorrect calculations. Because of the importance of this problem, 
it is described below, being based on Refs. [5,9,12,63,176-178].    

The ratio of the operator dispersion to its average value quantifies
the intensity of the system response to the variation of the 
considered observable. This response has to be finite in order that 
the system would be stable with respect to the observable-quantity 
fluctuations. That is, this ratio has to satisfy the stability 
condition [12,63,176-178]
\be
\label{507}
 0 \leq \frac{\Dlt^2(\hat A)}{| \lgl \hat A \rgl |} <
\infty \; .
\ee
This condition must hold for all observables and for any statistical 
system, including thermodynamic limit. For extensive observables, 
to be considered below, $\lgl\hat{A}\rgl\propto N$. The limiting ratio 
\be
\label{508}
\chi(\hat A) \equiv \lim_{N\ra\infty}\;
\frac{\Dlt^2(\hat A)}{N}
\ee
has the meaning of the response function related to the variation 
of the observable represented by the operator $\hat{A}$, and can be 
called {\it fluctuation susceptibility}. Therefore, another form of 
the stability condition is
\be
\label{509}
0 \leq \chi(\hat A) < \infty \;  .
\ee 

The number of particles is the observable represented by the number 
operator $\hat{N}$. Hence, the stability condition with respect to 
particle fluctuations is
\be
\label{510}
0 \leq \chi(\hat N) < \infty \;   .
\ee

>From the general relations of statistical mechanics and thermodynamics 
that can be found in almost any course [49,50,59,60,62,79,81,88,89,90], 
it is easy to show that quantity (508) is really proportional to some 
physical susceptibility. Being interested in particle fluctuations, 
one has to consider the dispersion $\Delta^2(\hat{N})$. The related
physical susceptibility is the isothermal compressibility. This can 
be defined in any statistical ensemble, as is shown below.

In the {\it canonical ensemble}, where the thermodynamic potential 
is the free energy $F = F(T,V,N)$, the isothermal compressibility is 
given by the derivatives
\be
\label{511}
\kappa_T = \frac{1}{V} \left ( \frac{\prt^2 F}{\prt V^2}
\right )^{-1}_{TN} = -\; \frac{1}{V} \left (
\frac{\prt P}{\prt V} \right )^{-1}_{TN} \;  .
\ee

In the {\it Gibbs ensemble}, with the Gibbs thermodynamic potential 
$G = G(T,P,N)$, the compressibility is
\be
\label{512}
 \kappa_T = -\; \frac{1}{V} \left ( 
\frac{\prt^2 G}{\prt P^2}\right )_{TN} = -\; \frac{1}{V} \left (
\frac{\prt V}{\prt P}\right )_{TN} \; .
\ee

And in the {\it grand canonical ensemble}, with the grand thermodynamic
potential $\Omega = \Omega(T,V,\mu)$, the compressibility becomes
\be
\label{513}
 \kappa_T = -\; \frac{1}{N\rho} \left ( 
\frac{\prt^2 \Om}{\prt\mu^2}\right )_{TV} = \frac{1}{N\rho} 
\left ( \frac{\prt N}{\prt\mu}\right )_{TV} \;   .
\ee
  
Of course, the value of the compressibility does not depend on the used 
ensemble, provided that it is correctly defined as a representative 
ensemble [54,63,64,71]. The fact that the compressibility is directly 
related to particle fluctuations is the most evident in the grand 
canonical ensemble, where
\be
\label{514}
 \kappa_T = \frac{\Dlt^2(\hat N)}{\rho TN} \;  .
\ee

The importance of correctly describing the particle fluctuations is 
caused by the fact that they define not only the compressibility, but
also are connected with several other observable quantities, such as  
the hydrodynamic sound velocity $s_T$,
\be
\label{515}
 s_T^2 \equiv \frac{1}{m} \left ( \frac{\prt P}{\prt \rho} 
\right )_T = \frac{1}{m\rho\kappa_T} = 
\frac{NT}{m\Dlt^2(\hat N)} \; ,
\ee
and the central structure factor
\be
\label{516}
 S(0) = \rho T \kappa_T = \frac{T}{ms_T^2} =
\frac{\Dlt^2(\hat N)}{N} \; .
\ee
As is seen, the susceptibility $\chi(\hat{N})$ coincides with the 
structure factor (516). 

Is is worth stressing that all expressions (511) to (516) are exact
thermodynamic relations that are valid for any stable equilibrium 
statistical system.

In stable statistical systems, the compressibility, as well as the 
structure factor, are finite. This is a very well known experimental 
fact. They can be divergent only at phase transition points, where,
as is known, the system is unstable. But everywhere outside of 
transition points, all these quantities must be finite.

\subsection{Fluctuation Theorem}

The stability condition (509) is necessary in order that the system 
would be stable with respect to the fluctuations of the observable 
quantity represented by the operator $\hat{A}$. But what can be said 
with regard to an observable represented by a composite operator
\be
\label{517}
 \hat A = \sum_i \hat A_i \; ,
\ee
given by a sum of several self-adjoint operators? How the fluctuations
for the total sum of $\hat{A}$ are connected with partial fluctuations
for $\hat{A}_i$? To formulate this question more precisely, let us give
some definitions.

\vskip 2mm

{\bf Definition}: {\it Thermodynamically normal fluctuations}  

\vskip 2mm

Fluctuations of an observable quantity, represented by a self-adjoint 
operator $\hat{A}$, are called thermodynamically normal if and only if
the stability condition (509) holds for this operator. Then the related
susceptibility (508) is also called thermodynamically normal.

\vskip 3mm
 
{\bf Definition}: {\it Thermodynamically anomalous fluctuations}

\vskip 2mm

Fluctuations of an observable quantity, represented by a self-adjoint
operator $\hat{A}$, are called thermodynamically anomalous if and only 
if the stability condition (509) does not hold for this operator. Then
the related susceptibility (508) is also termed thermodynamically 
anomalous.

\vskip 2mm

The question of interest is how the total fluctuation susceptibility 
$\chi(\hat{A})$ is connected with the partial fluctuation 
susceptibilities $\chi(\hat{A}_i)$? Or, in physical terminology, 
can it happen that the total susceptibility be finite, while some of 
the partial susceptibilities be infinite? The answer to this question 
is given by the following theorem on fluctuations of composite 
observables.  

\vskip 2mm

{\bf Fluctuation Theorem}.(Yukalov [63,177])

\vskip 2mm

Let the observable quantity be represented by a composite operator 
(517) that is a sum of self-adjoint operators. Then the dispersion of 
this operator is
\be
\label{518}
 \Dlt^2\left ( \sum_i \hat A_i \right ) = \sum_i
\Dlt^2 ( \hat A_i) \; + \; \sum_{i\neq j}
\lbd_{ij} \; \sqrt{\Dlt^2(\hat A_i) \Dlt^2(\hat A_j) } \; ,
\ee
where $|\lambda_{ij}|<1$, hence the total fluctuation susceptibility 
reads as
\be
\label{519}
\chi\left ( \sum_i \hat A_i \right ) =
\sum_i \chi(\hat A_i ) \; + \; \sum_{i\neq j} \lbd_{ij} \;
\sqrt{\chi(\hat A_i) \chi(\hat A_j ) } \;  .
\ee
From here it follows that the total fluctuation susceptibility is 
normal
if and only if all partial fluctuation susceptibilities are normal. And
the total fluctuation susceptibility is anomalous if and only if at least
one of the partial fluctuation susceptibilities is anomalous.

\subsection{Ideal-Gas Instability}

For illustrative purpose, one often considers the ideal Bose gas. The 
grand Hamiltonian for noninteracting particles is a particular case of
Hamiltonian (58), where the energy Hamiltonian is
\be
\label{520}
 \hat H = \int \hat\psi^\dgr(\br) \left ( -\; \frac{\nabla^2}{2m}
+ U \right ) \hat\psi(\br) \; d\br \; .
\ee
Substituting here the Bogolubov shift (48) yields the grand Hamiltonian
\be
\label{521} 
H = \int \eta^*(\br) \left ( -\; \frac{\nabla^2}{2m}
+ U -\mu_0 \right ) \eta(\br) \; d\br \; + \;
\int \psi_1^\dgr(\br) \left ( -\; \frac{\nabla^2}{2m}
+ U -\mu_1 \right ) \psi_1(\br) \; d\br \;  .
\ee
The equations of motion become
$$
i \; \frac{\prt}{\prt t} \; \eta(\br,t) = 
\left ( -\; \frac{\nabla^2}{2m} + U -\mu_0 
\right ) \eta(\br,t) \; ,
$$
\be
\label{522}
i \; \frac{\prt}{\prt t} \; \psi_1(\br,t) = 
\left ( -\; \frac{\nabla^2}{2m} + U -\mu_1 
\right ) \psi_1(\br,t) \;   .
\ee

Let us pass to the uniform gas, when there is no external potential, 
$U = 0$. Then, in equilibrium, the condensate function is constant,
$\eta({\bf r},t) = \eta = const$. The equation for the condensate 
function gives $\mu_0 = 0$. 

In the momentum representation, Hamiltonian (521), with $U=0$, 
reduces to
\be
\label{523}
 H = \sum_{k\neq 0} \left ( \frac{k^2}{2m} - \mu_1 \right )
a_k^\dgr a_k \; .
\ee
The condition of the condensate existence (15), as well as the 
Hugenholtz-Pines relation (203), result in $\mu_1 = 0$ for 
temperatures below the condensation temperature
\be
\label{524}
 T_c = \frac{2\pi}{m} \left [ \frac{\rho}{\zeta(d/2)}
\right ]^{d/2} \; ,
\ee
written here for a $d$-dimensional space. Expression (524) shows that
positive $T_c$ does not exist for $d = 1$, since $\zeta(1/2) = - 1.460$
and that $T_c = 0$ for $d = 2$, since $\zeta(1) = \infty$. Positive 
$T_c$ exists only for $d > 2$.

For the number operator $\hat{N} = N_0 + \hat{N}_1$, taking into 
account that ${\rm cov}(N_0, \hat{N}_1) = 0$ and $\Delta^2(N_0) = 0$, 
one finds
\be
\label{525}
 \Dlt^2(\hat N) = \Dlt^2 (\hat N_1) \;  .
\ee 

Let us emphasize that the condensate fraction does not fluctuate at 
all and that the total fluctuations are caused solely by the uncondensed 
particles. The number operator of the latter is
\be
\label{526}
 \hat N_1 \equiv \int \psi_1^\dgr(\br) \psi_1(\br) \; d\br
= \sum_{k\neq 0} a_k^\dgr a_k \; .
\ee
Invoking the commutation relations and the Wick theorem, one has
$$
\lgl a_k^\dgr a_k a_p^\dgr a_p \rgl = 
\lgl a_k^\dgr a_p^\dgr a_k a_p \rgl + \dlt_{kp} n_k \; ,
$$
$$
\lgl a_k^\dgr a_p^\dgr a_k a_p \rgl = n_k n_p +
\dlt_{kp} n_k^2 \; ,
$$
where the momentum distribution is 
$$
n_k \equiv \lgl a_k^\dgr a_k \rgl = \left [
\exp \left ( \frac{\bt k^2}{2m} \right ) -1 \right ]^{-1} \;  .
$$
This leads to
$$
\lgl \hat N_1^2 \rgl = N_1^2 + 
\sum_{k\neq 0} n_k ( 1 + n_k) \;  .
$$
Therefore particle fluctuations are characterized by the dispersion
\be
\label{527}
 \Dlt^2(\hat N) = \Dlt^2(\hat N_1) = 
\sum_{k\neq 0} n_k ( 1 + n_k) \; .
\ee

\vskip 2mm

{\bf Remark}. In some works, the authors forget that Bose-Einstein 
condensation necessarily requires broken gauge symmetry. Forgetting 
this, one extends the sum in Eq. (526) to $k = 0$. Then, separating 
the term with $k = 0$, one gets the condensate fluctuations described 
by the dispersion $\Delta^2(N_0)$ proportional to $N^2_0$. One blames 
the grand canonical ensemble to be guilty for this unreasonable result,
naming this "grand canonical catastrophe". However, as is clear, 
there is no any catastrophe here and not the grand ensemble is guilty, 
but the authors doing incorrect calculations. One should not forget 
that, if the gauge symmetry is not broken, then $N_0 \equiv 0$.    

\vskip 2mm

Summing the right-hand side of Eq. (527) yields
\be
\label{528}
 \Dlt^2(\hat N_1) = \left ( \frac{mT}{\pi} \right )^2
V^{4/3} \; .
\ee
This gives the fluctuation susceptibility
\be
\label{529}
 \chi(\hat N) = \chi(\hat N_1) = \lim_{N\ra\infty}
\left ( \frac{ma^2T}{\pi} \right )^2 N^{1/3} = \infty \; .
\ee
Consequently, the stability condition (510) does not hold. This 
means that the ideal uniform Bose-condensed gas is not stable. 
It is a pathological object that cannot exist in reality.

\subsection{Trapped Atoms}

But maybe the ideal Bose-condensed gas could be stabilized by 
confining it inside a trap formed by a trapping external potential? 
A general expression for such a trapping potential is given by the 
power-law form
\be
\label{530}
 U(\br) = \sum_{\al=1}^d \frac{\om_\al}{2} \left |
\frac{r_\al}{l_\al} \right |^{n_\al} \; ,
\ee
which is written here in the $d$-dimensional space. The trapping 
frequency $\omega_\alpha$ and the trapping length $l_\alpha$ are 
connected by the relations
\be
\label{531}
 \om_\al = \frac{1}{ml_\al^2} \; , \qquad 
l_\al = \frac{1}{\sqrt{m\om_\al}} \; .
\ee
It is also useful to introduce the effective frequency and effective 
length by the geometric averages
\be
\label{532}
 \om_0 \equiv \left ( \prod_{\al=1}^d \om_\al 
\right )^{1/d} = \frac{1}{ml_0^2} \; , \qquad 
l_0 \equiv \left ( \prod_{\al=1}^d l_\al 
\right )^{1/d} = \frac{1}{\sqrt{m\om_0}} \; .
\ee

Let us define the {\it confining dimension} [51]
\be
\label{533}
  s \equiv \frac{d}{2} + \sum_{\al=1}^d \frac{1}{n_\al} \; .
\ee
Passing from the trapping potential to the uniform case implies the 
limits
$$
 n_\al \ra \infty \; ,  \qquad l_0 \ra \frac{L}{2} \; ,
\qquad \prod_{\al=1}^d 2 l_\al \ra L^d \; ,
$$
where $L$ is the linear size of the system volume $V = L^d$. As a result, 
$s \ra d/2$, that is, $s$ becomes semi-dimension.

Bose-Einstein condensation of the ideal Bose gas in the trapping potential 
(530) can be described employing the generalized quasi-classical 
approximation [51]. The condensation temperature reads as
\be
\label{534}
 T_c = \left [ \frac{bN}{g_s(1) } \right ]^{1/s} \; ,
\ee
where we use the notation
$$
 b \equiv \frac{\pi^{d/2}}{2^s} \prod_{\al=1}^d
\frac{\om_\al^{1/2+1/n_\al}}{\Gm(1+1/n_\al)}     
$$ 
and introduce the generalized Bose function
\be
\label{535}
 g_s(z) \equiv \frac{1}{\Gm(s)} \int_{u_0}^\infty
\frac{zu^{s-1}}{e^u-z} \; du \; ,
\ee
in which the integration is limited from below by the value
\be
\label{536}
 u_0 \equiv \frac{\om_0}{2T} \; .
\ee
Recall that in the standard Bose function, the integration starts 
from zero.

The value $g_s(1)$ of the generalized function (535) is finite for 
all $s$ on the complex plane, since $\Gamma(s) \neq 0$, so that 
$1/\Gamma(s)$ is an entire function. But $\Gamma(s)$ can be negative 
for $s < 0$, e.g., it is negative in the interval $-1 < s < 0$. 
Therefore, $g_s(1)$ is positive and finite for all $s>0$. Contrary 
to this, the standard Bose function would diverge for $s \leq 1$, 
and there would be no finite condensation temperatures for these $s$. 
While, in the case of the generalized function (535), finite 
condensation temperatures formally exist for any positive $s$. 
Below $T_c$, and for $s > 0$, the condensate fraction is
\be
\label{537}
 n_0 = 1 - \left ( \frac{T}{T_c} \right )^s \qquad
( T \leq T_c ) \;  .
\ee
   
The most often studied trapping potential is the harmonic potential, 
for which $n_\alpha = 2$ and $s = d$. Then the condensation 
temperatures are
$$
T_c = \frac{N\om_0}{\ln(2N)} \qquad (d=1) \; ,
$$
\be
\label{538}
 T_c = \om_0 \left [ \frac{N}{\zeta(d)} \right ]^{1/d}
\qquad (d\geq 2) \; .
\ee

The condensation temperature (534) is finite for any finite $N$. But 
it is necessary to check whether it is finite in thermodynamic limit,
when $N \ra \infty$. For confined systems, the effective thermodynamic 
limit is defined [51] in Eq. (13). As an extensive observable, we can 
take the internal energy that, in the present case, below $T_c$, is
\be
\label{539}
 E_N = \frac{s}{b} \; g_{1+s}(1) T^{1+s} \; .
\ee
Then, the thermodynamic limit (13) reads as
\be
\label{540}
 N \ra \infty \; , \qquad E_N \ra \infty \; , \qquad
\frac{E_N}{N} \ra const \; .
\ee
The value $g_{1+s}$ is finite for $N \ra \infty$ at all $s > 0$. 
Hence, Eq. (540) can be rewritten as the limit
\be
\label{541}
  N \ra \infty \; , \qquad b \ra 0 \; , \qquad
bN \ra const \;  .
\ee
For the equipower traps, for which $n_\alpha = n$, the effective 
thermodynamic limit (541) takes the form
\be
\label{542}
  N \ra \infty \; , \qquad \om_0 \ra 0 \; , \qquad
N \om_0^s \ra const \;  .
\ee

Considering the thermodynamic limit for the condensation temperature 
(534), we have to take into account that the generalized function (535) 
yields
$$
g_s(1) \cong \frac{1}{(1-s)\Gm(s)} \left (
\frac{2T}{\om_0} \right )^{1-s} \qquad ( 0 < s < 1) \; ,
$$
\be
\label{543}
 g_s(1) \cong \ln \left ( \frac{2T}{\om_0} \right )
\qquad ( s = 1 ) \; .
\ee
Consequently, for the condensation temperature (534), as $N\ra\infty$, 
we find
$$
T_c \propto \frac{1}{N^{(1-s)/s} } \ra 0 \qquad
(0 < s < 1) \; ,
$$
$$
T_c \propto \frac{1}{\ln N } \ra 0 \qquad
(s = 1) \;
$$
\be
\label{544}
 T_c \ra const \qquad ( s > 1) \; .
\ee
Therefore, finite condensation temperatures exist only for $s > 1$. 
This implies that for harmonic traps, for which $s = d$, the finite 
condensation temperature occurs only for $d\geq 2$. Bose-Einstein 
condensation cannot happen in one-dimensional harmonic traps at finite 
temperature.

But this is not yet the whole story. As we know from Sec. 10.3, 
a finite condensation temperature can formally occur, however, the 
condensed system in reality is unstable, thus, cannot exist. To check 
the stability, it is necessary to consider the system susceptibilities. 
Specific heat for the Bose-condensed trapped gas is finite at all 
temperatures, displaying a jump at the transition point [51]. We need 
to consider the isothermic compressibility (514) that shows the system 
response with respect to particle fluctuations. The dispersion for the 
number operators behaves as
\be
\label{545}
 \Dlt^2(N_0) = 0 \; , \qquad \Dlt^2(\hat N) = \Dlt^2(\hat N_1) \; .
\ee

It is convenient to introduce the finite-$N$ susceptibility
\be
\label{546}
 \chi_N \equiv \frac{\Dlt^2(\hat N)}{N} \;   ,   
\ee
whose limit
$$
\lim_{N\ra\infty} \chi_N = \chi(\hat N)
$$
yields the susceptibility defined in Eq. (508). Below $T_c$, we obtain
\be
\label{547} 
 \chi_N = \frac{g_{s-1}(1)}{g_s(1)} \left (
\frac{T}{T_c} \right )^s \; .
\ee

Susceptibility (547) is negative for $s < 1$ and does not satisfy 
the stability condition (510). For the values of $s \geq 1$, we have
$$
\chi_N = \frac{2}{N} \left ( \frac{T}{T_c} \right )^2 
\qquad (s = 1) \; ,
$$
$$
\chi_N = \frac{1}{(2-s)\zeta(s)\Gm(s-1)} 
\left ( \frac{2T}{\om_0} \right )^{2-s} 
\left ( \frac{T}{T_c} \right )^2 \; , 
\qquad (1 < s < 2) \; ,
$$
$$
\chi_N = \frac{1}{\zeta(2)} \left ( \frac{T}{T_c} \right )^2 
\ln \left ( \frac{2T}{\om_0} \right ) \qquad ( s = 2) \; ,
$$
\be
\label{548}
 \chi_N = \frac{\zeta(s-1)}{\zeta(s)} 
\left ( \frac{T}{T_c} \right )^s \qquad ( s > 2) \; .
\ee
 
For asymptotically large $N$, we get
$$
\chi_N \propto N \qquad ( s = 1 ) \; ,
$$
$$
\chi_N \propto N^{(2-s)/s} \qquad (1 < s < 2 ) \; ,
$$
$$
\chi_N \propto \ln N \qquad ( s = 2 ) \; ,
$$
\be
\label{549}
 \chi_N \propto const \qquad ( s > 2 ) \; .
\ee
This shows that the trapped Bose gas is stable only for $s > 2$, when
the stability condition (510) is satisfied, that is, when
\be
\label{550}
 s \equiv \frac{d}{2} + \sum_{\al=1}^d \frac{1}{n_\al} > 2 \; .
\ee

In particular, for harmonic traps, for which $s = d$ and $b = \omega^d$,
one finds
$$
 \chi_N = \frac{2}{N} \left ( \frac{T}{\om_0} \right )^2
\qquad ( d = 1) \; .
$$
$$
 \chi_N = \frac{1}{N} \left ( \frac{T}{\om_0} \right )^2
\ln \left ( \frac{2T}{\om_0} \right )
\qquad ( d = 2) \; .
$$
$$
 \chi_N = \frac{\pi^2}{6\zeta(3)} \left ( \frac{T}{T_c} \right )^3
\qquad ( d = 3) \; .
$$
For large $N$, this gives
$$
 \chi_N \propto N \qquad ( d = 1 ) \;  .
$$
$$
 \chi_N \propto \ln N \qquad ( d = 2 ) \;  .
$$
$$
 \chi_N \propto const \qquad ( d = 3 ) \;  .
$$
Thus, the Bose-condensed gas in a harmonic trap is stable only in 
the three-dimensional space, $d = 3$. 

The above analysis demonstrates that confining the ideal Bose gas 
in a trap may stabilize it, which, however, depends on the confining 
dimension $s$, defined in Eq. (533). The occurrence of a formal 
expression for the critical temperature $T_c$ is not yet sufficient 
for claiming the possibility of Bose-Einstein condensation in a 
trapped gas, but it is also necessary to check the system stability. 
For example, in the case of the power-law trapping potentials, the 
condensation temperature formally exists for $s > 1$. But the trapped
condensed gas can be stable only for $s > 2$. One- and two-dimensional 
harmonic traps are not able to stabilize the condensate. Only the 
three-dimensional harmonic trap is able to host the ideal 
Bose-condensed gas.

\subsection{Interacting Systems}

Ideal gases are, actually, rather artificial objects, since there always
exist particle interactions, though, maybe, weak. Now we pass to studying
the stability of interacting systems.

To a great surprise, there have been published many papers, in which the 
authors claim that interacting Bose-condensed systems, both uniform as 
well as trapped, exhibit thermodynamically anomalous particle fluctuations
of the same kind as the ideal Bose gas, with the number-operator dispersion 
(528). By the Bogolubov theorem, one always has $\Dlt^2(\hat{N}_0) = 0$, 
hence, $\Dlt^2(\hat{N}) = \Dlt^2(\hat{N}_1)$. Then the thermodynamically 
anomalous dispersion $\Delta^2(\hat{N}) \propto N^{4/3}$ would lead to  
$\chi_N \propto N^{1/3}$ and to the divergence of $\chi(\hat{N})\ra\infty$.
In that case, the stability condition (510) is not satisfied, and the 
behavior of all physical quantities would be rather wild. Then the 
isothermal compressibility (514) would diverge, the sound velocity (515) 
would be zero, and the structure factor (516) would be infinite. That is, 
the system would be absolutely unstable.

Moreover, this would mean that any system with spontaneously broken gauge 
symmetry would not exist. Clearly, such a strange conclusion would 
contradict all known experiments observing Bose-Einstein condensed trapped 
gases. Superfluid helium is also the system with broken gauge symmetry, 
hence, it also would not be able to exist, which is evidently absurd.

In Refs. [5,9,12,63,176-178], it has been explained that the occurrence, 
in some works, of thermodynamically anomalous fluctuations is caused by 
incorrect calculations. One calculates the dispersion $\Dlt^2(\hat{N}_1)$ 
invoking the Bogolubov approximation that is a second-order approximation 
with respect to the operators of uncondensed particles. But the expression 
$\hat{N}_1^2$ is of fourth order with respect to these operators. 
Calculating the fourth-order terms in the second-order approximation, 
strictly speaking, is not self-consistent and can lead to unreasonable 
results, such as the occurrence of thermodynamically anomalous particle 
fluctuations. 

The correct calculation of the dispersion $\Delta^2(\hat{N})$ and, hence,
of susceptibility (546), can be done as follows. From the definition of 
the particle dispersion $\Delta^2(\hat{N})$, one has the exact expression
\be
\label{551}
\chi_N = 1 + \frac{1}{N} \int \rho(\br) \rho(\br')
[ g(\br,\br') - 1 ] \; d\br d\br' \; ,
\ee
which is valid for any system, whether uniform or nonuniform, equilibrium 
or not [176-178]. Here, 
$$
\rho(\br) = \rho_0(\br) + \rho_1(\br)
$$
is the total particle density and
$$
g(\br,\br') \equiv 
\frac{\lgl\hat\psi^\dgr(\br)\hat\psi^\dgr(\br')
\hat\psi(\br')\hat\psi(\br)\rgl}{\rho(\br)\rho(\br')}
$$
is the pair correlation function.

In the HFB approximation, analogously to the Bogolubov approximation, 
one has to retain in Eq. (551) the terms up to the second-order with respect 
to the operators of uncondensed particles. For nonuniform systems, one can 
employ the local-density approximation of Sec. 7.4. Then Eq. (551) 
reduces to
\be
\label{552}
\chi_N = 1 + \frac{2}{N} \int \rho(\br) \lim_{k\ra 0}
[ n(\bk,\br) + \sgm(\bk,\br) ] \; d\br \;   .
\ee
Using the formulas of Sec. 7.4 gives
\be
\label{553}
 \chi_N = \frac{T}{mN} 
\int \frac{\rho(\br)}{c^2(\br)} \; d\br \; .
\ee
The same Eq. (553) represents the structure factor (516). Equation (515),
defining the hydrodynamic sound velocity, leads to 
\be
\label{554}
 s_T^2 = \left [ 
\frac{1}{N} \int \frac{\rho(\br)}{c^2(\br)} \; d\br 
\right ]^{-1}  .
\ee
And the isothermal compressibility (514) becomes
\be
\label{555}
  \kappa_T = \frac{1}{m\rho N} \int 
\frac{\rho(\br)}{c^2(\br)} \; d\br \; ,
\ee  
provided the average density $\rho$ is defined.

For a uniform system, the above formulas reduce to
\be
\label{556}
 \chi_N \equiv \frac{\Dlt^2(\hat N)}{N} = S(0) =
\frac{T}{mc^2} \; , \qquad s_T = c \; , \qquad 
\kappa_T = \frac{1}{m\rho c^2} \;  .
\ee
 
It is important to emphasize the necessity of taking into account 
the gauge symmetry breaking in the above calculations. If the 
symmetry would not be broken, or if the anomalous average $\sigma$ 
would be omitted, one would get the divergence of expressions (553) 
and (555), which would mean the system instability [179]. 

In some works on particle fluctuations, one also makes the following 
mistake. One writes that, in the canonical ensemble, the condensate 
fluctuations are given by $\Delta(\hat{N}_0)$ that is equal to
$\Delta(\hat{N}_1)$, and one calculates the latter in the second 
quantization representation. However, this representation uses the 
field operators defined on the Fock space and, by construction, it is 
introduced for the grand canonical ensemble. So, $\Delta(\hat{N}_1)$
has nothing to do with condensate fluctuations that, by the Bogolubov
theorem correspond to $\Delta(\hat{N}_0) = 0$.

In this way, correct calculations lead to no anomalous thermodynamic 
particle fluctuations. The latter arise only in incorrect calculations.
There are no anomalous fluctuations neither in correctly employed 
Bogolubov or HFB approximations [5,9,12,63,176-179] nor in the 
renormalization group approach [180].    

If thermodynamically anomalous fluctuations would not be caused by 
calculational defects, but would be real, then not merely equilibrium 
Bose-condensed gas and superfluid helium would not exist, but the 
situation would be even more dramatic. This is because the systems 
with gauge symmetry $U(1)$ are just a particular case of systems 
with continuous symmetry, all such systems having general properties 
connected with their continuous symmetry and the symmetry breaking 
[181]. Therefore all such systems exhibiting thermodynamically 
anomalous fluctuations would not exist. We mean here only equilibrium
statistical systems, since nonequilibrium systems can possess strong 
fluctuations making them unstable [182-184].  

For example, many magnetic systems exhibit continuous symmetry
connected with spin rotation. The appearance of magnetic order in 
such magnetic systems implies the spontaneous breaking of the spin 
rotational symmetry. If the continuous symmetry breaking would lead 
to the appearance of thermodynamically anomalous fluctuations of 
the order parameter, then, in magnetic systems, this would mean the 
occurrence of thermodynamically anomalous magnetic susceptibility, 
hence, instability. Then there would be no stable equilibrium magnetic 
systems with continuous symmetry breaking, which is again absurd.   

To show that the spontaneous breaking of the spin-rotation symmetry 
does not lead to thermodynamically anomalous magnetic fluctuations 
[12], let us consider the Hesenberg model, with the Hamiltonian
\be
\label{557}
 \hat H = - \sum_{i\neq j} J_{ij} \bS_i \cdot \bS_j \; - \;
\sum_i \bB \cdot \bS_i \; ,
\ee
in which ${\bf S}_i$ is a spin operator on the $i$-lattice site, 
$J_{ij}$ is an exchange interaction potential, and ${\bf B}$ is an 
external magnetic field. The Hamiltonian enjoys the spin rotation 
symmetry in the absence of the external field ${\bf B}$.

The Gibbs potential is defined as 
\be
\label{558}
 G = - T \ln {\rm Tr} e^{-\bt\hat H} = G(T,N,\bB) \; .
\ee
The system magnetic moment 
\be
\label{559}
{\bf M} \equiv \frac{\prt G}{\prt \bB} = - \;
\lgl \frac{\prt\hat H}{\prt\bB} \rgl \equiv 
\lgl \hat{\bf M} \rgl
\ee
can be represented as the average of the magnetic-moment operator
\be
\label{560}
 \hat{\bf M} \equiv -\; \frac{\prt\hat H}{\prt\bB} =
\sum_i \bS_i \; .
\ee

The magnetic susceptibility tensor is given by the elements
\be
\label{561}
 \chi_{\al\bt} \equiv \frac{1}{N} \; 
\frac{\prt M_\bt}{\prt B_\al}  = - \; \frac{1}{N} \; 
\frac{\prt^2 G}{\prt B_\al \prt B_\bt} \; .
\ee
Direct calculations yield
\be
\label{562}
\chi_{\al\bt} = 
\frac{1}{NT} \; {\rm cov} (\hat M_\al, \hat M_\bt) \;  ,
\ee
which shows that the diagonal elements
\be
\label{563}
\chi_{\al\al} = \frac{\Dlt^2(\hat M_\al)}{NT}
\ee
are expressed through the dispersion of the components of the  
magnetic-moment operator (560).

The HFB approximation for the field operators of Bose systems is 
equivalent to the mean-field approximation for the spin operators
of magnetic systems. Therefore, it is reasonable to resort here 
to the mean-field approximation, although the results are 
qualitatively the same if we invoke more elaborate techniques.
In the mean-field approximation, Hamiltonian (557) reads as
\be
\label{564}
 \hat H = -\sum_i \bH \cdot \bS_i \; + \; 
NJ \lgl \bS_i \rgl^2 \; ,
\ee
in which the notation is used for the effective field 
\be
\label{565}
 \bH \equiv 2J \lgl \bS_i \rgl + \bB  
\ee
and the effective interaction
\be
\label{566}
 J \equiv \frac{1}{N} \sum_{i\neq j} J_{ij} \; .
\ee

For concreteness, let us consider spin one-half. Then the Gibbs 
potential (558) becomes
\be
\label{567}
 G = - NT \ln \left [ 2 \cosh \left ( \frac{H_0}{2T} \right )
\right ] + NJ \lgl \bS_i \rgl^2 \;  ,
\ee
where the ideality of the lattice is implied and 
$$
H_0 \equiv |\bH | = \sqrt{ \sum_\al H_\al^2} \;  .
$$

The average spin is defined by the extremization condition
\be
\label{568}
  \frac{\prt G}{\prt\lgl \bS_i \rgl} = 0 \; ,
\ee
which is equivalent to the equation
\be
\label{569}
 \lgl \bS_i \rgl = - \; 
\frac{1}{N} \; \frac{\prt G}{\prt\bB} \;.
\ee
As a result, one finds
\be
\label{570}
 \lgl \bS_i \rgl = \frac{\bH}{2H_0} \; \tanh \left (
\frac{H_0}{2T} \right ) \;  .
\ee
This defines the magnetic susceptibility (561) as
\be
\label{571}
 \chi_{\al\bt} = \frac{\prt}{\prt B_\al} \;
\lgl S_i^\bt \rgl \;  .
\ee

Let us define the {\it order parameter}
\be
\label{572}
 \eta \equiv 2 | \lgl \bS_i \rgl | \; .
\ee
In view of Eq. (570), this reads as
\be
\label{573}
 \eta = \tanh \left ( \frac{H_0}{2T} \right ) \; .
\ee

Susceptibility (571) takes the form
$$
\chi_{\al\bt} = \frac{\eta}{2H_0} 
( \dlt_{\al\bt} + 2J \chi_{\al\bt} ) \; +
$$
\be
\label{574}
+ \; \frac{H_\bt}{2H_0^2} \left ( 
H_\al + 2J \sum_\gm \chi_{\al\gm} H_\gm\right )  
\left ( \frac{1-\eta^2}{2T} - \frac{\eta}{H_0} \right ) \; .
\ee
  
Directing the external magnetic field along the axis $z$, so that
\be
\label{575}
 B_x = B_y = 0 \; , \qquad B_z \equiv h \; ,
\ee
yields 
$$
H_x = H_y = 0 \; , \qquad 
H_z = 2J \lgl S_i^z \rgl + B_z \;  ,
$$
which can be rewritten as
\be
\label{576}
 H_\al = \dlt_{\al z} H_0 \; , \qquad 
H_0 = H_z = J \eta + h \; .
\ee
The average spin components become 
\be
\label{577}
 \lgl S_i^x \rgl = \lgl S_i^y \rgl = 0 \; , \qquad
\lgl S_i^z \rgl = \frac{\eta}{2} \;  .
\ee

The susceptibility tensor (574) leads to the equation
\be
\label{578}
 \chi_{\al\bt} = 
\frac{1}{2} ( \dlt_{\al\bt} + 2J\chi_{\al\bt} ) 
\left [ \frac{\eta}{H_0} + \dlt_{\bt z} \left (
\frac{1-\eta^2}{2T} - \frac{\eta}{H_0} \right ) \right ] \;,
\ee
with the order parameter
\be
\label{579}
 \eta = \tanh \left ( \frac{J\eta + h}{2T} \right ) \;  .
\ee
Equation (578) shows that the nondiagonal elements are zero:
\be
\label{580}
 \chi_{xy} = \chi_{xz} = \chi_{yz} = 0 \;,
\ee
while the diagonal elements give the transverse components
\be
\label{581}
\chi_{xx} = \chi_{yy} = \frac{\eta}{2h}
\ee
and the longitudinal component
\be
\label{582}
 \chi_{zz} = \frac{1-\eta^2}{2[2T-J(1-\eta^2)] } \; .
\ee
The latter, with the notation for the critical temperature 
$T_c\equiv J/2$, can be represented as
\be
\label{583}
  \chi_{zz} = \frac{1-\eta^2}{4[T-T_c(1-\eta^2)] } \;   .
\ee

At low temperature, and $h \ra 0$, the order parameter (579) 
behaves as
\be
\label{584}
 \eta \simeq 1 - 2 \exp \left ( -\; \frac{T_c}{T} \right )
\qquad ( T \ll T_c )  
\ee
and susceptibility (583), as
\be
\label{585}
 \chi_{zz} \simeq 
\frac{1}{T} \exp \left ( -\; \frac{T_c}{T} \right ) 
\qquad ( T \ll T_c ) \; .
\ee
At high temperature, and $h \ra 0$, susceptibility (583) acquires the
Curie-Weiss law
\be
\label{586}
 \chi_{zz} \simeq  \frac{1}{4(T-T_c)} \qquad ( T \ge T_c ) \; .
\ee
The latter susceptibility diverges at the critical point $T_c$. However, 
this divergence has nothing to do with the thermodynamically anomalous 
behavior, since this is the divergence with respect to temperature $T$, 
but not with respect to the number of particles $N$. In addition, the 
phase transition point is the point of system instability, where the 
system becomes nonequilibrium and fluctuations have right to infinitely 
rise. 

One introduces the transverse susceptibility
\be
\label{587}
\chi_\perp \equiv \chi_{xx} = 
\frac{\Dlt^2(\hat M_x)}{NT}
\ee
and the longitudinal susceptibility
\be
\label{588}
\chi_{||} \equiv \chi_{zz} = \frac{\Dlt^2(\hat M_z)}{NT} \; ,
\ee
where relation (563) is taken into account. In view of Eqs. (581) and 
(582), one finds the dispersions for the magnetic-moment operator (560),
characterizing the transverse fluctuations,
\be
\label{589}
 \frac{\Dlt^2(\hat M_x)}{N} = T \chi_\perp =
\frac{\eta T}{2h} \; ,
\ee
and the longitudinal fluctuations,
\be
\label{590}
\frac{\Dlt^2(\hat M_z)}{N} = T \chi_{||} =
\frac{T(1-\eta^2)}{4[T-T_c(1-\eta^2)]} \;  ,
\ee
of the system magnetic moment. 

The longitudinal fluctuations are always thermodynamically normal, if 
calculated in a self-consistent way, as it should be, according to the 
stability condition (509). In some papers, one finds thermodynamically 
anomalous magnetic fluctuations of the same type as for Bose systems, 
with $\Delta(\hat{M}_z)/N \propto N^{1/3}$. But, as has been explained 
in Ref. [5], this is due to the same mistake as one does when dealing 
with Bose systems. One approximates Hamiltonian (557) by a second-order 
form, with respect to small deviations from the average magnetic moment. 
And then, one considers the fourth-order form calculating the dispersion 
of $\hat{M}_z$. Going outside of the region of applicability of the 
chosen approximation leads to the appearance of meaningless results. 
But self-consistent calculations, as is shown above, always give normal 
longitudinal fluctuations.   

The transverse fluctuations are known [185] to be much larger than the
longitudinal ones. Formally, Eq. (589) diverges when $h \ra 0$. This, 
however, does not make the transverse magnetic fluctuations 
thermodynamically anomalous. To be thermodynamically anomalous, 
expression (589) should diverge with respect to the number of particles 
$N$, or, what is the same, with respect to the system volume $V$. But 
here, it is the divergence with respect to $h$. 

Moreover, one should not forget that below the transition temperature 
$T_c$ the spin-rotation symmetry is broken. The symmetry breaking is 
described by switching on a small external magnetic field $h \neq 0$.
But then Eq. (589) is finite. Switching off this field restores the 
symmetry, as a result of which Eq. (589) would diverge, similarly to
how the compressibility of a Bose-condensed system would diverge being
incorrectly calculated without the gauge symmetry breaking. Therefore, 
as soon as the spin-rotation symmetry has been broken, when $h \neq 0$, 
all fluctuations are thermodynamically normal. And above $T_c$, where
the symmetry is not broken, one has $\eta \cong h/2T$, hence 
$\Delta^2(\hat{M}_x)/N \cong 1/4$, which is again finite for any $h$.

When some symmetry in a system is broken, the mathematically correct 
definition of statistical averages is understood in the sense of the 
Bogolubov quasiaverages [16]. Then, as is well known, one has, first, 
to accomplish the thermodynamic limit, with $N \ra \infty$ and, only 
after this, to consider the limit $h \ra 0$. In that sense, there is 
no any thermodynamically anomalous fluctuations. 

Note that real magnetic systems always possess magnetic anisotropy. 
This can be small, but never exactly zero, which corresponds to the 
presence of a finite $h$. Consequently, in real equilibrium magnetic 
systems, there are no thermodynamically anomalous fluctuations. And 
there are no thermodynamically anomalous fluctuations in any 
equilibrium system with the spontaneous breaking of any continuous 
symmetry. In the other case, such a system would be unstable and 
could not be in equilibrium.

\section{Nonground-State Condensates}

\subsection{Coherent Modes}

First of all, it is necessary to concretize what is meant under 
nonground-state condensates. The stationary equation (335) for the 
condensate function can be treated as an eigenproblem. Generally,
an eigenproblem can yield a spectrum of possible eigenvalues and 
a set of the related eigenfunctions. So, generally, the eigenproblem,
corresponding to Eq. (335), can be represented in the form
$$
\left [ -\; \frac{\nabla^2}{2m} + U(\br) 
\right ] \eta_n(\br) \; +
$$
\be
\label{591}
 +\; \Phi_0 \left [ | \eta_n(\br) |^2 \eta_n(\br) +
2\rho_1(\br) \eta_n(\br) + \sgm_1(\br) \eta_n^*(\br) + 
\xi(\br) \right ] = E_n \eta_n(\br) \;  ,
\ee
in which the minimal eigenvalue defines the chemical potential
\be
\label{592}
 \mu_0 = \min_n E_n \;  .
\ee   
When $E_n = \mu_0$, Eq. (591) corresponds to the standard ground-state
Bose-Einstein condensate, while, for higher eigenvalues $E_n$, this 
equation corresponds to nonground-state condensates. The values of 
$\rho_1({\bf r})$, $\sigma_1({\bf r})$, and $\xi({\bf r})$ depend
on the index $n$, but for short, this dependence is not shown 
explicitly.  

The condensate function describes the coherent part of the system.
In the limit of asymptotically weak interactions ($\Phi_0 \ra 0$) and 
low temperature ($T \ra 0$), when the whole system is in the coherent 
state, Eq. (591) reduces to the nonlinear Scr\"odinger equation
\be
\label{593}
\left [ - \; \frac{\nabla^2}{2m} + U(\br) \right ] \eta_n(\br)
 + \Phi_0 | \eta_n(\br)|^2 \eta_n(\br) =
E_n \eta_n(\br) \;  .
\ee
In the particular case, for $E_n = \mu_0$, it is called the 
Gross-Pitaevskii equation.

The condensate function $\eta_n({\bf r})$ is normalized to the number 
of condensed particles, as in Eq. (52). It is convenient to introduce 
the function $\varphi_n({\bf r})$ by the relation
\be
\label{594}
  \eta_n(\br) = \sqrt{N_0} \; \vp_n(\br) \; ,
\ee
so that $\varphi_n({\bf r})$ be normalized to one,
\be
\label{595}
\int | \vp_n(\br) |^2 d\br = 1 \; .
\ee
Then Eq. (591) transforms into
$$
\left [ -\; \frac{\nabla^2}{2m} + U(\br) 
\right ] \vp_n(\br) \; +
$$
\be
\label{596}
 + \; \Phi_0 \left [ N_0 | \vp_n(\br)|^2 \vp_n(\br) +
2\rho_1(\br) \vp_n(\br) + \sgm_1(\br) \vp_n^*(\br)
+ \frac{\xi(\br)}{\sqrt{N_0}} \right ] = 
E_n \vp_n(\br) \; ,
\ee
while Eq. (593), into
\be
\label{597}
 \left [ -\; \frac{\nabla^2}{2m} + U(\br) 
\right ] \vp_n(\br) + \Phi_0 N_0 | \vp_n(\br)|^2 \vp_n(\br) 
= E_n \vp_n(\br) \;  .
\ee

The solutions to Eqs. (591) and (596) define the coherent modes in 
the general case [66] and Eqs. (593) and (597), the coherent modes for
asymptotically weak interactions and temperature [186]. These coherent 
modes, first introduced in Ref. [186], correspond to nonground-state 
condensates. The properties of such coherent modes and the methods of 
their generation have been studied in a series of papers [66, 186-215]. 
A dipole coherent mode was excited in experiment [216]. These coherent 
modes are also called topological, since the nonground-state condensates,
corresponding to different coherent modes, describe particle densities 
with different spatial topology.

\subsection{Trap Modulation}

There are several requirements that are necessary for creating a 
nonground-state condensate. First of all, it is clear that such a 
condensate cannot be equilibrium. Hence, its creation requires the action 
of external time-dependent fields. Second, the system of Bose particles 
has to possess a discrete spectrum in order that it would be possible to 
distinguish the usual ground-state Bose-Einstein condensate from a 
nonground-state condensate. This means that the system is to be placed
inside a trapping potential. And, third, to transfer particles from their
ground-state to a chosen excited state, it is necessary, either to employ
a resonant field or to use rather strong pumping.   

A straightforward way of imposing external alternating fields is by 
modulating the trapping potential. Let the confining potential be composed
of two parts,
\be
\label{598}
 U(\br,t) = U(\br) + V(\br,t) \;  ,
\ee
in which the first term is a trapping potential and the second term
\be
\label{599}
V(\br,t) = V_1(\br) \cos \om t + V_2(\br) \sin \om t
\ee
realizes the modulation of this potential with frequency $\omega$. 

There exists one limitation on the spatial dependence of the modulated 
trapping potential (598). In Refs. [206,207], the {\it shape-conservation 
theorem} has been proved, showing that the trap modulation moves the 
whole condensate without changing its shape if and only if the trapping 
potential $U({\bf r})$ is harmonic, while the modulation term (599) is 
linear with respect to the spatial variables. In that case, the trap 
modulation would not be able to produce excited coherent modes. So, to 
generate these modes, one has to avoid this particular case of spatial 
dependence.    

Suppose that at the initial time $t=0$ the system has been completely 
condensed, being in the energy state $E_0 = \mu_0$. Then, to transfer 
the system to an energy state $E_n$, one has to use the alternating 
field with a frequency $\omega$ close to the transition frequency 
$\omega_n = E_n - \mu_0$. Under this {\it resonance condition}
\be
\label{600}
\left | \frac{\Dlt\om}{\om} \right | \ll 1 \qquad 
(\Dlt\om \equiv \om - \om_n ) \;  ,
\ee
it is sufficient to invoke the pumping fields of small amplitudes.

The time-dependent equation (323) for the condensate function, after
the substitution of the relation
\be
\label{601}
\eta(\br,t) = \sqrt{ N_0} \; \vp(\br,t) \; ,
\ee
similar to Eq. (594), transforms into
$$
 i \; \frac{\prt}{\prt t} \; \vp(\br,t) =
\left [ - \; \frac{\nabla^2}{2m} + U(\br,t) - \mu_0 \right ]
\vp(\br,t) \; +
$$
\be
\label{602}
 +\;  \Phi_0 \left [ N_0 \rho_0(\br,t) \vp(\br,t)
+ 2 \rho_1(\br,t) \vp(\br,t) + \sgm_1(\br,t) \vp^*(\br,t)
+ \frac{\xi(\br,t)}{\sqrt{N_0} } \right ] \; .
\ee
      
We can look for the solution to this equation represented 
[66,186,187,198] as an expansion over the coherent modes,
\be
\label{603}
\vp(\br,t) = \sum_n C_n(t) \vp_n(\br) e^{-i\om_n t} \;  ,
\ee
so that the coefficient function $C_n(t)$ be slow as compared to the 
fast oscillating exponential functions:
\be
\label{604}
 \frac{1}{\om_n} \left | \frac{d C_n}{dt} \right | \ll 1 \; .
\ee

Let us introduce the matrix elements corresponding to particle 
interactions,
\be
\label{605}  
 \al_{mn} \equiv \Phi_0 N_0 \int | \vp_m(\br)|^2 \left [
2 | \vp_n(\br) |^2 - | \vp_m(\br) |^2 \right ] d\br \; ,
\ee
and to the action of the modulating field,
\be
\label{606}
 \bt_{mn} \equiv \int \vp_m^*(\br) \left [ V_1(\br) -
i V_2(\br) \right ] \vp_n(\br) d \br \; .
\ee
Also, let us define the expression
$$
 \ep_n(t) \equiv \al_{nn} \; -
$$
\be
\label{607}
 - \; \Phi_0 \int \vp_n^*(\br) \left [
2\rho_1^{(n)}(\br) \vp_n(\br) - 2\rho_1(\br,t) \vp_n(\br) +
\sgm_1^{(n)}(\br) \vp_n^*(\br) + \frac{\xi^{(n)}(\br)}{\sqrt{N_0}}
\right ] d\br \; ,
\ee
in which the functions with the upper index $n$ correspond to the
stationary solutions characterized by the condensate function 
$\varphi_n({\bf r})$ and where  
$$
 \al_{nn} = \Phi_0 N_0 \int | \vp_n(\br) |^4 d\br \; .
$$

The trap modulation produces not only the required coherent mode 
but it also destroys the condensate by transferring particles from 
the condensate to the fraction of uncondensed particles. Therefore,
the generation of the coherent mode can be effectively done only 
during a finite {\it depletion time} $t_{dep}$, when the transfer 
from the condensate to the uncondensed fraction is yet negligible.
During this time, the variation of quantity (607) is small, such 
that
\be
\label{608}
 \left | \frac{t}{\ep_n} \; \frac{d\ep_n}{dt} \right | \ll 1
\qquad (t < t_{dep} ) \; .
\ee
It is convenient to make the change
\be
\label{609}
 C_n(t) = c_n(t) \exp [ - i \ep_n(t) t ] \;  ,
\ee
in which $\varepsilon_n = \varepsilon_n(t)$ is treated as a slow 
function of time, in the sense of inequality (608). We may notice 
that in the limit of a completely coherent system, when the fraction 
of uncondensed particles is negligibly small, Eq. (607) does not 
depend on time.

Then, we substitute expansion (603) into Eq. (602), employ the above 
notations, and invoke the averaging techniques [114,117,119,121,217-219], 
based on the existence of different time scales [220,221]. As initial 
conditions, we assume nonzero $c_n(0)$ and $c_0(0)$, while all other
coefficient functions $c_j(0) = 0$ for $j \neq 0,n$. This procedure
yields [66,186,195] the equations
$$
i \; \frac{dc_0}{dt} = 
\al_{0n} | c_n|^2 c_0 + 
\frac{1}{2}\; \bt_{0n} c_n e^{i\Dlt\om t} \; , 
$$
\be
\label{610}
 i \; \frac{dc_n}{dt} = 
\al_{n0} | c_0|^2 c_n + 
\frac{1}{2}\; \bt^*_{0n} c_0 e^{-i\Dlt\om t} \;  .
\ee
Solving these equations gives the {\it fractional mode populations}
\be
\label{611}
 p_n(t) \equiv | c_n(t) |^2 \;  .
\ee

As a concluding remark to this section, it is worth emphasizing that
expansion (603) corresponds to the diabatic representation [66,215], and 
one should not confuse it with the adiabatic representation [222], which 
is not suitable for the studied resonant process.

\subsection{Interaction Modulation}

Another way of exciting the cloud of particles confined inside a trap is 
by varying the particle interactions by means of the Feshbach-resonance 
techniques [1-3,5,34,223]. This method can also be used for generating the 
coherent modes, as has been mentioned in Refs. [206,207,210], and analyzed 
in detail in Refs. [214,215].  

Let the scattering length be modulated so that the particle interaction 
becomes time-dependent according to the law
\be
\label{612}
 \Phi(t) = \Phi_0 + \Phi_1 \cos(\om t) + \Phi_2 \sin(\om t) \; .
\ee
Following the same procedure as in the case of the trap modulation and 
introducing the notation 
\be
\label{613}
 \gm_m \equiv N_0 ( \Phi_1 - i \Phi_2) \int \vp_0^*(\br)
| \vp_m(\br)|^2 \vp_n(\br) \; d\br \; ,
\ee
in which $n$ is fixed and $m = 0,n$, we get the equations
$$
i\; \frac{dc_0}{dt} = \al_{0n} | c_n|^2 c_0 + 
\left ( \gm_0 | c_0|^2
+ \frac{1}{2}\; \gm_n | c_n|^2 \right ) c_n e^{i\Dlt\om t} +
\frac{1}{2} \; \gm_0^* c_n^* c_0^2 e^{-i\Dlt\om t} \; ,
$$
\be
\label{614}
 i\; \frac{dc_n}{dt} = \al_{n0} | c_0|^2 c_n + 
\left ( \gm_n^* | c_n|^2
+ \frac{1}{2}\; \gm_0^* | c_0|^2 \right ) c_0 e^{-i\Dlt\om t} +
\frac{1}{2} \; \gm_n c_0^* c_n^2 e^{i\Dlt\om t} \;   .
\ee

Both these ways of modulating either the trapping potential or particle 
interactions can be used for generating excited coherent modes.

Nonequilibrium systems with the generated coherent modes, representing
nonground-state condensates, possess a variety of interesting properties.
We can mention the following effects: interference patterns and 
interference currents [194,195,198], mode locking [186,198,199], dynamical
phase transitions and critical phenomena [190,194,195,198], chaotic motion
[206,207], atomic squeezing [198,201,202], Ramsey fringes [211-213], and 
entanglement production [224-226] that can be quantified by a general 
measure of entanglement production [227-229].

The above-mentioned effects can be realized by resonant alternating 
fields of rather low amplitudes. When increasing the amplitude of the 
pumping field, it becomes feasible to generate the excited coherent 
modes with the frequencies of the alternating fields, which are not 
exactly in resonance with the transition frequencies. Thus, the 
transition between the coherent modes, characterized by the transition 
frequency $\omega_{12}$, can be done by means of the harmonic generation 
and parametric conversion [206,207].   

{\it Harmonic generation} occurs, when the driving frequency $\omega$ 
satisfies the condition
\be
\label{615}
 n\om = \om_{12} \qquad (n=1,2,\ldots ) \;  .
\ee
{\it Parametric conversion} requires the use of two alternating 
fields, with the driving frequencies $\omega_1$ and $\omega_2$, such 
that
\be
\label{616}
\om_1 \pm \om_2 = \om_{12} \;  .
\ee
In the case of two pumping fields, there exists the {\it combined 
resonance} under the condition
\be
\label{617}
 n_1 \om_1 + n_2 \om_2 = \om_{12} \qquad
(n_i = \pm 1, \pm 2, \ldots ) \;  .
\ee
And, generally, the application of several external alternating 
fields, with the driving frequencies $\omega_i$, can generate 
coherent modes under the condition of {\it generalized resonance}, 
when
\be
\label{618}
 \sum_i n_i \om_i = \om_{12} \qquad 
(n_i = \pm 1, \pm 2, \ldots ) \; .
\ee 

The amplitudes of external alternating fields can be made arbitrarily 
strong. Therefore, all these effects can be realized in experiments. 
The particle interactions can also be varied in a wide range. For 
instance, employing the Feshbach resonance techniques, it is possible 
to tune the interactions of $^7$Li atoms over seven orders of magnitude 
[230]. Hence, the generation of nonground-state condensates can be 
done by the interaction modulation as well.

\subsection{Turbulent Superfluid}

As follows from the previous sections, increasing the modulation 
amplitude results in the generation of more and more coherent modes, 
whose excitation becomes more and more easy, especially when several 
alternating fields are involved. This is because it is sufficient 
that the frequencies of the modulating fields be such that one of 
the above conditions be approximately satisfied. As soon as this 
happens, the related coherent modes become excited. Intensive field 
modulation generates simultaneously several coherent modes.

When alternating fields are applied creating an oscillating anisotropy, 
with local rotation moments, then the prevailing coherent modes will 
be quantum vortices. An important feature of the vortex creation by 
means of the anisotropic trap modulation, contrary to the vortex 
creation by means of rotation, is the generation of vortices as 
well as antivortices, that is, the generation of the vortices with 
opposite rotation velocities. The oppositely rotating vortices repel 
each other and diffuse in space, separating from each other. At the 
beginning, when the amplitude modulation is not yet too strong, there 
should arise just a small number of vortices having the standard 
properties [1-3,231,232], except that vortices and antivortices both 
are present. In the case, when the whole system is uniformly rotated, 
the increased rotation frequency induces a vortex lattice [231,232]. 
Contrary to this, when the trapped system is subject to the action of 
alternating fields, nonuniformly and anisotropically shaking the 
trapped particle cloud, the created vortices possess different axes 
of rotation and different rotation velocities. Therefore no vortex 
lattice is possible. Then, increasing the amplitude of the alternating 
fields produces a large number of vortices with randomly distributed 
vorticities. Such a tangle of quantized vortices forms what is called 
{\it quantum turbulence}, and the whole system is said to be in the 
state of {\it turbulent superfluid}.

The problem of turbulent superfluid has been addressed in a number of 
works. The related literature has been reviewed in articles [233-237]
(see also recent Refs. [238-240]). There are plenty of experiments 
observing quantum turbulence in liquid $^3$He and $^4$He. Quantum 
turbulence in trapped gases has also been observed [241]. The 
description of turbulent superfluids as continuous vortex mixtures 
has been advanced [237].

\subsection{Heterophase Fluid}

Increasing further the amplitude of the alternating field breaks 
the turbulent superfluid into spatially separated pieces, with 
Bose-condensed droplets separated by normal, nonsuperfluid, spatial
regions. Such a state reminds the Bose glass, or granular condensate,
considered in Sec. 9.5. But now it is a highly nonequilibrium state. 
This state is analogous to heterophase mixtures consisting of several 
randomly intermixed phases [54]. Thence, it is called 
{\it heterophase fluid}. 

An external modulating field acts on the system similarly to the 
action of a spatial random potential [110], such as treated in Sec. 9. 
The possibility of mapping the system with a time-dependent modulation 
to the system with a spatially random potential is very important, 
since it allows us to understand the behavior of modulated 
nonequilibrium systems by comparing them with equilibrium random 
systems. The proof of this mapping is as follows.  

Let the system Hamiltonian
\be
\label{619}
H(t) = H_0 + \hat V(t)
\ee
consist of the usual term $H_0$, containing no time-dependent fields,
and a term
\be
\label{620}
 \hat V(t) = \int \hat\psi^\dgr(\br) V(\br,t) 
\hat\psi(\br) \; d\br \;  ,
\ee
with an external potential depending on time. The characteristic 
variation time $t_{mod}$ of the modulating potential $V({\bf r}, t)$
is assumed to be much longer than the local-equilibrium time $t_{loc}$,
but much shorter than the time of experiment $t_{exp}$,
\be
\label{621}
 t_{loc} \ll t_{mod} \ll t_{exp} \; .
\ee

The modulating potential pumps energy into the system that can be 
associated with the effective temperature
\be
\label{622}
T^* \equiv \frac{1}{N} \int_0^{t_{exp}} \left |
\left \lgl \frac{\prt\hat V(t)}{\prt t} \right \rgl \right | \; dt \; .
\ee
If the pumping potential is periodically oscillating with a frequency 
$\omega$ and period $t_{mod} = 2 \pi / \omega$, then
$$
\frac{\prt\hat V(t)}{\prt t} = \om \hat V(t) = 
\frac{2\pi}{t_{mod}}\;  \hat V(t) \; .
$$
In this case, the effective temperature (622) is
\be
\label{623}
 T^* = \frac{2\pi}{Nt_{mod}} 
\int_0^{t_{exp}} | \;\lgl \hat V(t) \rgl\; | \; dt \; .
\ee
Denoting the amplitude of the modulating potential $V({\bf r}, t)$ as
$V_{mod}$, we have
$$
|\; \lgl \hat V(t) \rgl\; | \approx N V_{mod} \; .
$$
Therefore, the effective temperature (623) becomes
\be
\label{624} 
T^* \approx 2\pi \; \frac{t_{exp}}{t_{mod}} \; V_{mod} \;  .
\ee
One may notice that the effective temperature depends on $t_{exp}$, 
though this dependence is week, in the sense that 
$$
\frac{t_{mod}}{T^*} \left | \frac{\prt T^*}{\prt t_{exp}} 
\right | \; \leq \; \frac{t_{mod}}{t_{exp}} \; \ll \; 1 \;  .
$$

Under the slow modulation, such that $t_{mod} \gg t_{loc}$, the system, 
at each moment of time, is in quasiequilibrium. Consequently, one can 
define the local in time thermodynamic potential
\be
\label{625}
 \Om(t) = - T^* \ln { \rm Tr } \exp\{ - \bt^* H(t) \} \; ,
\ee
where $\beta^* = 1/ T^*$. Because $t_{exp} \gg t_{mod}$, we are
interested not in the local potential (625) but in the coarse-grained
potential
\be
\label{626}
\Om = \frac{1}{t_{mod}} \int_0^{t_{mod}} \Om(t) \; dt \;  ,
\ee
averaged over oscillations that are fast as compared to $t_{exp}$.

At each moment of time $t$ the potential $V({\bf r}, t)$ describes 
a spatial potential. This can be characterized by the relation
\be
\label{627}
 V(\br,t) = \xi(\br) \; ,
\ee
which defines the functional
\be
\label{628}
t = t[\xi(\br) ] \;  .
\ee    
Equations (627) and (628) symbolize the fact that for each time $t$
there corresponds a spatial potential $\xi({\bf r})$ and vice versa, 
a potential $\xi({\bf r})$ is ascribed to time $t$. The relation between
the interval $[0,t_{mod}]$ and the topological space $\{\xi({\bf r})\}$,
without much loss of generality, can be taken as homeomorphic. 

The variation of time is equivalent to the variation of the spatial 
potential, so that
$$
 dt = \frac{\dlt t[\xi(\br)]}{\dlt\xi(\br)} \; \dlt\xi(\br) \; .
$$
With relation (628), Hamiltonian (619) becomes the functional
\be
\label{629}
H[\xi(\br)] = H_0 + 
\int \hat\psi^\dgr(\br) \xi(\br) \hat\psi(\br) \; d\br 
\ee
of the spatial field  $\xi({\bf r})$. Therefore, the averaged 
thermodynamic potential (626) takes the form
\be
\label{630}
\Om = - T^* \int \ln {\rm Tr} \exp \{ - \bt^* H[\xi(\br)] \} \;
{\cal D}\xi(\br)  .
\ee
The latter is equivalent to the thermodynamic potential of an equilibrium 
system in a random external field.
   
If the external alternating field has an amplitude $V_{mod}$ and the whole 
trapped system is subject to the modulation, then the modulation amplitude 
$V_{mod}$ plays the role of the correlation amplitude $V_R$ and the 
effective trap length $l_0$, of the correlation length $l_R$ in Eq. (459). 
The effective trap length $l_0 = 1/\sqrt{m \omega_0}$ and the effective 
trap frequency $\omega_0$ are defined in Eq. (532).

More strictly, the system state, produced by the trap modulation, 
depends not merely on the modulation amplitude $V_{mod}$, but on the 
amount of the total energy pumped into the system, playing the role of 
the effective temperature in the nonequilibrium system [237]. For an 
alternating field, with the driving frequency $\omega$ and the related 
period $t_{mod} = 2\pi/\omega$, acting on the system during the total
time of experiment $t_{exp}$, the pumped energy is associated with the
effective temperature (624). This energy should be treated as the effective 
modulation amplitude. Thus, instead of the localization length (460), we get
\be
\label{631}
 l_{loc} = \frac{\hbar^4}{m^2 V_0^2 l_0^3} =
\left ( \frac{\hbar\om_0}{V_0} \right )^2 l_0 \; ,
\ee
where, for clarity, the Planck constant is restored.
 
Depending on the relation between the localization length (631) and 
the trap length $l_0$, there can exist the following states:
\begin{eqnarray}
\label{632}
\begin{array}{ll}
l_{loc} > l_0     & \quad (superfluid) \; , \\
a < l_{loc} < l_0 & \quad (heterophase \; fluid)\; , \\
 l_{loc} \leq a   & \quad (chaotic\; fluid ) \; .
\end{array}
\end{eqnarray}
The superfluid state here includes all types of superfluids, the regular 
superfluid having no vortices, the vortex superfluid with a small number 
of vortices, and the turbulent superfluid with a random tangle of many 
vortices. This classification, in terms of the pumped energy, reads as 
follows: 
\begin{eqnarray}
\label{633}
\begin{array}{ll}
V_0 < \hbar\om_0     & \quad (superfluid) \; , \\
\hbar \om_0 < V_0 < \hbar \om_0  \sqrt{ l_0/a} & 
\quad (heterophase \; fluid)\; , \\
V_0 \geq \hbar\om_0 \sqrt{l_0/a}  & 
\quad (chaotic\; fluid ) \; .
\end{array}
\end{eqnarray}

Chaotic fluid is a strongly fluctuating system having neither long-range
order nor even local order. It resembles the state of weak turbulence 
[242] or the chaotic state [243]. Qualitatively, the overall scheme, 
representing the sequence of states arising under the action of an 
alternating field, with respect to the amount of the pumped energy $V_0$, 
is shown [237] in Figure 1.

After the external modulation field is switched off, a finite quantum 
system relaxes to its equilibrium state during the relaxation time 
defined by particle collisions, the trap size, and trap shape [244].
The relaxation time becomes quite long for quasi-one-dimensional traps, 
where it may last, without equilibration for thousands of collisions 
between the oscillating Bose-condensed droplets [245]. This is because 
the one-dimensional system with local interactions is the integrable 
Lieb-Liniger system [246,247]. And quasi-one-dimensional systems, 
approaching integrability, display very long equilibration times.

In a quasi-one-dimensional trap, collisions, restricted to the motion 
of particles in the axial direction, with the particles remaining in 
the same transverse ground state, are not accompanied by energy change, 
hence, do not lead to thermalization. For such two-body collisions, 
equilibration and thermalization occurs only under transverse excitations. 
The corresponding rate of populating the radially excited modes by 
pairwise collisions, can be estimated from the Fermi golden rule that,
at low temperature $T < \omega_{\bot}$, gives [248,249] the rate
\be
\label{634}
\Gm_2 \approx 2.8 \om_\perp \zeta \exp ( -2 \om_\perp/T) \; ,
\ee
where the dimensionless parameter 
$$
\zeta \equiv \rho_{1d}\; \frac{a_s^2}{l_\perp}
$$
is expressed through the three-dimensional scattering length $a_s$, 
transverse oscillator length $l_{\bot}$, and the one-dimensional density
$$
 \rho_{1d} = \rho\pi l^2_\perp = \frac{N}{2l_z} \; .
$$
At temperature tending to zero, this rate is exponentially suppressed.
However, it can be essential for finite temperatures.

For quasi-one-dimensional traps at low temperatures, the three-body 
collision rate [249] can become important,
\be
\label{635}
 \Gm_3 \approx 6.9 \zeta^2 \om_\perp \; .
\ee

The states, described above, have been experimentally realized by 
means of the trap modulation [241,250]. The whole diagram, showing 
the dependence of the produced states on the modulation amplitude 
and modulation time has been presented [250], starting from the 
regular superfluid, through vortex superfluid, to turbulent superfluid, 
and to heterophase fluid.

\section{Conclusions}

In this review, the basic theoretical problems have been considered,
arising in the description of systems with Bose-Einstein condensate.  
The solutions to these problems are elucidated. The main conclusions 
can be briefly summarized as follows.

(i) The global gauge symmetry breaking is the necessary and sufficient 
condition for Bose-Einstein condensation. This is an exact mathematical 
fact. The symmetry breaking results in the appearance of both, the 
condensate fraction and anomalous averages. The latter cannot be 
neglected without destroying the theory self-consistency. Omitting the 
anomalous averages is principally wrong, yielding unreliable and often
unreasonable results.

(ii) The Hohenberg-Martin dilemma of conserving versus gapless theories
is resolved by introducing two Lagrange multipliers guaranteeing the 
validity of two normalization conditions, for the numbers of condensed 
and uncondensed particles. The use of these two Lagrange multipliers is 
necessary as soon as the global gauge symmetry has been broken. 

(iii) Bose-Einstein condensed systems in strong spatially random 
potentials can be described by means of the method of stochastic 
decoupling. Perturbation theory with respect to the strength of disorder
can fail, leading to incorrect conclusions.

(iv) Thermodynamically anomalous fluctuations of any observable 
quantities are strictly prohibited in all equilibrium statistical 
systems, irrespectively of the used representative statistical ensemble.
Thermodynamically anomalous particle fluctuations, of either condensed 
or uncondensed particles, cannot exist in Bose-condensed systems. The 
occurrence of thermodynamically anomalous fluctuations can be due only 
to calculational mistakes.

(v) The method has been suggested of generating nonground-state 
condensates of trapped particles. The method can be realized by applying
alternating external fields modulating either the trapping potential or 
particle interactions. This makes it possible to create different types 
on nonground-state condensates, such as coherent modes, turbulent 
superfluids, and heterophase fluids.

\vskip 5mm
{\bf Acknowledgments}

\vskip 2mm
I am very much grateful for many useful discussions and permanent collaboration
to V.S. Bagnato and E.P. Yukalova. Financial support from the Russian 
Foundation for Basic Research is appreciated.

\newpage

\begin{figure}[h]
\centerline{\includegraphics[width=16cm]{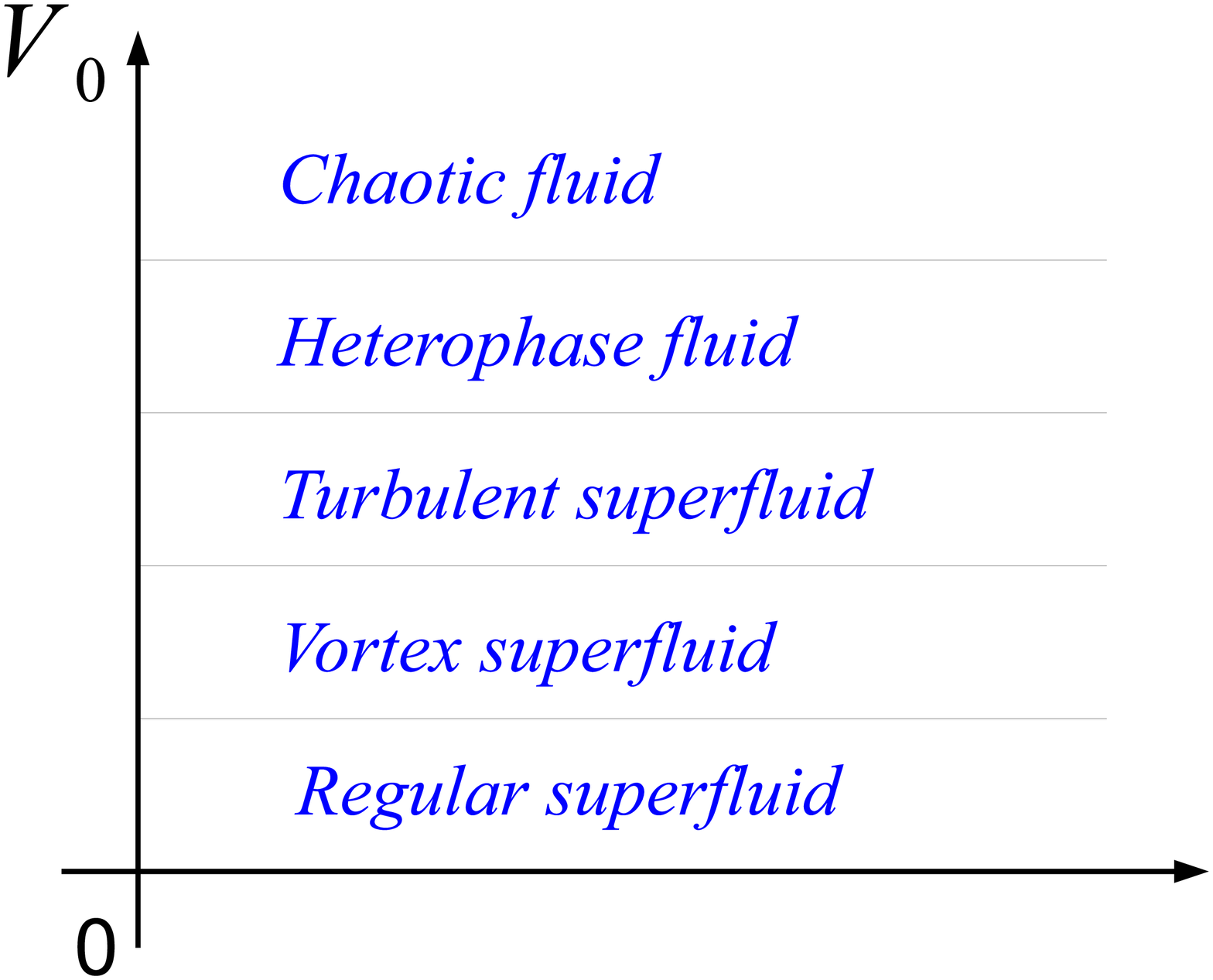}}
\caption{Scheme of the sequence of states for a trapped Bose-condensed
system subject to the action of an alternating external field, with
the increasing pumped energy $V_0$.}
\label{fig:Fig.1}
\end{figure}

\end{document}